%% file: paper.tex
\documentclass[twocolumn,5p,times]{article}
\usepackage{graphicx}
\usepackage{lscape,graphicx}
\usepackage{hhline}
\usepackage{multicol}
\usepackage{colordvi}
\usepackage{dcolumn}
\usepackage[fleqn]{amsmath}
\usepackage[mathscr]{euscript}
\usepackage[usenames, dvipsnames]{color}
\usepackage[usenames,dvipsnames,svgnames,table]{xcolor}
\usepackage[linktocpage]{hyperref}
\usepackage[a4paper,bindingoffset=0.2in,left=0.5in,right=0.5in,top=1.0in,bottom=1.0in,footskip=0.25in]{geometry}
\newcommand*{\doi}[1]{\href{http://dx.doi.org/#1}{doi: #1}}
\usepackage[numbers]{natbib}
\makeatletter
\renewcommand{\section}{\@startsection{section}{1}{\z@}{-3.5ex \@plus -1ex \@minus -.2ex}{1.3ex \@plus.2ex}{\normalfont\small\bfseries\boldmath}}
\makeatother

\makeatletter
\renewcommand{\subsection}{\@startsection{subsection}{2}{\z@}{-3.5ex \@plus -1ex \@minus -.2ex}{1.3ex \@plus.2ex}{\normalfont\small\bfseries\boldmath}}
\makeatother

\makeatletter
\renewcommand{\subsubsection}{\@startsection{subsubsection}{3}{\z@}{-3.5ex \@plus -1ex \@minus -.2ex}{1.3ex \@plus.2ex}{\normalfont\small\bfseries\boldmath}}
\makeatother

\makeatletter
\everymath{\if@display\else\thickmuskip=0mu plus 0mu\fi}
\makeatother

\title{\large \bf {\tt ee$\in$MC}: Simulation of $\bf e^{+}e^{-} \to \mu^{+} \mu^{-} (\gamma) $ and $\bf e^{+}e^{-} \to \tau^{+}\tau^{-} (\gamma) $ Events }
\date{}

\author{\normalsize Ian M. Nugent$^{*}$ \\ \normalsize Victoria, B.C., Canada}

\begin{document}
\twocolumn[
  \begin{@twocolumnfalse}
    \maketitle
\begin{abstract}
We present a new Monte-Carlo generator, {\tt ee$\in$MC}, for the simulation of $e^{+}e^{-} \to \mu^{+} \mu^{-} (\gamma) $, 
$e^{+}e^{-} \to \tau^{+}\tau^{-} (\gamma) $ and $\tau$  lepton decays which is suitable for $e^{+}e^{-}$ 
colliders from threshold up to energy of $\sim 10.58GeV$. The $e^{+}e^{-} \to \mu^{+} \mu^{-}(\gamma) $, $e^{+}e^{-} \to \tau^{+}\tau^{-}(\gamma)$ 
interactions are computed up to the exact LO matrix element
where the running of $\alpha_{QED}$ includes both the LO leptonic and hadronic vacuum polarization.  
Infra-red divergences are investigated within the Yennie-Frautschi-Suura Exponentiation formalism,
where several well-known approximations and a formulation based on the exact LO contribution to the soft and virtual photons are applied.
% It is found that at low energies,
%the theoretical predictions are sensitive to the approximations applied which leads to important consequences for the predicted cross-sections and
%the hadronic vacuum polarization contribution to $g-2$.
%The soft and virtual photon contribution in the Yennie-Frautschi-Suura Exponentiation Form-Factor 
%is computed at exact LO for spin $\frac{1}{2}$ particles along with the more common ultra-relativistic and the scalar-QED approximations. 
The simulation of the $\tau$ lepton consists of the leptonic decay modes at Born level and theoretical models based on Flux-Tube Breaking Models and 
Chiral Resonance Lagrangian Models as well as experimental models for the majority of semi-leptonic decay modes. This includes hadronic models for
investigating the origin of low mass scalar sector recently observed in decays of the $\tau$ lepton and their association with the origin of the constituent-quark mass
through Chiral-symmetry Breaking. 
\end{abstract}
\end{@twocolumnfalse}
]
%\begin{keyword}
%Tau Lepton \sep Electron-positron Collider \sep Monte-Carlo Simulation
%\end{keyword}

%\end{frontmatter}
%\linenumbers
\renewcommand{\thefootnote}{\fnsymbol{footnote}}
\footnotetext[1]{Corresponding Author \\ \indent   \ \ {\it Email:} inugent.physics@outlook.com}
\renewcommand{\thefootnote}{\arabic{footnote}}
\section{Introduction}

Monte-Carlo (MC) simulations play a pivotal role both as a technique to compare theoretical predictions to experiments and to facilitate the interpretation of the measured 
experimental results in particle physics. Consequently, the physics phenomena which can be probed experimentally depend on the theoretical construct and physics phenomena 
included in the Monte-Carlo simulation. 
On most experiments, the $\tau$ lepton is typically simulated with {\tt TAUOLA}  \cite{Jadach:1990mz} and on $e^{+}e^{-}$ colliders
in conjunction with {\tt KORALB}  \cite{Jadach:1990mv} or {\tt KK2F}  \cite{kk2f} using primarily Chiral Resonance Lagrangian (ChRL) based models. These generators were 
originally developed over 30 years ago for the LEP experiments which probed an invariant mass around the Z boson where the ultra-relativistic approximation is
valid for $\tau$-pair production  \footnote{ More recently, 
the {\tt Pythia8} collaboration has added the $\tau$ decays to their MC generator emulating the decay modes in {\tt TAUOLA} \cite{Ilten:2014}, where the spin correlations
are included through the helicity density matrix by means of the algorithms developed in  \cite{Collins:1987cp,Knowles:1988hu,Richardson_2001}.}.
 {\tt TAUOLA}  \cite{Jadach:1990mz} has supported the community since its inception and has played a pivotal role in  polarization measurements of the Z-boson 
 \cite{2006257,Stahl:2000aq} to hadronic  \cite{Braaten:1991qm}, Electro-weak couplings  \cite{2006257,Stahl:2000aq} and renormalization and the running of the 
strong coupling constant $\alpha_{S}$  \cite{Davier:2008sk}. More recently, the spin dynamics of {\tt TAUOLA}  \cite{Jadach:1990mz} were essential for the 
measurements of the spin and CP of the Higgs  \cite{2020135426,PhysRevD.100.112002,Czyczula:2012} at hadron collider experiments. 
The MC generator presented here is designed to facilitate measurements for a broad range of physics phenomena accessible through
$e^{+}e^{-} \to \mu^{+} \mu^{-} (\gamma) $, $e^{+}e^{-} \to \tau^{+}\tau^{-} (\gamma) $ and decays of the $\tau$ lepton at electron-positron colliders with energies from
threshold up to a centre-of-mass energy of $\sim 10.58GeV$ where the produced $\tau$ leptons are not entirely ultra-relativistic.
The $e^{+}e^{-} \to \mu^{+} \mu^{-} \gamma$ interaction is essential for the Initial-State-Radiation Method 
used in determining the hadronic contribution of the muon's anomalous magnetic moment  \cite{Binner:1999bt,Czyz:2000wh,Jegerlehner,Lees:2012cj,Babusci_2013,Campanario:2013uea,2016626,Davier:2017zfy}. This MC generator incorporates many of the physics phenomena required to probe
polarization in $e^{+}e^{-} \to \tau^{+}\tau^{-} (\gamma) $  \cite{2006257,Stahl:2000aq}, to measure the hadronic vacuum polarization  \cite{Jegerlehner,Alemanu_1998,Davier:1997vd,Davier:2002dy,Benayoun:2015gxa,LopezCastro:2015cja,Pich:2013lsa,Bruno:2018ono,Miranda:2020wdg,Aoyama:2020ynm}, hadronic and Electro-weak couplings
 \cite{Stahl:2000aq,Braaten:1991qm,Davier:2008sk,Maltman_2001}, renormalization and the running of the strong coupling constant $\alpha_{S}$
 \cite{Braaten:1991qm,Davier:2008sk,Maltman_2001}, which have been a focus of research 
involving the $\tau$ lepton. The physics phenomena and modelling has been extended, in particular, for hadronic interactions to allow for the investigation of 
additional phenomena and answer questions about the physical structure of the strong interaction. 
The $\tau$ lepton has a mass situated near the perturbative non-perturbative threshold for QCD
 \cite{Stahl:2000aq,Braaten:1991qm,Davier:2008sk} making it a unique probe for experimentally exploring the dynamics of low energy QCD.
Within the non-perturbative regime, the dynamics of the hadronic decay structure can not be calculated perturbatively and are not fully
understood\footnote{There has been impressive progress with
Lattice calculations over recent years, however there has been little interaction between Lattice calculation and experimental $\tau$ physics because there is no
direct method to connect theoretical calculations to the experimentally accessible distributions such as spectral-density function distribution or the decay structure of
the resonances in the non-perturbative regime. The construct of the Flux-Tube Breaking Model \cite{Isgur:1988vm} provides an opportunity, albeit model dependent,
to connect the recent advances from Lattice QCD to the experimental measurements.}.
As a consequence of having no clear methodology to connect the theoretical description of the strong force to the experimentally accessible measurements in the
non-perturbative regime there are many phenomenological models which are constructed on a range of axioms. Between these models there are 
divergences in the interpretations of the resonances, the available decay modes and even what the composition and quantum numbers are or how to interpret 
some of the states observed.
Hadronic decays of the $\tau$ lepton, owing to the suppression of second class currents, $SU(3)_{f}$ flavour symmetry  \cite{Suzuki:1993,Asner:2000nx}, provide access to
a mixture of hadronic resonance states which are relatively highly suppressed in hadronic colliders
 \cite{Alexeev:2719624,PhysRevLett.115.082001} and thereby offer a complementary means to probe these hadronic resonances and their mixing  \cite{Suzuki:1993,Asner:2000nx}.
One of the fundamental unanswered questions is what is the
origin of the constituent-quark mass, which makes up the bulk of the mass in hadrons, and how is it generated through the strong interaction?
Recently, significant contributions which have been interpreted in terms of low mass hadronic scalars, $\sigma/f_{0}(500)$ and $\kappa/K_{0}^{*}(700)$,
have been observed in $\tau$ decay  \cite{CLEO3pi,Edwards:1999fj,BONDAR2002139,PS2010,Nugent:2013hxa}. 
These low mass scalar states have been associated with the origin of the constituent-quark mass,
%\footnote{The meson and baryon mass come from two components: 
%the quark mass from the ``Higgs'' mechanism in Electro-weak symmetry breaking;
%and the constituent-quark mass which is the mass contribution associated with binding energy from the strong interaction \cite{PDG2020}.}, 
however this interpretation is model dependant and there
is debate on the exact nature of these states  \cite{PDG2020,PhysRevD.59.074001,PEL_EZ_2004,Dabado_1997,Oller_2003,Tornqvist_1999,Napsuciale_2004,Napsuciale_2004b,Ishida_1999,Sadron_1999,Fariborz_2014,Black_1999}.
The Chiral Lagrangian Models \cite{PhysRevD.59.074001,PEL_EZ_2004,Dabado_1997,Oller_2003} and Linear Sigma Models 
  \cite{Tornqvist_1999,Napsuciale_2004,Napsuciale_2004b,Ishida_1999,Sadron_1999,Fariborz_2014,Black_1999} are examples of models where low mass scalar resonances 
play a crucial role in the formation of the constituent-quark mass
  \cite{PDG2020,PhysRevD.59.074001,PEL_EZ_2004,Dabado_1997,Oller_2003,Tornqvist_1999,Napsuciale_2004,Napsuciale_2004b,Ishida_1999,Sadron_1999,Fariborz_2014,Black_1999}.
Non-Linear Sigma Models  \cite{GellMann:1960}, including the Skyrme Model \cite{Skyrme:1962vq,Skyrme:1962vh,Speight:2018} and Nambu-Jona-Lasinio Model  \cite{Nambu:1961}
also generate mass through a Chiral-symmetry Breaking mechanism which is associated with the low lying scalar ``$\sigma$'' meson but can incorporate the vector and 
pseudo-scalar states into the symmetry breaking  \cite{PDG2020,Speight:2018}.
While, in other theoretical constructs the mass generation is not necessarily directly related to the $\sigma$ meson or the Chiral-symmetry Breaking mechanism, 
for example the non-relativistic quark models 
 \cite{Godfrey:1985xj,Kokoski:1985is,Mitra:1966,Capstick:2000qj}, Bag Model  \cite{Chodos:1974a,Chodos:1974b,DeGrand:1975} and String Models \cite{Maldacena:1997re}. 
Many of these models, including ${^{3}P_{0}}$ quark models, do not even interpret the low mass scalars as resonant states.
A comparison of the alternative models and theoretical predictions with experimental data is vital to elucidate the
underlying structure in nature and to determine the correct interpretation of these ``scalar'' hadronic contributions observed in $\tau$ decays and if they are
related to the origin of the constituent-quark mass.
While that latter question is more broad, even less is known about the resonance structures when investigated in finer detail.
Hadrons are composite objects with a finite size and shape, this impacts the decay structure \cite{Isgur:1988vm,Asner:2000nx} of the mesons. 
This consequently has implications for the application of time-order perturbation theory and the assumptions on which Z-graphs are dominant in the formation 
of the meson resonances \cite{Isgur:1988vm} including whether the hadronic resonances propagators can be described as ``on-shell'' or ``off-shell'' resonances 
 \cite{Isgur:1988vm,Finkemeier:1995sr}\footnote{For narrow resonances, the ``on-shell'' or ``off-shell'' provide similar predictions since variations occur far 
away from the physical resonance mass. The approximation that a hadronic resonance can be treated as a narrow resonance is not necessarily valid for broad 
resonances such as the $a_{1}(1260)$ or $K_{1}(1270/1400)$ states.}. 
The structure of the hadronic resonances, in particular for broad resonances, is directly related to the dependency of the decay 
width on the opening of available decay channels as a function of 
$\sqrt{s}$. This connection between the resonance structure and the mass dependence of the width also has implications for the effective mass of the resonance 
through causality. Although causality is implemented through dispersion relations, within the literature, there is no unique formalism for the decay width dependence on 
available decay channels and the causal relation  \cite{Isgur:1988vm,Gounaris:1968mw,Kuhn:1990ad,Tornqvist:1987ch} but instead they depend on the assumption 
made on either a point-like or finite meson size/coupling.  The assumption of point-like or finite meson size also impacts the  emission of Final-State-Radiation in the hadronic interactions \cite{LopezCastro:2001apj,FloresTlalpa:2006gz,FloresTlalpa:2006gf,FloresTlalpa:2006gs,Lees:2015qna}.
With the large dataset at the next generation B-factories, BELLE-II in Japan, which is expected to accumulate more than 45 
billion $\tau$-pair events  \cite{abe2010belle,Kou_2019}, an improved Monte-Carlo generator which includes these phenomena is essential for probing and interpreting these 
questions in the data. 

The Monte-Carlo generator is constructed as a stand-alone package using an object-oriented design in C++ containing the
random number generators, the phase-space simulation and the physic models.  The random number generators included
within the generator are: the Merseene Twister 32bit  \cite{MT32} and 64bit  \cite{MT64Tab,MT64F2} and
Marsaglia's {\tt xorshift64} and {\tt xorshift1064} generators  \cite{xorshift} where the seed for the random number
generators are initialized using  \cite{Knuth:1981}. The inclusion of 64bit random number generators, in addition to the more typical 32bit routines used in
high energy physics, allows for finer granularity in the pseudo-random number generation more inline with the floating point precision employed in the 
simulation.
The phase-space simulation and sampling are described in Section \ref{sec:PS}. The physics modelling for the QED interactions 
$e^{+}e^{-} \to \mu^{+} \mu^{-} (\gamma)$, $e^{+}e^{-} \to \tau^{+}\tau^{-} (\gamma)$ 
events is presented in Section \ref{sec:QED}, while Section \ref{sec:Tau} covers the Electro-weak and hadronic models corresponding to the 
leptonic and semi-leptonic $\tau$ decays. The impact of radiative corrections and the polarization measurements are discussed
in Section \ref{sec:QEDSpin}.  The conclusions are summarized in Section \ref{sec:Conclusion}.

\section{Phase-Space \label{sec:PS}}
  The differential phase-space distributions, $dPS$, is simulated with the recursive mass formulation 
from  \cite{Byckling:1969} modified to 
include the Jacobian normalization factors for computing the cross-section or decay rate\footnote{The phase-space is programmed to ensure there 
are no numerical issues related to massless particles allowing for an arbitrary ordering of particle masses.}. 
The efficiency of the simulation is optimized by embedding importance sampling \cite{Lopes:2006,Gelman:2014} into the 
phase-space generator to modify the acceptance envelope to approximate the structure of the physics model. 
This optimization yields a stable acceptance envelope with an efficiency of $\sim 10-20\%$ for events with a hard ISR/FSR photon, 
a invariant mass $\sqrt{s}$ from threshold to $10.58GeV/c^{2}$ and a 
soft-photon cut-off energy of $0.1-0.001GeV$ for $e^{+}e^{-} \to \mu^{+} \mu^{-} \gamma$ and $e^{+}e^{-} \to \tau^{+}\tau^{-} \gamma$ interactions\footnote{The 
leptons are generated for the full phase-space, while the hard photon at LO is generated for the full phase-space excluding the region forbidden by the soft-photon cut-off. At higher orders, the full phase-space is simulated and the soft-photon cut-off is applied for each individual Feynman diagram.}\footnote{The exact efficiency depends on the numerical padding applied to the pre-sampling for determining the acceptance envelope.}. 
For narrow resonances, an alternative procedure is required for efficiently determining an ordered vector of particle masses.
Instead of generating a random vector for the masses as in  \cite{Byckling:1969}, the masses are generated within the upper and lower mass extrema 
where a mass of the narrow resonance is sampled using a Breit-Wigner distribution by 
means of the integral method  \cite{Lopes:2006}, $x=M+M\Gamma \tan\left(\pi\left(u-\frac{1}{2}\right)\right)$ 
 \cite{Jadach:1993hs}.  % \cite[Eq. 17]{Jadach:1993hs}
This means that the efficiency is reduced by the ratio of allowable kinematic phase-space $(N-3)!$, 
where $N$ is the number of particles, instead of the width of narrow resonance. 

\section{QED Interactions: $e^{+}e^{-} \to l^{-} l^{+} (\gamma) $ where $l=\mu$ or $\tau$\label{sec:QED}}

The QED interactions $e^{+}e^{-} \to l^{-} l^{+} (\gamma)$ where $l=\mu$ or $\tau$ are calculated for the Born level and the LO
 ISR and FSR diagrams  \cite{Mandl:1985bg,Peskin:1995ev} with interference.  The differential QED cross section can be expressed as  
$d\sigma= \frac{|\bar{{\mathcal M}}|^{2}}{4(|\vec{P}_{e^{-}}|E_{e^{+}}+E_{e^{-}}|\vec{P}_{e^{+}}|)}\times dPS$ 
 \cite{Halzen:1984mc}, %\cite[4.29]{Halzen:1984mc}
 where $dPS$ is the differential phase-space and the spin-averaged sum matrix element, $|\bar{{\mathcal M}}|^{2}$, is constructed explicitly in terms of the 
perturbative QED expansion of the Feynman Diagram. The Feynman Diagrams are implemented in terms of the
Dirac-Spinor and $\gamma$-Matrices using an object orientated formalism for the mathematic structures\footnote{The 
object-oriented formulation was inspired by  \cite{Ilten:2014}.}. The exponentiation, which subtracts off the divergencies from radiative 
corrections, see Section \ref{sec:YFS}, is applied as a multiplicative factor to the matrix element.

\subsection{Spinor and Polarization Vector Formalism} 

The perturbative QED calculations for $e^{+}e^{-} \to l^{-} l^{+} (\gamma)$ are written in terms of the Feynman Diagram for Born level and LO radiative corrections
 \cite{Peskin:1995ev,Halzen:1984mc,Ryder:1985wq,steane:2013wra}
constructed directly from the $\gamma$-Matrices and Dirac-Spinor. Both Weyl and Dirac-Pauli representations are 
included along with the rotation between representations \cite{Manogue}.
%\footnote{To allow for a straight forward implementation of the polarization for each Dirac-Spinor, the
%spinors are constructed in the Weyl representation and then rotated to the Dirac Representation if required.}. 
In the Weyl representations, the Dirac-Spinors can be
represented in terms of the transformation properties \cite{Ryder:1985wq} and normalized zero momentum free wave helicity states \cite{Peskin:1995ev}, 

\begin{equation}
\resizebox{0.35\textwidth}{!}{$ 
u(p)
=
\left(\begin{array}{l}
\left[I \sqrt{\frac{E+M}{2}} - \vec{\mathbf \sigma} \cdot \hat{\mathbf P} \sqrt{\frac{E-M}{2}}\right] \chi\\
\left[I \sqrt{\frac{E+M}{2}} + \vec{\mathbf \sigma} \cdot \hat{\mathbf P} \sqrt{\frac{E-M}{2}}\right] \chi
\end{array} \right)
$}
\end{equation}

\begin{equation}
\resizebox{0.35\textwidth}{!}{$ 
v(p)
=
\left(\begin{array}{l}
\left[I \sqrt{\frac{E+M}{2}} - \vec{\mathbf \sigma} \cdot \hat{\mathbf P} \sqrt{\frac{E-M}{2}}\right] \chi\\
-\left[I \sqrt{\frac{E+M}{2}} + \vec{\mathbf \sigma} \cdot \hat{\mathbf P} \sqrt{\frac{E-M}{2}}\right] \chi
\end{array} \right)
$}
\end{equation}

\noindent where

\begin{equation}
\resizebox{0.45\textwidth}{!}{$ 
\chi(\uparrow)=\left(\begin{array}{l}
cos\left(\frac{\theta}{2}\right)e^{-\imath\frac{\phi}{2}} \\
sin\left(\frac{\theta}{2}\right)e^{\imath\frac{\phi}{2}}
\end{array} \right),\mbox{  } \chi(\downarrow)=\left(\begin{array}{l}
-sin\left(\frac{\theta}{2}\right)e^{-\imath\frac{\phi}{2}} \\
cos\left(\frac{\theta}{2}\right)e^{\imath\frac{\phi}{2}}
\end{array} \right)
$}
\label{eq:spinorRyder}
\end{equation}

\noindent \cite{steane:2013wra}. $\sigma^{\mu}$ is a Lorentz-vector constructed from the Pauli-matrices $(I,\vec{\mathbf\sigma})$ and $\hat{\mathbf {P}}$ is the unit normal for the momentum vector \cite{Peskin:1995ev,Ryder:1985wq} for both massive and massless Dirac-particles. 
The Dirac-Spinor formulation
enables the initial state to be configured and summed over for an arbitrary polarization state, allowing 
for the simulation of both polarized and unpolarized beams.
The simulated $\sigma(e^{+}e^{-}\to\mu^{+}\mu^{-})$  cross-section is compared to theoretical prediction  \cite{Smith1994117} in
Figure \ref{fig:CS_mumu}, demonstrating an accuracy of the spinor formalism in the Monte-Carlo simulation at the per-mill level. 
The photons are constructed from the circular polarization vectors corresponding to the photon helicity states in the non-covariant Coulomb Gauge 
\cite{Mandl:1985bg,Peskin:1995ev,Halzen:1984mc,Walecka:2013}. 
%The $\sigma(e^{+}e^{-}\to\mu^{+}\mu^{-}(\gamma))$ cross-section as a function of $\mu^{+}\mu^{-}$ invariant mass is presented in Figure \ref{fig:CS_mumu_Rad}.  

\subsection{Renormalization and $\alpha_{QED}$\label{sec:Renormilization}}
Renormalization is included through the running of the coupling                          
constant  \cite{Jegerlehner,Sturm_2013} by means of Wards Identity \cite{Mandl:1985bg} % \cite[pg 197-198]{Mandl:1985bg} 
for the first order leptonic  \cite{Sturm_2013}           
and hadronic vacuum  \cite{Jegerlehner} polarizations within the ``on-shell'' renormalization scheme. The hadronic vacuum polarization 
is decomposed in the perturbative and non-perturbative contributions. The perturbative contributions to the hadronic vacuum polarization are determined using  
the leading light fermions, 
\begin{equation}
\resizebox{0.4\textwidth}{!}{$ 
\delta\alpha_{QED}^{pert-had}(s)=\frac{\alpha}{3\pi}\sum_{f=u,s,d,c}Q_{f}^{2}N_{cf}\left(\ln\left(\frac{s}{m_{f}^{2}}\right)-\frac{5}{3}\right) \label{eq:pertQDC}
$}
\end{equation} % \cite[Eq. 2.235]{Jegerlehner}  
\noindent   \cite{Jegerlehner}, while the  hadronic vacuum polarization in the non-perturbative regime is determined from experimental data  \cite{PDG2020} 
using dispersion relations, 
\begin{equation}
\resizebox{0.4\textwidth}{!}{$
\delta\alpha_{QED}^{nonpert-had}=-\frac{\alpha s}{3\pi}\left[P\int_{m_{threshold}}^{s_{cut}}\frac{ds' R_{data}(s')}{s'(s'-s)}+P\int_{s_{cut}}^{\infty} \frac{ds' Q_{f}^{2}N_{cf}}{s'(s'-s)} \right]
$}
\end{equation} %\cite[Eq. 5.25]{Jegerlehner}
  \cite{Jegerlehner}. Non-perturbative contributions from higher mass narrow resonances neglected by the perturbative 
formulation  \cite{Jegerlehner}, % \cite[Eq. 5.25]{Jegerlehner}
$J/\psi$, $\psi(2s)$, $\psi(3s)$, $\Upsilon(1s)$, $\Upsilon(2s)$, $\Upsilon(3s)$ and $\Upsilon(4s)$ are approximated 
as Breit-Wigners resonances  % \cite[Eq. 2.266]{Jegerlehner}
\cite{Jegerlehner}\footnote{The masses of the $J/\psi$, $\psi(2s)$, $\psi(3s)$, $\Upsilon(1s)$, $\Upsilon(2s)$, $\Upsilon(3s)$ 
and $\Upsilon(4s)$ resonances are taken from  \cite{PDG2018}.}\footnote{For numerical reasons, when $|\delta\alpha|$ associated with Equation \ref{eq:HadvacBW} is $\approx {\mathcal O}(\alpha)$ it is truncated. This approximation is justifiable for $\sqrt{s\pm\epsilon}\gg\Gamma$ and $\sqrt{\epsilon} < \Gamma$ since the truncated components approximately cancel.},

\begin{equation}
\resizebox{0.3\textwidth}{!}{$ 
\Pi_{\gamma ren}^{'}(s)\simeq \frac{-3 \Gamma_{e^{+}e^{-}}s\left(s-M_{R}^{2}-\Gamma^{2}\right)}{\alpha M_{R}\left[\left(s-M_{R}^{2}\right)^{2}+M_{R}^{2}\Gamma^{2}\right]}
$}
\label{eq:HadvacBW}
\end{equation}.  

\noindent   \cite{Jegerlehner}. 
The shift in $\alpha_{QED}$ as a function of $s$, $\delta \alpha_{QED}(s)$, is displayed in Figure \ref{fig:RunningOfAlpha} for both the 
leptonic and hadronic vacuum polarization contribution. The hadronic vacuum polarization contribution 
to $\delta \alpha_{QED}$ is consistent with predictions from \cite{Jegerlehner,MC:2010} at the $5-10\%$ level depending on the $\sqrt{s}$ and 
which narrow resonances are included, where the leptonic contribution to  $\delta \alpha_{QED}$ is in 
agreement to better than $1\%$. Moreover, \cite[Fig. 87]{MC:2010} illustrates that there is a non-negligible uncertainty between the compilations of the 
hadronic vacuum polarization to  $\delta\alpha_{QED}$, $\sim 2-10\%$ depending on $\sqrt{s}$.
From the impact of $\delta \alpha_{QED}$ in Figure \ref{fig:CSvsMass}, the uncertainty related to the vacuum-polarization on the 
total cross-section can be conservatively estimated to be $\sim 0.3\%$ at $\sqrt{s}=10.58GeV$.

\subsection{Summation of Infra-red Divergences \label{sec:YFS}}
The summation of the infra-red divergences \cite{Bloch:1937} are subtracted by means of applying the exponentiation procedure  \cite{Yennie:1961} to the hard-scattering.
The cross-section with the multiplicative subtraction can then be written as:

\begin{equation}
\resizebox{0.375\textwidth}{!}{$ 
d\sigma= \frac{\sum_{n=0}^{\infty} Y_{i}(Q_{i}^{2})Y_{f}(Q_{f}^{2})|\sum_{k=1}^{\infty}\bar{{\mathcal M_{n}^{k}}}|^{2} dPS_{n}^{\delta M}}{4(|\vec{P}_{e^{-}}|E_{e^{+}}+E_{e^{-}}|\vec{P}_{e^{+}}|)}
$}
\label{eq:YSR}
\end{equation}

\noindent \cite{kk2f,Peskin:1995ev,Yennie:1961}\footnote{Based on the discussion in \cite{Yennie:1961} we do not include the YFS cross-terms in contrast to \cite{kk2f}.}
%\footnote{It is important to note, that the perturbative matrix element series, $M_{n}^{k}$, with the Type-IV exponentiation correspond 
%to the full QED calculations for spin 1/2 particles interacting within a vector field. This is not a scalar QED approximation.} 
where $|\sum_{k=1}^{\infty} \bar{{\mathcal M_{n}^{k}}}|^{2}$ is the spin-averaged sum of all matrix elements including the interferences with $n$ real photons and $k$ photon exchanges, where 
the symmetrisation factor for $n$ real photons has been implicitly included, and $dPS_{n}^{\delta M}$ is the corresponding phase-space with a soft-photon cut-off 
defined in terms of the Lorentz invariant quantity $M^{\prime}-M=\delta M$ \cite{Schwinger:1998}\footnote{The YFS Exponentiation Form-Factors representation in terms of the soft-photon cut-off energy, $E_{l},$ as presented in this section, is a ``special case'' of the generalized YFS Exponential Form-Factor in the $e^{+}e^{-}$ centre-of-mass reference frame \cite{Yennie:1961}.  The generalized YFS Exponentiation Form-Factor is Lorentz invariant, 
gauge invariant and independent of renormalization \cite{Yennie:1961}. Given that the total cross-section, YFS Exponentiation Form-Factors and matrix element are Lorentz invariant, it follows that there must exist a Lorentz invariant representation of the single variable 
upon which the YFS Exponentiation Form-Factor depends and the soft-photon cut-off associated with this term must also necessarily be Lorentz invariant. From \cite{Schwinger:1998}, this can be defined in terms of the Lorentz invariant 
quantity $M^{\prime}-M=\delta M$, instead of the more common ``special case'' soft-photon energy cut-off, $E_{l}$. $\delta M$ is defined to be consistent with the 1 
photon limit at $M^{\prime}=\sqrt{s}$ where $E_{l}=\frac{s-(\sqrt{s}-\delta M)^{2}}{2\sqrt{s}}$.  This produces the isotropic photon distribution in the $M^{\prime}$ centre-of-mass frame as predicted by the Lorentz frame in \cite{Yennie:1961}. To have a  physically meaningful interpretation of the soft-photon cut-off in terms of the minimum detector energy, $E_{min}$, in the Laboratory frame it is recommended that $\delta M\ll E_{min}$, so that the number of soft-photons converting into hard-photons when boosting into the Laboratory frame is negligible. In this case, the hard-photons which do not satisfy the $E_{\gamma}>E_{min}$ criteria will effectively be treated as soft-photons in the experiment.   The association of the infra-red divergences to the fermion self-energy 
or vertex graph is gauge dependant \cite{Peskin:1995ev,Halzen:1984mc,Yennie:1961}, therefore for each Feynman diagram the cut-off criteria are applied to the photon 
based on their associations with the vertices to remove the 
ambiguities. Ordering is taken into account for multiple photon emissions from a common fermion.}, applied to each hard photon. All permutations of hard photons and
their associated polarization states are explicitly summed over.   
$Y_{i}(Q_{i}^{2})$ and $Y_{f}(Q_{f}^{2})$ are the 
Yennie-Frautschi-Suura (YFS) Exponentiation Form-Factors for the initial and final state. 
The explicit form of the YFS Exponentiation Form-Factor is implemented using:
\begin{itemize}
\item Type I: The Yennie-Frautschi-Suura Exponentiation Form-Factor from  \cite{Yennie:1961} %\cite[Eq. 2.31+2.32]{Yennie:1961}.
\begin{equation}
\resizebox{0.375\textwidth}{!}{$ 
Y_{Type-I}=e^{-\frac{\alpha}{\pi}\left(\ln\frac{s-2m_{i}^{2}}{m_{i}^{2}}-1\right)\ln\left(\frac{s}{4E_{soft}^{2}}\right)+\frac{\alpha}{2\pi}\ln\frac{s-2m_{i}^{2}}{m_{i}^{2}}}
$}
\end{equation}
\noindent where $E_{soft}$ is defined as the soft-photon cut-off energy defined in the $e^{+}e^{-}$ centre-of-mass frame and $m_{i}$ is the mass of the leptons.
\item Type II: The YFS Exponentiation Form-Factor implemented in the {\tt KK2F} Generator \cite{kk2f}.  %\cite[Eq. 18]{kk2f}.
\begin{equation}
\resizebox{0.375\textwidth}{!}{$
Y_{Type-II}=e^{\frac{2\alpha}{\pi}\left(\ln\frac{s-2m_{i}^{2}}{m_{i}^{2}}-1\right)\ln\left(\frac{2E_{soft}}{\sqrt{s}}\right)+\frac{\alpha}{2\pi}\left(\ln\frac{s-2m_{i}^{2}}{m_{i}^{2}}-1\right)+\frac{\alpha}{\pi}\left(-\frac{1}{2}+\frac{\pi^{2}}{3}\right)}
.$}
\end{equation}
\item Type III: The YFS Exponentiation Form-Factor determined in \cite{Peskin:1995ev} % \cite[Eq. 6.84]{Peskin:1995ev} 
from the divergent component of the soft-photon and virtual photon 
contributions, also referred to as the Sudakov Form-Factor \cite{Peskin:1995ev}.
\begin{equation}
\resizebox{0.3\textwidth}{!}{$ 
Y_{Type-III}=e^{-\frac{\alpha}{2\pi}F_{IR}(q^{2})\ln\left(\frac{-q^{2}}{E_{soft}}\right)},
$}
\end{equation}
\noindent where the coefficient for the divergent logarithm is defined as,
\begin{equation}
\resizebox{0.275\textwidth}{!}{$ 
F_{IR}(q^{2})=\int_{0}^{1}\frac{m^{2}-q^{2}/2}{m^{2}-q^{2}\zeta(1-\zeta)}d\zeta -1.
$}
\end{equation}
 \noindent \footnote{The numerical calculations of the Form-Factor are determined to a precision of $\ll 1\times10^{-4}\%$  for an invariant mass $0<s<12^{2}GeV^{2}/c^{4}$.}This YFS
 Exponentiation Form-Factor satisfies
 the physical boundary conditions of $[0,1]$  \cite{Peskin:1995ev} and in the limit $-q^{2}\gg m_{i}^{2}$ reproduces the semi-classical calculation for the number of 
radiated photons  \cite{Peskin:1995ev}.
\item  Type IV: The YFS Exponentiation Form-Factor is determined from the leading order soft-photons and virtual contributions  \cite{Schwinger:1998} by equating them to 
the ${\mathcal O}(\alpha)$ term in the exponential series $e^{-{\mathcal O}(\alpha)}=1-{\mathcal O}(\alpha)+\frac{{\mathcal O}(\alpha)^{2}}{2!}...$  \cite{Peskin:1995ev}. 
This calculation includes contributions from the magnetic Form-Factor that are specific to spin $1/2$ particles, which are neglected in the scalar calculation, and the 
non-divergent component of the vacuum polarization \cite{Yennie:1961}. The contribution to the leading order soft-photons and virtual contributions from the leading order 
term for the Coulomb attraction between charges is removed \cite{Smith1994117} to provide a YFS Exponentiation Form-Factor consistent with the physical boundary 
conditions for $0\le Y_{Type-IV}\le 1$. The Coulomb attraction, 
\begin{equation}
F_{c}=\frac{\frac{\pi\alpha}{v}}{1-e^{-\left(\frac{\pi\alpha}{v}\right)}}, 
\end{equation}
\noindent where $v=\sqrt{1-4m_{i}^{2}/s}$,  \cite{Smith1994117,Schwinger:1998} is applied in addition to the YFS Exponentiation Form-Factor to account for the removal of 
the leading order term  \cite{Smith1994117}\footnote{The correction from the Uehling-Seber effect  \cite{Smith1994117} is not included since the running of $\alpha(s)$ 
has already been included in the calculation of the Coulomb factor $F_{c}$ and hard scattering following the procedure in Section \ref{sec:Renormilization}.}. The YFS Exponentiation Form-Factor is then defined as,
\begin{equation}
\resizebox{0.4\textwidth}{!}{$
\begin{array}{ll}
Y_{Type-IV}=&e^{\frac{2\alpha}{\pi}\left(\left[\left(1+v^{2}\right)\chi(v)-1\right]\left[\ln\left(\frac{\delta M}{m_{l}}\right)+1\right]+\chi(v)-\frac{(1+v)}{v}\int_{0}^{v}\frac{dv'\chi(v')}{1-v'^{2}} \right)} \times \\
&e^{-\frac{\alpha}{\pi}\left(P\int_{0}^{1}\frac{dv'\left(1+v'^{2}\right)\ln\frac{v'^{2}}{1-v'^{2}}}{v^{2}-v'^{2}}+\chi(v)+\frac{3(1-v^{2})}{(3-v^{2})}\chi(v)\right)-F_{c}|_{{\mathcal O}(\alpha)}}
\end{array}
$}
\end{equation}
\noindent where $\chi(v)=\frac{1}{2v}\ln\left(\frac{1+v}{1-v}\right)$ and $\delta M=\sqrt{s+E_{soft}^{2}}-\sqrt{s}+E_{soft}$.
%  Based on the integral calculated in 
%\cite{Schwinger:1998}, the virtual photon contribution to the ${\mathcal O}(\alpha)$ term in the 
%exponential series is inconsistent with the calculation in  \cite{PhysRevD_72_114019}\footnote{The results from  \cite{Schwinger:1998} were confirmed numerically.}.
\end{itemize}

A comparison of the YFS Exponentiation Form-Factor for the FSR only can be found in Figure \ref{fig:YFS}. It can be seen that near threshold,
the exponential factor exceeds the physical boundary $[0,1]$ for the YSF Form-Factors Type-I   \cite{Yennie:1961} and Type-II  \cite{kk2f} which is not unexpected
for YSR determined in the ultra-relativistic limit. The deviation of the YFS Form-Factors Type-III, which only includes the infra-red divergent components of
the virtual and soft photons from the other YFS Form-Factors shows the importance of including the non-divergent contribution from the soft and virtual photons.
The YFS Form-Factors Type-IV includes the most complete description of the soft and virtual photons, including the magnetic Form-Factor for spin $1/2$ particles,
for $0<v<1$.

The mass dependence of the simulated $e^{+}e^{-}\to\mu^{+}\mu^{-}(\gamma)$ and $e^{+}e^{-}\to\tau^{+}\tau^{-}(\gamma)$ cross-sections is summarized in 
Figure \ref{fig:CSvsMass}, while
the dependence on the soft-photon cut-off due to the truncation of the perturbative series within the YFS Exponentiation formalism can be seen in Figure \ref{fig:CSvsEl}.
From Figure \ref{fig:CSvsEl}, it can be seen that the ultra-relativistic approximation, Type-II, has a statistically significant difference from the exponentiation
based on the complete leading order soft-photons and virtual contributions
\cite{Schwinger:1998}, Type-IV. This suggests that there may be additional uncertainties neglected in \cite{Banerjee_2008}
and other generators that use the
ultra-relativistic approximations (or Sudakov Form-Factor \cite{Peskin:1995ev}) for soft and virtual photons \cite{Balossini_2006,CERN89-08}.
The $e^{+}e^{-}\to\mu^{+}\mu^{-}(\gamma)$ cross-section prediction is of particular importance for the experimental luminosity 
measurement at the B-Factories \cite{BABAR:2013agn}\footnote{The luminosity measurement at Belle-II is now using
the BaBaYaga@NLO generator for Bhabha events \cite{Belle-II:2019usr}. } and given Figure \ref{fig:CSvsEl} may be related to the absolute branching fractions 
in $\tau$ decays at the B-Factories tending to be lower than at LEP \cite{PDG2020}.
The impact of the exponentiation, for Type III and Type IV, and the soft-photon cut-off on the differential cross-section 
$e^{+}e^{-}\to\mu^{+}\mu^{-}(\gamma)$ ($e^{+}e^{-}\to\tau^{+}\tau^{-}(\gamma)$)
as a function of $\mu^{+}\mu^{-}$ ($\tau^{+}\tau^{-}$) invariant mass can be seen in Figure \ref{fig:CS_mumu_Rad}. 
The simulation in Figure \ref{fig:CS_mumu_Rad}
is reasonably
consistent with the theoretical approximation using the radiator function from  \cite{Aubert_2004} where multiplicative corrections for the presence of additional
initial and final state radiation and the Coulomb potential through the Sommerfeld-Sakharov factor \cite{PhysRevD_72_114019,PhysRevLett.103.231801} are included.
The LO leptonic and hadronic vacuum polarizations
have been included in the running of the electro-magnetic coupling, $\alpha$, in both the analytical theoretical approximation and in the simulation. However, the narrow
resonances are not observable in the simulated distributions due to the sampling range and limited number of simulated statistics.
There is a clear dependence on the soft-photon cut-off for the 
differential distribution of the 
$e^{+}e^{-}\to\mu^{+}\mu^{-}(\gamma)$ as a function of $\mu^{+}\mu^{-}$ invariant mass.%, particularly for the most energetic mass bin where the 
The dependency of the cross-section on the soft-photon cut-off can also be seen 
in Figure \ref{fig:CSvsEl} for both 
the $e^{+}e^{-}\to\mu^{+}\mu^{-}(\gamma)$ and $e^{+}e^{-}\to\tau^{+}\tau^{-}(\gamma)$ cross-sections in all of the exponentiation methods.
The inclusion of higher order of perturbation theory in the simulation is expected to reduce this dependency on the soft-photon cut-off 
in Figures \ref{fig:CSvsEl} and  \ref{fig:CS_mumu_Rad}.
%is expected to be reduced when higher order contributions are included in the simulation.
This is a consequence of 
the exponentiation formalism where the 
 positive definite multiplicative subtraction, $[0,1]$, of the infra-red divergencies which sums over all orders in $\alpha$, results in a infinite series 
constructed from the remaining terms in the perturbative Feynman series from above the soft-photon cut-off. This asymptotically 
converges to the total QED calculation as the order of additional terms included in the calculation goes to $\infty$. The choice of 
exponentiation for the infra-red summations was made to allow for the consistent inclusion of additional photons above the soft-photon cut-off in the simulation.
The truncation of the higher order terms in the Feynman series is the most significant uncertainty for the theoretical predications and can be estimated as the
change in the cross-section between a given soft-photon cut-off relative to the nominal limit $E_{l}\to\sqrt{s}$ as seen in Figure \ref{fig:CSvsEl}. 
Additional higher order terms are required to reach the expected precision at Belle-II. 
In addition to higher orders in the QED perturbative diagrams, further improvements can be made through the inclusion of contributions from the neutral
weak interaction by means of Electro-weak mixing\footnote{Both \cite{Banerjee_2008} and \cite{Balossini_2008} estimate the electro-weak contribution to be 
$\le0.1\%$ at B-Factory energies. This can be taken as a theoretical uncertainty on neglecting the contribution from the weak interaction for the total cross-section. }.  

\section{$\tau$ Lepton Decays \label{sec:Tau}}

For $\tau$ decays the differential decay width is determined at Born level $d\Gamma=\frac{1}{2M_{\tau}}\left|\mathcal{M}\right|^{2}\times dPS $
where the matrix element is defined as:  

\begin{equation}
\resizebox{0.4\textwidth}{!}{$
\mathcal{M}=\frac{-\imath g^{2}}{2}\left[\bar{u}_{1}\gamma^{\mu}\frac{(I-\gamma_{5})}{2}u_{3}\right]\left[\frac{-G_{\mu\nu}+\frac{q_{\mu}q_{\nu}}{M_{W}^{2}}}{q^{2}-M_{W}^{2}}\right]\left[\bar{v}_{2}\gamma^{\nu}\frac{(I-\gamma_{5})}{2}u_{4}\right]
$}\label{eq:taulep}
\end{equation}
\normalsize
\noindent for the leptonic decays and 

\begin{equation}
\resizebox{0.35\textwidth}{!}{$
\mathcal{M}=\frac{-\imath g^{2}}{4}\left[\bar{u}_{1}\gamma^{\mu}\frac{(I-\gamma_{5})}{2}u_{3}\right]\left[\frac{-G_{\mu\nu}+\frac{q_{\mu}q_{\nu}}{M_{W}^{2}}}{q^{2}-M_{W}^{2}}\right]J^{\nu} \label{eq:tauhad}
$}
\end{equation}

\noindent for the hadronic decays  \cite{PhysRevD.4.2821,Renton:1990td}, where $M_{W}^{2}=\frac{g^{2}\sqrt{2}}{8G_{F}}$. The hadronic models are included 
into the algebraic formalism for the Dirac-Spinors and $\gamma$-Matrices through the Lorentz-invariant hadronic current, $J^{\mu}$. This particular 
representation was implemented to enable the inclusion of Feynman Diagrams with photons above the soft-photon cut-off in conjunction 
with the remaining Electro-weak corrections. 
The hadronic current associated with each of the hadronic decay modes\footnote{Charge conservation is implied through this paper unless otherwise stated.}
are constructed in terms of the covariant amplitudes and Form-Factors of the given hadronic models determined from the
sum over the final spin-states  \cite{Jadach:1993hs,Renton:1990td}.
The Dirac-Spinor 
corresponding to the $\tau$ lepton can be constructed and decayed in either an unpolarized or a predefined polarization state 
while all helicity states of the Dirac-Spinors for the neutrinos (and final-state leptons) 
are summed over\footnote{Equation \ref{eq:taulep} and \ref{eq:tauhad} are constructed in terms of the 
left-handed helicity projection operators, $\frac{(I-\gamma_{5})}{2}$.}. The Lorentz invariant matrix element corresponding to the $\tau$ lepton decay is computed 
in the reference frame that the $\tau$ polarization is defined. This solution is chosen to 
facilitate the interfacing of spin dynamics with the QED 
interactions described in Section \ref{sec:QEDSpin}. An alternative solution is implemented in  \cite{Jadach:1993hs,Kuhn:1992nz,Was:1984pd,Jadach:1998wp,Jadach:1999vf}, 
where the spin
dynamics are calculated in the rest frame of the hadronic decay products. 
 
\subsection{Overview of Hadronic Models}
In the one meson semi-leptonic decays of the $\tau$ lepton, the Lorentz invariant hadronic current is simply defined in terms of a scalar Form-Factor 
 \cite{PhysRevD.4.2821}, which can be 
determined from lattice QCD  \cite{Aoki:2016frl},
$J^{\mu}=F_{0}Q^{\mu}$  \cite{Stahl:2000aq,PhysRevD.4.2821}. The higher 
multiplicity decay modes have a mass dependent Form-Factor and are implemented with phenomenological hadronic models, 
$J^{\mu}=F_{1}(s,\cdots)\epsilon^{\mu}$ for vector and axial-vector mesons and $J^{\mu}=F_{0}(s)Q^{\mu}$ for scalar and pseudo-scalar mesons 
 \cite{Stahl:2000aq,PhysRevD.4.2821}. In order to compare the physics predictions between separate 
theoretical constructs and to previous results, the hadronic models are implemented for several separate frame-works: Chiral-Resonance-Lagrangian (ChRL) 
following the approach in   \cite{Finkemeier:1995sr,Kuhn:1990ad,Jadach:1993hs,Decker:1992kj} (Section \ref{sec:ChRL}); 
phenomenological models  \cite{CLEO3pi,Edwards:1999fj,Feindt:1990ev}, including the Vector-Dominance Model  \cite{BONDAR2002139,Bondar:1999}(Section \ref{sec:Phem});
as well as a ``Chromoelectric Flux-Tube Breaking Model'' constructed in the context on the ``strong coupling lattice formulation''
  \cite{Kokoski:1985is,Isgur:1983wj,Isgur:1984bm} combined with a ``revitalized [${^{3}P_{0}}$] quark model'' 
 \cite{Isgur:1988vm,Godfrey:1985xj,Kokoski:1985is,Isgur:1983wj,Isgur:1984bm} 
(Section \ref{sec:FluxTube}).
These models cover the majority of hadronic decay modes containing up to five mesons. For $\tau$ decays with a higher number of mesons or for decay modes 
which are not implemented in the generator, we provide a generic scalar phase-space, $J^{\mu}=Q^{\mu}$  \cite{Stahl:2000aq,PhysRevD.4.2821}.

\subsection{Chiral-Resonance-Lagrangian Models\label{sec:ChRL}}

The majority of hadronic decay modes included in  \cite{Jadach:1993hs} are included in this generator by means of the ChRL Models in 
 \cite{Finkemeier:1995sr,Kuhn:1990ad,Decker:1992kj}, where the covariant amplitudes associated with the hadronic current are constructed using the formalism in 
 \cite{Kuhn:1992nz}. 
The $\tau^{-}\to\rho^{0}\nu_{\tau}\to\pi\pi^{0}\nu_{\tau}$, $\tau^{-}\to a_{1}(1260)\nu_{\tau}\to\pi^{-}\pi^{0}\pi^{0}\nu_{\tau}$
$\tau^{-}\to a_{1}(1260)\nu_{\tau}\to\pi^{-}\pi^{-}\pi^{+}\nu_{\tau}$ are implemented in terms of the ChRL Model augmented with
contributions from the Vector-Dominance Model, referred to as the K\"uhn-Santamaria Model  \cite{Kuhn:1990ad}. The K\"uhn-Santamaria Model, has been implemented with several
variations based on the work of  \cite{Lees:2012cj} and   \cite{Anderson:1999ui}. In   \cite{Lees:2012cj}, an improved $\rho$ spectral function was measured using the
Gounaris-Sakuria  Model \cite{Gounaris:1968mw},  a vector dominance based Model, from which the dispersion relations in  \cite{Kuhn:1990ad} were adapted following  \cite{Anderson:1999ui}, while  \cite{Anderson:1999ui} included
the Blatt-Weisskopf-Barrier Centrifugal-factor, $F_{BWB}=\frac{1+R^{2}|P(\sqrt{s}=m_{\rho})|^{2}}{1+R^{2}|P(\sqrt{s})|^{2}}$, to investigate possible modifications to the
P-wave behaviour of the $\rho$ mesons related to an additional energy dependence in the width  \cite{Blatt::1952}. The spectral function from   \cite{Lees:2012cj} %\cite[Eq. 26]{Lees:2012cj} 
is taken for the parameterization of the $\rho(770)$, $\rho^{'}(1450)$ and $\rho^{''}(1700)$ and is available for \footnote{The Chiral Breit-Wigner described 
in \cite{Kuhn:1990ad} % \cite[Eq. 2.5]{Kuhn:1990ad} 
is included in the  \cite{Decker:1992kj} model. We include a correction to the normalization, 
$2\times$  \cite{Kuhn:1990ad,Jadach:1993hs,Decker:1992kj}, or more precisely $\sqrt{2}\times$ the axial-vector Form-Factor.}  the Breit-Wigner with the dispersive relation
in %\cite[Eq. 2.6-2.9]{Kuhn:1990ad}
\cite{Gounaris:1968mw} and  % \cite[Eq. 6-16]{Gounaris:1968mw}
\cite{Kuhn:1990ad}, where
the amplitudes of the $\rho^{'}(1450)$ and $\rho^{''}(1700)$  are defined relative to the $\rho(770)$ and not $\rho(770)+\omega(782)$ following an analogous procedure
to   \cite{Anderson:1999ui}. 
For the
$\tau^{-}\to a_{1}(1260)\nu_{\tau}\to\pi^{-}\pi^{0}\pi^{0}\nu_{\tau}$ and
$\tau^{-}\to a_{1}(1260)\nu_{\tau}\to\pi^{-}\pi^{-}\pi^{+}\nu_{\tau}$ modes, the K\"uhn-Santamaria Model  \cite{Kuhn:1990ad} is implemented numerically by means of the
decay width formulated in % \cite[Eq. 3.8-3.9]{Kuhn:1990ad} 
  \cite{Kuhn:1990ad}  and the dispersive relations  % \cite[Eq. 3.19-3.20]{Kuhn:1990ad}
 \cite{Kuhn:1990ad} for each of the four previously mentioned
variations of $\pi\pi$ Form-Factors.

The three hadron decay modes from  \cite{Finkemeier:1995sr,Decker:1992kj} are implemented in the simulation, where the  \cite{Finkemeier:1995sr} model has been extended to 
include additional decay modes including the $\tau^{-}\to K^{-}K^{-}K^{+}\nu_{\tau}$ decay mode measured in 
 \cite{Aubert:2007mh,Lee:2010tc} and the  scalar $\tau^{-}\to K^{-}\eta\nu_{\tau}$ and $\tau^{-}\to K^{-}\eta^{'}(958)\nu_{\tau}$,  \cite{Finkemeier_1996}. The 
$\tau^{-}\to K^{-}K^{-}K^{+}\nu_{\tau}$ is assumed to proceed through the $K_{1}(1650)$ resonance which decays through the 
$\omega$ and $\phi$ resonances, where the $\omega(782)$ contribution is highly suppressed due to the limited phase-space. A possible vector contribution is included
through the $K^{*}$ states again proceeding through the $\omega$ and $\phi$. The two axial-vector and the vector Form-Factors are defined as:

\begin{equation}
\resizebox{0.35\textwidth}{!}{$
\begin{array}{lcl}
F_{1}(s,s_{1},s_{2}) &=& -V_{us} \frac{2\sqrt{2}}{3f_{\pi}}T_{K_{1}}^{(3)}(s)T_{\omega}(s_{2}) \\
\end{array}
$}
\end{equation}
\begin{equation}
\resizebox{0.35\textwidth}{!}{$
\begin{array}{lcl}
F_{2}(s,s_{1},s_{2}) &=& -V_{us}\frac{2\sqrt{2}}{3f_{\pi}}T_{K_{1}}^{(3)}(s)T_{\omega}(s_{1}) \\
\end{array}
$}
\end{equation}
\begin{equation}
\resizebox{0.4\textwidth}{!}{$
\begin{array}{lcl}
F_{3}(s,s_{1},s_{2}) &=& V_{us}\frac{1}{2\sqrt{2}\pi^{2}f_{\pi}^{3}}T_{K^{*}}^{(2)}(s)\left[T_{\omega}(s_{2})+T_{\omega}(s_{1})\right] 
\end{array}
$}
\end{equation}

\noindent using the convention for the hadronic current from  \cite{Kuhn:1992nz}. %\cite[Eq. 32]{Kuhn:1992nz}.
Where $T_{K_{1}}^{(3)}(s)$, $T_{K^{*}}^{(2)}(s)$  %\cite[Eq. 42]{Finkemeier:1995sr} 
and $T_{\omega}(s)$  %\cite[Eq. 36]{Finkemeier:1995sr} 
can be written: 

\begin{equation}
\resizebox{0.375\textwidth}{!}{$
T_{K_{1}}^{(3)}(s)=\frac{\left(BW_{K_{1}(1400)}(s) +\lambda_{K_{1}} BW_{K_{1}(1270)}(s) + \mu_{K_{1}}BW_{K_{1}(1650)}(s)  \right)}{1+\lambda_{K_{1}}+\mu_{K_{1}}}
$}
\end{equation}
\begin{equation}
\resizebox{0.375\textwidth}{!}{$
T_{K^{*}}^{(2)}(s)=\frac{\left(BW_{K^{*}(892)}(s) +\lambda_{K^{*}} BW_{K^{*}(1410)}(s) + \mu_{K^{*}}BW_{K^{*}(1680)}(s)  \right)}{1+\lambda_{K^{*}}+\mu_{K^{*}}}
$}
\end{equation}
\begin{equation}
\resizebox{0.3\textwidth}{!}{$ 
T_{\omega}(s)=\frac{\left(BW_{\omega(782)}(s) +\epsilon BW_{\phi(1020)}(s)\right)}{1+\epsilon}.
$}
\end{equation}

\noindent \cite{Finkemeier:1995sr} where $\lambda_{K_{1}}=(0.33+\imath 0)$  \cite{Finkemeier:1995sr}, $\mu_{K_{1}}=(0.35+\imath 0)$, $\lambda_{K^{*}}=(\frac{-6.5}{26}+\imath 0)$ 
 \cite{Finkemeier:1995sr}, $\mu_{K^{*}}=(\frac{-1}{26}+\imath 0)$  \cite{Finkemeier:1995sr} and
$\epsilon=(0.05+\imath 0)$. $BW_{K_{1}(1400)}(s)$ and  $BW_{K_{1}(1270)}(s)$ are the fixed-width Breit-Wigner 
resonances for the $K_{1}(1270)$ and $K_{1}(1400)$ states from 
 \cite{Finkemeier:1995sr} and $BW_{K_{1}(1650)}(s)$ is the fixed width Breit-Wigner   \cite{Finkemeier:1995sr} %\cite[Eq. 34]{Finkemeier:1995sr}
with $M_{K_{1}(1650)}=1.672GeV/c^{2}$ and width $\Gamma_{K_{1}(1650)}=0.158GeV$. 

The ChRL Models from  \cite{Finkemeier:1995sr} are extended to include $\tau^{-}\to K^{-}\eta\nu_{\tau}$ and $\tau^{-}\to K^{-}\eta'(958)\nu_{\tau}$
decays, where the scalar contribution from the $K_{0}^{*}(1430)$ is estimated from the formalism in  \cite{Finkemeier_1996}. The corresponding Form-Factor for
$\tau^{-}\to K^{-}\eta\nu_{\tau}$ and $\tau^{-}\to K^{-}\eta'(958)\nu_{\tau}$ are  $F_{V}(s)=T_{K^{*}}^{(1)}$   \cite{Finkemeier:1995sr} %\cite[Eq. 10]{Finkemeier:1995sr} 
and

\begin{equation}
\resizebox{0.425\textwidth}{!}{$ 
  \begin{array}{lcl}
F_{S}(s)&=&c_{s}\left(\frac{m_{K}^{2}-m_{\eta/\eta'}^{2}}{M_{K_{0}^{*}(1430)}^{2}}\right)\left(\frac{M_{K_{0}^{*}(1430)}^{2}}{(M_{K_{0}^{*}(1430)}^{2}-s)-\imath s^{1/2}\Gamma(s,0)}\right)
\end{array}
$}
\end{equation}

\noindent where we approximate $c_{v}\approx c_{s} \approx 1$ and $\Gamma(s,L)=\Gamma\frac{M^{2}}{s}\left(\frac{P(s)}{P(M^{2})}\right)^{2L+1}$
  \cite{Finkemeier:1995sr,Decker:1992kj,Finkemeier_1996}. 

The $5\pi$ modes are modelled with the K\"uhn-W\c as Model  \cite{Kuhn:2006nw}\footnote{Assumptions are made about the units in order to obtain a meaningful
decay-rate.}. Within the narrow-width approximation  \cite{Kuhn:2006nw}, we obtain a ratio of 
decay rates similar to  \cite{Kuhn:2006nw}. %\cite[Eq. 20]{Kuhn:2006nw}
In order to better reflect the measured values in  \cite{PDG2020}, we apply weights to cancel these ratios.

A summary of the simulated ChRL branching ratios can be found in Table \ref{table:ChRLBR}. While reproducing the results
from {\tt TAUOLA}  \cite{Finkemeier:1995sr,Jadach:1993hs,Decker:1992kj} we found a common discrepancy in the normalization for the 
3-body axial-vector current ($\sqrt{2}$) current and vector current ($2$) (see \cite{Georgi:2009}).
%These differences in the vector current could be related to the difference between 
%\cite[Eq. 21]{Jadach:1993hs} and  \cite[Eq. 26]{Decker:1992kj}. 
In addition to the common normalization discrepancy, two mode specific issues where identified.
Firstly, for the non-strange vector current in \cite{Finkemeier:1995sr}, there is an additional correction for the normalization of $1/(\sqrt{2}f_{\pi})$, 
which appears to be related to the $f_{\pi}^{2}$ and $f_{\pi}^{3}$ difference between \cite[Eq. 21]{Jadach:1993hs} 
and  \cite[Eq. 26]{Decker:1992kj}\footnote{The correction to non-strange decay modes
in  \cite{Finkemeier:1995sr} reproduce the published values in  \cite[Table III]{Finkemeier:1995sr} associated with the Wess-Zumino term. Unfortunately, for the 
non-strange decay mode $\tau^{-}\to \eta\pi^{-}\pi^{0}\nu_{\tau}$, even with corrections, the Wess-Zumino term is 
$\sim\sqrt{\frac{2}{3}}$ of the values quoted in  \cite{Finkemeier:1995sr,Decker:1992kj}. }. 
Secondly, in the $\pi^{0}\pi^{0}K^{-}$ axial-vector decay modes from \cite[Table I]{Jadach:1993hs} and  \cite[Table I]{Decker:1992kj} there is an additional factor of 
$1/2$ that has to be taken into account. These numerical
corrections for the Form-Factors have been implemented as a run-time option in the simulation and are included in the results presented here.

\subsection{Phenomenological Models\label{sec:Phem}}

The phenomenological models implemented in this generator are primarily based on models developed by experimental collaborations and which were fit to the experimental data.
These phenomenological models are constructed employing attributes from theoretical models, for example ChRL Models and Vector-Dominance Models 
and adapted to measure experimental quantities and test theoretical assumptions. 
In the late 1990's, the $\tau\to 3\pi\nu_{\tau}$ decay model was investigated by a model which was constructed phenomenologically and fit to the data by the CLEO 
experiment in  \cite{CLEO3pi}. This model was primarily based on Chiral-Resonance-Lagrangian
Models, with modifications to check the assumptions of Flux-Tube Breaking Model   \cite{Isgur:1988vm,CLEO3pi,Godfrey:1985xj,Kokoski:1985is}.
One of the main conclusions of the experimental comparison of the hadronic models was the importance in including the D-wave 
component of the $a_{1}(1260)$ decay predicted in   \cite{Isgur:1988vm,Godfrey:1985xj,Kokoski:1985is} and the inclusion of 
additional resonances, namely the $\sigma$, $f_{0}(1370)$ and $f_{2}(1270)$ resonances  \cite{Stahl:2000aq,CLEO3pi,Anderson:1999ui}. 
Evidence for additional resonances \cite{Nugent:2013hxa,Ketzer:2019wmd} can be found in the literature.
Additional variations to the nominal model investigate the impact of the dispersion relation on the running of the mass for 
the $a_{1}(1260)$ incorporating the non-zero meson size\footnote{Within the CLEO model, it was found that the inclusion of the non-zero meson size was required to 
have a meaningful running of the mass for the $a_{1}(1260)$  \cite{CLEO3pi}.}
\footnote{When comparing the Flux-Tube Breaking Model  \cite{Isgur:1988vm} with the CLEO Model  \cite{CLEO3pi}, one finds that there is a difference in the
formalism applied in the calculate of the $a_{1}(1260)$ width because the $a_{1}(1260)$ propagator is reduced to the Breit-Wigner term neglecting the polarization vectors.
Moreover, a factor of $s$ must be applied in the denominator of Eq. 12 in   \cite{CLEO3pi} to reproduce the results in   \cite[Fig. 9b]{CLEO3pi}.}, 
and the statistical significance of the $a_{1}^{\prime}(1640)$.
Unfortunately, the CLEO experiment did not find sufficient statistical significance
to separate between the nominal fit, the fitted model with the inclusion of the $a_{1}^{\prime}(1640)$ (CLEO Model [with $a_{1}^{\prime}(1640)$]) and the fitted model 
with the non-zero meson size (CLEO Model [R=1.4])\footnote{The running of the $a_{1}(1260)$ mass was not explicitly included in either of these
 models  \cite{CLEO3pi}.}  \cite{CLEO3pi}. 

The ChRL Model \cite{Finkemeier:1995sr,Kuhn:1990ad,Kuhn:1992nz,Decker:1992kj} and K\"uhn-Santamaria Model  \cite{Lees:2012cj,Kuhn:1990ad},
are constructed using current conservation, $Q^{\mu}J_{\mu}=0$ \cite{Feindt:1990ev}, for the formal structure of the vertices using the Form-Factor approach
\cite{Isgur:1988vm,Peskin:1995ev} to describe the vertices in the covariant amplitude for the hadronic current\footnote{ 
 Within the Form-Factor approach for the formal structure of the vertices
\cite{Isgur:1988vm,Peskin:1995ev}, the individual vertices are constructed to satisfy a particular set of conserved properties associated for the given
interaction \cite{Isgur:1988vm,Peskin:1995ev,Feindt:1990ev}. As with the ChRL \cite{Finkemeier:1995sr,Kuhn:1990ad,Kuhn:1992nz,Decker:1992kj}, 
the particular choice of conservation criteria, the implementation and the corresponding Form-Factors are model dependant.}.
The approach in these models is related to the construction of the amplitudes under the assumption that the lowest dimensional 
Born terms are dominate \cite{Feindt:1990ev,Kuhn198416}.  Alternatively, the vertex may be constructed from the amplitudes corresponding to the 
lowest order angular momentum amplitudes \cite{Feindt:1990ev}. These two fundamentally different assumptions are both entirely ``justifiable'' descriptions of the strong 
interaction \cite{Feindt:1990ev}. By relating the helicity amplitudes ratios in the final state helicity frame to the spherical harmonics the  vertex Form-Factors can be constructed in terms of the angular momentum amplitudes \cite{Isgur:1988vm,Feindt:1990ev}. This was implemented for the $a_{1}(1260)$  
in \cite{Feindt:1990ev} by explicitly constructing the hadronic current in terms of the helicity amplitudes corresponding to the independent Born terms 
for the vertex formalism in \cite{Pilkuhn} for the $A \to V+P$ interaction under the assumption $\epsilon_{a_{1}}\cdot p_{a_{1}}=\epsilon_{\rho}\cdot p_{\rho}=0$ 
\cite{Isgur:1988vm}.
This is realised using a generic formalism, without applying simplifications, for the covariant amplitude constructed explicitly from the Feynman calculus for the given
diagram in terms of interchangeable propagators and vertex projectors based on the Form-Factor methodology. This generic formalism will become more essential 
for incorporating more sophisticated models later on. The hadronic current in terms of this formalism 
for the $a_{1}(1260)$ in the Feindt Model \cite{Feindt:1990ev} can be expressed as:

\begin{equation}
\resizebox{0.4\textwidth}{!}{$
J^{\mu}=V_{ij}F_{a_{1}(1260)}P_{a_{1}(1260)}^{\mu\nu}V_{\nu\alpha}^{A\to VP(S/D-wave)}P_{\rho}^{\alpha\beta}V_{\beta}^{V\to PP}
$}
\end{equation}

\noindent where $V_{ij}$ is the corresponding CKM matrix element, and $F_{a_{1260}(1260)}$ is the coupling constant for the $a_{1}(1260)$ resonance. 
Under the narrow-width approximation, the propagator is defined as:

\begin{equation}
\resizebox{0.2\textwidth}{!}{$ 
P_{Res}^{\mu\nu}=T^{\mu\nu}BW_{Res}(s_{Res})
$}
\end{equation}

\noindent where $T^{\mu\nu}=\left(-g^{\mu\nu}+\frac{q^{\mu}q^{\nu}}{Q^{2}}\right)$ and the normalized vertex parameterizations are

\begin{equation}
\resizebox{0.325\textwidth}{!}{$
V_{\mu\nu}^{A\to VP (S-wave)}=g_{\mu\nu}-\frac{k_{\mu}Q_{\nu}}{k\cdot Q +\sqrt{Q^{2}k^{2}}}
$}
\end{equation}
\begin{equation}
\resizebox{0.375\textwidth}{!}{$
V_{\mu\nu}^{A\to VP (D-wave)}=g_{\mu\nu}\frac{\left( Q\cdot k \right)^{2}-Q^{2} k^{2}}{Q^{2}}-k_{\mu}Q_{\nu}\frac{k\cdot Q +2\sqrt{Q^{2}k^{2}}}{Q^{2}}
$}
\end{equation}
\noindent  and
\begin{equation}
\resizebox{0.175\textwidth}{!}{$
V_{\mu}^{V\to PP} = (q_{1}-q_{2})_{\mu}
$}
\end{equation}
\noindent  \cite{Feindt:1990ev}.  % \cite[Eq. 5,  7 \& 8]{Feindt:1990ev}
 $Q^{\mu}$, $k^{\mu}$  are the Lorentz Vector for the $a_{1}(1260)$ and $\rho$ while $q_{1}^{\mu}$ and $q_{2}^{\mu}$ are the Lorentz Vectors for the
 pions resulting from the decay of the $\rho$. The $a_{1}(1260)$ and $\rho(770)$ Breit-Wigner are described with 
$\frac{g(s)}{M_{res}^{2}-s-\imath M_{res}\Gamma_{res}}$ and $\frac{1}{M_{res}^{2}-s-\imath M_{res}\Gamma_{res}}$ where $g(s)=(\sqrt{s}/m_{a_{1}(1260)})^{n}$ for $n=-0.5$ 
in the Bowler Model  \cite{Bowler:1987bj}.  The formalism for generic propagators and vertices allows for the replacement of the $a_{1}(1260)$ Form-Factors 
with those from more complete models to allow for a direct comparison between the two assumptions for the vertex projectors.  

In addition to the $\tau^{-}\to (3\pi)^{-}\nu_{\tau}$ Models from the CLEO experiment  \cite{CLEO3pi}, we have incorporated the 
$\tau^{-}\to K^{-}\pi^{-}\pi^{-}\nu_{\tau}$ Model by the CLEO experiment \cite{Asner:2000nx} which was the first model employed to extract the 
mixing angle between the singlet and triplet $K_{1}$ states using $\tau$ decays. The model was based on Chiral Perturbation Theory (ChPT) supplemented with 
Charged-Vector-Conservation (CVC) to describe the $\tau^{-}\to K^{-}\pi^{-}\pi^{-}\nu_{\tau}$  decay in terms of the intermediate states $K\rho$ and $K^{*}(892)$, 
the $K_{1}$  mixing angle $\theta_{K_{1}}$ and $SU(3)_{f}$ suppression factor 
$|\delta_{K_{1}}|=\left|\frac{\left(m_{s}-m_{u}\right)}{\sqrt{2}\left(m_{s}+m_{u}\right)}\right|=0.18$ \cite{Asner:2000nx} where $m_{s}$ and $m_{u}$ are the $s$ and 
$u$-quark masses within the QCD potential\footnote{The definition of $|\delta_{K_{1}}|=\left|\frac{\left(m_{s}-m_{u}\right)}{\sqrt{2}\left(m_{s}+m_{u}\right)}\right|$ 
is a relative factor that appears in the mixing between the ${}^1P_{1}$ and ${}^{3}P_{1}$ weak decay amplitudes for the $K_{1}$ states in the non-relativistic static 
limit  \cite{Suzuki:1993,Blundell:1995au}.  The numerical value $|\delta_{K_{1}}|=0.18$ corresponds to the quark masses, $m_{u}=0.33GeV/c^{2}$ and $m_{s}=0.55GeV/c^{2}$, for a simple-harmonic oscillator wave-function in a Coulomb potential with a linear confining term within the non-relativistic static limit where the $\sqrt{2}$ 
factor is related to the relative $K_{A}$ and $K_{B}$ amplitudes in the quark model \cite{Blundell:1995au}. }. This simplified model is a relevant benchmark when 
discussing more sophisticated models for the extraction of the 
mixing angle between the singlet and triplet $K_{1}$ states described in Section \ref{sec:FluxTube}. To reproduce the ratio of $K^{*}(892)\pi$ to 
$K\rho(770)$ in the CLEO Model \cite{CLEO3pi}, the difference between the current convention in \cite{CLEO3pi} and \cite{Jadach:1993hs} must be taken into account.

Although the $\tau\to 4\pi\nu_{\tau}$ was measured by ARGUS  \cite{Albrecht:1986kz}, the first experimental analysis to investigate the
decay structure of $\tau\to 4\pi\nu_{\tau}$ was by
the CLEO experiment \cite{Edwards:1999fj}. Subsequently, the Vector-Dominance based Novosibirsk parameterization for $e^{+}e^{-}\to4\pi$ was 
applied to $\tau^{-}\to(4\pi)^{-}\nu_{\tau}$ by means of Charge-Vector-Conservation (CVC)  \cite{BONDAR2002139} and compared to $\tau$ decay through
CVC  \cite{PhysRevD.4.2821}.  In addition to the current formalism from the CLEO Model  \cite{Edwards:1999fj,Jadach:1993hs} and
Novosibirsk Model \cite{BONDAR2002139} we also implement a more generic current formalism with interchangeable vertex and propagators.
The CLEO Model for $\tau\to \pi\omega\nu_{\tau}$  \cite{Edwards:1999fj,Jadach:1993hs} is modified to include the generic vertex and propagator:

\begin{equation}
T^{\mu\nu}\epsilon_{\nu abc}q_{123}^{b}q_{4}^{c}P^{a\beta}\epsilon_{\beta lmn}q_{12}^{m}q_{3}^{n}P^{l\gamma}(q_{1}-q_{2})_{\gamma}
\end{equation}

\noindent following the convention from \cite{Decker:1992jy}.

Based on the formalism applied for the $\tau^{-}\to(4\pi)^{-}\nu_{\tau}$ decays, new phenomenological decay models 
are developed for the $\tau^{-}\to K^{-}\pi^{-}\pi^{-}\pi^{0}\nu_{\tau}$ and %
$\tau^{-}\to K^{-}\pi^{-}K^{-}\pi^{0}\nu_{\tau}$ 
decays. The  $\tau^{-}\to K^{-}\pi^{-}\pi^{-}\pi^{0}\nu_{\tau}$ decay through the
$K_{1}^{-}(1270)\to K^{-}\omega$ and $K^{*}(1410/1680)\to K_{1}(1270/1400)\pi$ intermediate states, where the $K^{*}\pi$ and $K\rho$ are included for the latter 
$K_{1}(1270)$ decays. 
From vector dominance it is assumed that the $\tau^{-}\to K^{-}\pi^{-}K^{-}\pi^{0}\nu_{\tau}$ mode is produced through an excited
$\rho$ meson which has a $J^{PC}=1^{--}$. 
Since charged light-mesons are not eigen-states of C-parity, it follows that the $\rho^{-}$ can decay into
$\rho^{-}\to a_{1}^{0}\pi^{-}$, $\rho^{-}\to h_{1}^{0}\pi^{-}$ and $\rho^{-}\to a_{1}^{-}\pi^{0}$. 
\footnote{The same argument of conservation of C-parity for the production of $h_{1}$ and $a_{1}$ states applies to
$\tau\to\rho(770)/\rho'(1450)^{-}/\rho''(1700)^{-}\nu_{\tau}\to\pi^{-}\pi^{-}\pi^{+}\pi^{0}\nu_{\tau}$ decays. This would have important implications for the
backgrounds when measuring the $a_{1}$ decay structure in $\tau^{-}\to\pi^{-}\pi^{-}\pi^{+}\nu_{\tau}$.
Moreover, the $b_{1}$ meson ($J^{PC}=1^{-+}$) could be produced through $a_{1}^{-}(1640)\to \pi^{-}b_{1}^{0}$ or $a_{1}^{-}(1640)\to \pi^{0}b_{1}^{-}$, where
the $b_{1}$ is known to decay into $\omega(782)\pi$  \cite{PDG2020}. Thus, the $b_{1}(1235)$ would be expected to be experimentally observable in the 
$\tau^{-}\to  3\pi 2\pi^{0}\nu_{\tau}$ decay with the $a_{1}(1640)$ resonance being produced at the W-vertex.}.
The inclusion of the  $h_{1}^{0}(1170)$ and
$h_{1}^{0}(1415)$ intermediate states in addition to the $ a_{1}(1260)$ are 
consistent with the predictions from the quark model  \cite{Kokoski:1985is}\footnote{We include an overall $e^{\imath\pi}$ phase difference between the 
$a_{1}$ and $h_{1}$ states.}. 
The $\tau^{-}\to K^{-}\pi^{-}\pi^{-}\pi^{0}\nu_{\tau}$ 
and $\tau^{-}\to K^{-}\pi^{-}K^{+}\pi^{0}\nu_{\tau}$ decays are 
modelled using the Feindt normalized S/D-wave parameterization formalism  \cite{Feindt:1990ev}, % \cite[Eq. 5 and 7]{Feindt:1990ev}, 

\begin{equation}
\resizebox{0.4\textwidth}{!}{$
J^{\mu}=\frac{4\sqrt{6}C_{g\otimes q}V_{ij}}{f_{\pi}^{2}}P_{1234}^{\mu\nu}A_{V\to AP}^{S/D}V_{\nu\alpha}^{Feindt,S/D}P_{123}^{\alpha\beta}A_{A\to VP}^{ S/D}V_{\beta\gamma}^{Feindt,S/D}P_{12}^{\gamma\delta}(q_{1}-q_{2})_{\delta},
$}
\end{equation}

\noindent  \cite{Jadach:1993hs} for $K^{*}\to K_{1}(1270/1400)\pi$ and $\rho\to a_{1}/h_{1}\pi$, where $C_{g\otimes q}$ represents the 
product of Clebsch-Gordan Coefficients combined with the amplitude factors from the quark model for 
each decay process and $V_{ij}$ is the corresponding CKM matrix element. The particle ordering is defined using the Levi-Civita so that $\epsilon_{i,j,k,l}=1$. 
The initial $K^{*}$ and $\rho$ states are modelled following the ChRL formalism,

\begin{equation}
\resizebox{0.38\textwidth}{!}{$
F_{\rho}= BW_{\rho(770)}(s)+A_{1}BW_{\rho(1450)}(s)+A_{2}BW_{\rho(1700)}(s)
$}
\end{equation}
\begin{equation}
\resizebox{0.39\textwidth}{!}{$
F_{K^{*}}=BW_{K^{*}(892)}(s)+B_{1}BW_{K^{*}(1410)}(s)+B_{2}BW_{K^{*}(1680)}(s)
$}
\end{equation}
\noindent with $A_{1}=-0.20$ and $A_{2}=-0.017$ from the CLEO Model description  \cite{Edwards:1999fj} and $B_{1}=-0.135$  \cite{Finkemeier:1995sr} 
with $B_{2}$ estimated from   
$B_{2}=B_{1}\frac{A_{2}}{A_{1}}\approx 0.011$. The intermediate  $K^{*}$ and $\rho$ states are defined in \cite{Finkemeier:1995sr}.  % \cite[Eq. 8 \& 10 ]{Finkemeier:1995sr}.  
The $h_{1}^{0}(1170)$ and $h_{1}^{0}(1415)$ are described as isolated Breit-Wigner distributions  \cite{Finkemeier:1995sr}, where the $h_{1}^{0}(1415)$ is suppressed by
$\delta_{h_{1}}=-0.175$ provides an overall suppression factor of $\approx 33\times$. The $K_{1}(1270/1400) \to K \omega$ is modelled with

\begin{equation}
\resizebox{0.4\textwidth}{!}{$
J^{\mu}=\frac{4V_{us}}{f_{\pi}^{2}}P_{1234}^{\mu\nu}A_{A\to VP}^{S/D}V_{\nu\alpha}^{Feindt,S/D}P_{123}^{\alpha\beta}
\epsilon_{\beta \gamma mn}q_{12}^{m}q_{3}^{n}
P_{12}^{\gamma\delta}(q_{1}-q_{2})_{\delta}.
$}
\end{equation}

\noindent
 The mixing factor $\epsilon$ between the $K_{1}(1270/1400)$ resonances from  \cite{Finkemeier:1995sr} is modified to match the world average for the 
branching ratios,  $K_{1}(1270)\to K\omega$ $[(11.0\pm 2.0)\%]$ and  $K_{1}(1400)\to K\omega$ $[(1.0\pm 1.0)\%]$   \cite{PDG2020}. The additional coefficients 
$A_{V\to AP}^{S/D}$ and $A_{A\to VP}^{ S/D}$ represent the relative amplitudes of S-wave and D-wave production for 
the interactions involving vectors ($V$), axial-vectors ($A$) and pseudo-scalar ($P$). To simplify the model, the relative S-wave and D-wave contributions to the 
$a_{1}(1260)$, $h_{1}^{0}(1170)$, $h_{1}^{0}(1415)$, $K_{1}(1270)$  and $K_{1}(1400)$ are fixed to the expected values \cite{PDG2020}\footnote{In most models,
the D/S-wave ratio in the $K_{1}(1270)$ and $K_{1}(1400)$ depends on the mixing angle and the relative suppression of the production mechanism in the model
\cite{Suzuki:1993,Asner:2000nx,Godfrey:1985xj,Blundell:1995au}. 
This means the mixing is not expected to be the same between kaon decays and
those produced directly at the W vertex in $\tau$ decays. Consequently, $K^{*}(1680)\to K_{1}(1270/1400)\pi^{0}$ decays where the $\pi^{0}$ is missed in 
the reconstruction will be an irreducible background for any $\tau^{-}\to K_{1}^{-}\nu_{\tau}$ analysis, in particular for measurements of the mixing angle 
 \cite{Asner:2000nx}.
}.
The  $\tau^{-}\to K^{-}\pi^{-}\pi^{-}\pi^{0}\nu_{\tau}$ and
$\tau^{-}\to K^{-}\pi^{-}K^{-}\pi^{0}\nu_{\tau}$ decays are primarily produced through the $K^{*}(1680)$ and $\rho^{''}(1700)$ which in the quark model have 
a D wave $q\bar{q}$ angular momentum, $1^{3}D_{1}$  \cite{Kokoski:1985is,Halzen:1984mc,Abele:2001pv}, and therefore
may have a significant D-wave fraction in the axial-vector and pseudo-scalar decays, particularly for the lower energy states. Moreover, unlike in the direct 
production from the W vertex in $\tau$ decays where second class currents are suppressed, in hadronic production the 
strange and non-strange singlet and triplet states are not suppressed by the same $SU(3)_{f}$ symmetry thereby the singlet states will play a more significant role 
in the decay structure.
%$K^{*}(1680)$  and $\rho^{''}(1700)$
%$C_{g}$ and $V_{ij}$ are the product of the Clebsch-Gordan Coefficients  \cite{PDG2020} and
%CKM Matrix element  \cite{PDG2020} for the individual modes.
The  $K_{1}(1270)$, $K_{1}(1400)$, $h_{1}^{0}(1170)$ and $h_{1}^{0}(1415)$ are modelled with a fixed width ChRL Breit-Wigner, 
\begin{equation}
\resizebox{0.23\textwidth}{!}{$ 
BW_{fixed}(s)=\frac{-M_{res}^{2}+\imath M_{res}\Gamma_{res}}{s-M_{res}^{2}+\imath M_{res}\Gamma_{res}}
$}
\end{equation}
\noindent  \cite{Finkemeier:1995sr}, where the $a_{1}(1260)$ is modelled using
\begin{equation}
\resizebox{0.3\textwidth}{!}{$ 
BW_{a_{1}(1260)}(s)\frac{M_{res}^{2}}{M_{res}^{2}-s-\imath M_{res}\Gamma_{res}g(s)/g(M_{res}^{2})}
$}
\end{equation}
\noindent the K\"uhn-Santamaria Model from  \cite{Kuhn:1990ad}  where $g(s)$ is defined in \cite{Kuhn:1990ad}.  %\cite[Eq. 3.16]{Kuhn:1990ad}.  
The vector resonances are described by the ChRL 
P-wave Breit-Wigner, 
\begin{equation}
\resizebox{0.3\textwidth}{!}{$ 
BW_{P-wave}(s)=\frac{M_{res}^{2}}{M_{res}^{2}-s+\imath \sqrt{s}\Gamma_{res}\frac{M_{res}^{2}}{s}\left(\frac{|P(s)|}{|P(M_{res}^{2})|}\right)^{3}}
$}
\end{equation} 
 \cite{Decker:1992kj}. The numerical values and S/D-wave fractions can be found in
Table \ref{table:SDwave}. This can be compared to the Flux-Tube Breaking Model where the excited $ K^{*}$ and $\rho$ resonances decay entirely in an S-wave configuration \cite{Kokoski:1985is}.  
The $\tau^{-}\to K^{-}\pi^{-}K^{-}\pi^{0}\nu_{\tau}$ invariant mass
distributions are presented in Figure \ref{fig:phemKpiKpi0} for an unsuppressed production of the $h_{1}^{0}(1170)\pi^{-}$ and
with a suppression of the $h_{1}^{0}(1415)\pi^{-}$ states. %, as presented in Figure \ref{fig:phemKpiKpi0}, these states should be observable at BELLE-II 
% \cite{abe2010belle,Kou_2019}. 
The relative 
suppression of the $h_{1}^{0}(1170)$ and $h_{1}^{0}(1415)$ states depends on the mixing between the $I=0$ singlet and octet states, where in the ideal mixing case the
mesons are composed of $\sqrt{\frac{1}{2}}(u\bar{u}+d\bar{d})$ and $s\bar{s}$ for the $h_{1}^{0}(1170)$ and $h_{1}^{0}(1415)$ respectively with 
the $h_{1}^{0}(1415)$ production 
 being OZI  \cite{Zweig:1964jf,Okubo:1963fa,Iizuka:1966} suppressed. Typically, the suppression factor for OZI  \cite{Zweig:1964jf,Okubo:1963fa,Iizuka:1966} 
modes is between $\approx 20-100\times$. The ideal mixing case is consistent with the measurement by the BES Experiment \cite{BES:2018}. Based on Figure \ref{fig:phemKpiKpi0}, which contains a suppression factor of $33\times$ for the $h_{1}^{0}(1415)$, 
it is expected that the BELLE-II Experiment  \cite{abe2010belle,Kou_2019} will be able to observe the $h_{1}^{0}(1415)$ state.
From Figure \ref{fig:phemKpiKpi0}, it can be seen that the production of the $h_{1}^{0}(1415)$ impacts the $\rho^{''}(1700)$ line-shape, 
the ratio of $K^{*,0}(892)$ to $K^{*,-}(892)$ and $K^{*,+}(892)$, the recoil in the 
$K^{-}\pi^{-}K^{+}$ invariant mass, in addition to the expected resonance peak in the $K^{-}\pi^{0}K^{+}$ distribution, 
providing a clean signature for its' observation.  
These channels also provide an opportunity to measure the D/S-wave ratio in the excited $K^{*}$ and $\rho$ $1^{3}D_{1}$ states.
The branching fractions are summarized in Table \ref{table:PhemBR}.

\subsection{Flux-Tube Breaking Model\label{sec:FluxTube}}

The Flux-Tube Breaking Model   \cite{Isgur:1988vm,Godfrey:1985xj,Kokoski:1985is,Isgur:1983wj,Isgur:1984bm} is implemented for both the non-strange and strange 
$\tau$ decays in two, three and some of the four meson systems.  The formalism is primarily based on   \cite{Isgur:1988vm},
with modifications to include additional resonances and the strange sector from   \cite{Godfrey:1985xj,Kokoski:1985is,Isgur:1983wj,Isgur:1984bm,Blundell:1995au}.
However, unlike the formalism in  \cite{Isgur:1988vm}, the hadronic currents and covariant amplitudes are constructed directly from the vertices and propagators. 
This allows for the use of the generic formalism for the hadronic current described in Section \ref{sec:Phem}, where the hadronic currents are defined to include the 
initial coupling and outgoing polarization vector compared to the covariant amplitude which describes the strong interaction.
The particle ordering of the covariant amplitudes follows the convention $\epsilon_{i,j,k}=1$ (or $\epsilon_{i,j,k,l}=1$) for the hadronic currents, 
instead of the convention from  \cite{Kuhn:1992nz}. 
The meson couplings in the hadronic current, displayed in Table \ref{table:IMRCouplingConstants} for each of 
the mesons produced at the $W^{\pm}$ vertex, are assumed to be constants  \cite{Isgur:1988vm}. The vertex amplitudes under the narrow-width approximation 
for the hadronic interactions, %\cite[Table II]{Kokoski:1985is}
 \cite{Kokoski:1985is}, are supplemented by Clebsch-Gordan coefficients for Isospin  \cite{PDG2020,Griffiths:1987tj}
to obtain the amplitude for the individual decay processes 
\footnote{For the Flux-Tube Breaking Model we choose the ordering convention $\epsilon_{i,j,k}=1$ (or $\epsilon_{i,j,k,l}=1$)  for the covariant amplitudes
to allow for direct application of Clebsch-Gordan coefficients for Isospin  \cite{PDG2020,Griffiths:1987tj} to obtain the amplitude for
individual decay modes. In the $3\pi$ modes, we directly impose the Bose symmetry.}.
The ``bare" propagator for the covariant amplitude is described in terms of an ``on-shell'' spin-one massive vector boson, the transverse vertices project out higher order corrections from 
effects such as one-particle-bubbles  \cite{Isgur:1988vm}. 
From this, it follows that depending on the dominant graph contribution to each individual resonance the propagator can be represented in terms
of the ``purely resonant time-ordered''  \cite{Isgur:1988vm} component of the propagator and non-resonant contribution,

\begin{equation}
\resizebox{0.375\textwidth}{!}{$
P(s)=\frac{\alpha}{s-\hat{m}^{2}(s)+\imath m_{0}\Gamma(s)}+\frac{1-\alpha}{2m_{0}\left(\sqrt{s}-\bar{m}(s)\right)+\imath m_{0}\Gamma(s)}
$}\label{eq:IMRpropogator}
\end{equation}

\noindent  \cite{Isgur:1988vm},  % \cite[Eq. 8-10]{Isgur:1988vm},
 where $\alpha$ is a free parameter related to the fraction of non-resonant to purely resonant contribution to the propagator.
This representation allows for the investigation of the dominant Z-graph contributions to each individual resonance.  The mass dependant decay widths for each resonance is 
constructed from the covariant amplitude, which begins with the incoming polarization vector of the initial meson resonance, to provide a width that in the
$\lim\limits_{s\to \infty} \Gamma(s)\to 0$ 
 \cite{Isgur:1988vm}\footnote{This definition for the mass-dependent width is model dependant  \cite{Isgur:1988vm} (and is not the same as in the CLEO model 
 \cite{CLEO3pi}). 
In models that use a point-like coupling, for example  \cite{Tornqvist:1987ch}, the mass dependant decay width increased indefinitely 
 \cite{Isgur:1988vm,Tornqvist:1987ch}. 
The choice of formalism for the mass-dependent width impacts not only the width of the resonance but also the observed $s$ dependent mass of the resonance.}.
The causal effect of the mass dependent decay widths on the mass of the resonance are taken into account in $\hat{m}^{2}(s)$ and $\bar{m}$ through the dispersion-relations,

\begin{equation}
\resizebox{0.375\textwidth}{!}{$                                                                                                                      
  \hat{m}^{2}(s)=m_{0}^{2}-\frac{1}{\pi}P\int_{s_{thres}}^{\infty}ds^{'} m_{0}\Gamma(s^{'})\left(\frac{1}{s^{'}-s}-\frac{1}{s^{'}-m_{0}^{2}} \right)  
$} \label{eq:IMRDisperMhat}
\end{equation}

\noindent  \cite{Isgur:1988vm} and    % \cite[Eq. A30]{Isgur:1988vm} and

\begin{equation}
\resizebox{0.375\textwidth}{!}{$
\bar{m}(s)=m_{0}-\frac{1}{2\pi}P\int_{m_{thres}}^{\infty}dm^{'} m_{0}\Gamma(m^{'2})\left(\frac{1}{m^{'}-\sqrt{s}}-\frac{1}{m^{'}-m_{0}} \right)
$} \label{eq:IMRDisperMbar}
\end{equation}

\noindent \cite{Isgur:1988vm}.  %\cite[Eq. A31]{Isgur:1988vm}. 

The transverse vertices projectors from \cite{Isgur:1988vm} are extended to include a more complete set of normalized vertices, which are summarized in
 Table \ref{table:IMRVertices}.
These are combined with the corresponding vertex amplitudes   \cite{Isgur:1988vm,Kokoski:1985is}\footnote{The mock-masses are obtained from 
%  \cite[Table V]{Kokoski:1985is}, 
 \cite{Kokoski:1985is}, the string-breaking constant, $\gamma_{0}$, is defined relative to the $\rho$ meson decay  \cite{Isgur:1988vm}  and $\beta_{A}$s
are approximated as $0.4$  \cite{Isgur:1988vm}, except for $\beta_{K_{A}}$ and $\beta_{K_{B}}$ which are $0.3$ based on  \cite{Blundell:1995au}, 
following the procedure for the nominal fit in  \cite{Isgur:1988vm}. } and the Isospin to obtain the vertex Form-Factor for the
individual interaction.

\subsubsection{Non-Strange Decay Modes\label{sec:IMRnonstrange}}
The non-strange vector and axial-vector contribution is assumed to be produced primarily through the $\rho(770)$ and $a_{1}(1260)$, 
where the pseudo-scalar contribution, $\pi(1300)$, and contribution from the second class currents are assumed to be negligible\footnote{Excited resonance of the 
$\rho(770)$ and $a_{1}(1260)$ state  \cite{PDG2020,Kokoski:1985is} are not incorporated in the current version of the generator.}.
In the vector current, the $\rho(770)$ state, both the  $\pi\pi$ and  $KK$ channels are taken into account.
The $a_{1}(1260)$ implementation is based on   \cite{Isgur:1988vm}\footnote{For the $a_{1}(1260)$, we find that a pole-mass larger than the world average \cite{PDG2020} 
is required, $m_{a_{1}(1260)}=1.26GeV/c^{2}$.}\footnote{The particle mass implemented in this instantiation of the Flux-Tube Breaking Model are based on the world average 
values \cite{PDG2020} unless otherwise stated. Although the Flux-Tube Breaking does predict the particle masses \cite{Godfrey:1985xj}, the world average values 
\cite{PDG2020} are typically more precise and allow for a more obvious association of the state in the Flux-Tube Breaking Model with the experimentally observed state.
Within the program, particle masses are user defined and not specific to the implementation.} with the intermediate state $\rho(770)$ and $K^{*}(892)$ and 
includes the decay final states: 
$\pi^{-}\pi^{-}\pi^{+}$, $\pi^{-}\pi^{0}\pi^{0}$, $K^{-}\pi^{-}K^{+}$,
$K^{0}\pi^{-}\bar{K}^{0}$ and $K^{-}\pi^{0}K^{0}$. In the CLEO measurement of $\tau\to (3\pi)^{-}\nu_{\tau}$  \cite{CLEO3pi} and more recently in the 
ChRL Model  \cite{Nugent:2013hxa}, 
the main sources of discrepancy between the theoretical predictions and experimental measurements was associated with additional resonances not included in the 
theoretical models  \cite{Stahl:2000aq,CLEO3pi,Nugent:2013hxa,Schmidtler}.  Therefore, we extend the Flux-Tube Breaking Model describing the $a_{1}(1260)$ decay  \cite{Isgur:1988vm} to include 
the lowest energy scalar intermediate state based on \cite{Kokoski:1985is}
%  \cite[Table III]{Kokoski:1985is}
 and Table \ref{table:IMRVertices}\footnote{Additional extensions to the $a_{1}(1260)$ model may be provided in a future publication.}, which is associated to the $f_{0}(1370)$ and includes the decays: $f_{0}(1370)\to \pi^{0/-}\pi^{0/+}$; $f_{0}(1370)\to K^{0/-}K^{0/+}$; and $f_{0}(1370)\to \eta\eta$\footnote{The $f_{0}(1370)\to \gamma \gamma$
 and $f_{0}(1370)\to 4\pi$  channels have been observed, but are not included in the current model.}. 
This is in contrast to the more conventional association of the lowest energy scalar state to the $f_{0}(500)$ in the Unitarized Quark Model \cite{Tornqvist:1995kr}\footnote{In \cite{Godfrey:1985xj}, the lowest energy scalar mass was predicted to be $~1.09GeV/c^{2}$, but they did not take this prediction seriously and explicitly 
did not associate it with the newly measured $f_{0}(1370)$ arguing the $f_{0}(1370)$ was a residual of the nearby $1^{++}$ state \cite[Ref. 9]{Godfrey:1985xj}. 
Instead, they associate it with a $qq\bar{q}\bar{q}$ state with $[m_{\pi\pi}\simeq 0.86GeV/c^{2},\Gamma_{\pi\pi}=0.85GeV]$ \cite{Godfrey:1985xj} which would be 
consistent with the $f_{0}(500)$.}.  
The choice of this extension will become apparent once the dynamics of the $f_{0}(1370)$ in this ${}^{3}P_{0}$ model are explained.
The two-body width formula %\cite[Eq. A16]{Isgur:1988vm}
\cite{Isgur:1988vm} has been modified to the more generic form,

\begin{equation}
\resizebox{0.375\textwidth}{!}{$ 
\Gamma_{A}(s_{i})=\Gamma_{A}(m_{0}^{2})\frac{\sum \frac{\left(C_{A\to CB}f_{A\to CB}(s_{i})\right)^{2}\left(k_{i}^{2}(s_{i})\right)^{\frac{2J+1}{2}}}{\sqrt{s_{i}}}}{
\sum \frac{\left(C_{A\to CB}f_{A\to CB}(m_{A}^{2})\right)^{2}\left(k_{i}^{2}(m_{A}^{2})\right)^{\frac{2J+1}{2}}}{\sqrt{m_{A}^{2}}}}
$}
\label{eq:IMRwidth}
\end{equation}

\noindent where
\begin{equation}
\resizebox{0.375\textwidth}{!}{$
k_{i}(s_{i})=\sqrt{\left(s_{i}-(m_{B}+m_{C})^{2}\right)\left(s_{i}-(m_{B}-m_{C})^{2}\right)}/\left(2\sqrt{s_{i}}\right),
$}
\end{equation}
\noindent $J$ is the spin, $C_{A\to CB}$ 
is the Clebsch-Gordan coefficient and the sum is conducted over all 2-body final states.
After including the running of the decay width, Equation \ref{eq:IMRwidth}, and mass from the dispersion relations, 
Equations \ref{eq:IMRDisperMhat} and/or \ref{eq:IMRDisperMbar}, 
in the propagator it is found that there is a significant distortion of the line shape of the scalar resonance $f_{0}(1370)$ and $K_{0}^{*}(1430)$, 
as seen in Figure \ref{fig:scalar_IMR}.
In particular for the $f_{0}(1370)$, threshold effects for the pair production of $K^{+}K^{-}$ and $K^{0}\bar{K}^{0}$ create peaking structures consistent
with the $f_{0}(980)$ and the pair production of $\pi^{+}\pi^{-}$ create a broad distribution consistent with the $\sigma/f_{0}(500)$. For the $K_{0}^{*}(1430)$,
the main observed threshold effects comes from the pair production of $K^{\pm}\pi^{0}$ and $K^{0}\pi^{\pm}$ for charged and $K^{0}\pi^{0}$ and $K^{\pm}\pi^{\mp}$ 
for neutral resonances, resulting in the a broad distribution consistent with the $\kappa/K_{0}^{*}(700)$. 
The $K\eta$ threshold produces an additional slight threshold peak, however, the amplitude is substantially smaller than the $K\pi$ threshold effect.
The $K\eta{\prime}(958)$ threshold occurs near the $K_{0}(1430)$ resonance peak causing a sudden enhancement of the decay width which significantly changes
the resonance line-shape.
 These large distortions are consistent with predictions from other $^{3}P_{0}$ 
models  \cite{Tornqvist_1995,Beveren,Boglione_2002} but are contrary 
to models such as the tetra-quark model  \cite{Alford_2000,Maiani_2004,_t_Hooft_2008}, 
linear $\sigma$ models  \cite{Tornqvist_1999,Napsuciale_2004,Napsuciale_2004b,Ishida_1999,Sadron_1999,Fariborz_2014,Black_1999}, Chiral-Lagrangian models  \cite{PhysRevD.59.074001,PEL_EZ_2004,Dabado_1997,Oller_2003} and molecular models  \cite{Phys.Rev.130.776,Kalashnikova_2005}
 that treat these low mass scalars as distinct resonances  \cite{PDG2020}. Alternatively, the COMPASS Experiment has associated the $f_{0}(980)$ 
production with a Triangular Singularity \cite{Landau:1959} generating a ``resonance-like state with axial-vector quantum numbers''  
\cite{Alexeev:2719624,PhysRevLett.115.082001}, where $\tau$ decays would provide a clean
environment to confirm this effect  \cite{Dai_2019}.
Although this prediction is sensitive to the amplitudes, vertex Form-Factor, $f_{S\to PP}(s_{i})$, and Z-graph contributions 
%\footnote{Within this Flux-Tube Breaking Model 
%the formalism for the dispersion relations are well defined, however, as mentioned earlier there is a model assumption about the point-like/compositeness of the meson 
%in the formalism and subtractions employed in the dispersion relations.} 
there are three distinguishing
features that can be used to discriminate this prediction for
the nature of the $f_{0}(500)/\sigma$, $f_{0}(980)$ and $K_{0}^{*}(700)/\kappa$ from the assumption that they are true resonances. 1) The $f_{0}(980)$ 
exhibits a double peak structure
corresponding to the two pair-production threshold that contribute to the peak, namely the $K^{+}K^{-}$ and $K^{0}\bar{K}^{0}$ pair production thresholds, and thereby 
has a strong $s\bar{s}$ coupling. 2) The
interference between the $f_{0}(500)/\sigma$ ($f_{0}(980)$ or $K_{0}^{*}(700)/\kappa$) with other resonances does not change sign at their peak as one 
would expect for a resonance,
for example see Equation \ref{eq:IMRpropogator}. 3) The $\sigma\to\eta\eta$ contribution generates a small peak related to the $\eta\eta$ threshold at 
$\sim 1.096GeV/c^{2}$ ($2\times m_{\eta}$). 
The nature of the low mass scalars is of particular interest because in many models, they play a significant role in
the constituent-quark mass of the mesons and baryons through 
Chiral-Symmetry-Breaking in models such as Chiral-Lagrangian   \cite{PhysRevD.59.074001,PEL_EZ_2004,Dabado_1997,Oller_2003} and linear $\sigma$ Models  \cite{Tornqvist_1999,Napsuciale_2004,Napsuciale_2004b,Ishida_1999,Sadron_1999,Fariborz_2014,Black_1999}, an analog of the ``Higgs'' mechanism in Electro-weak symmetry breaking  \cite{PDG2020}. In the Flux-Tube Breaking Model, the coupling of the scalars are not dependant on mass of the outgoing mesons as one would expect if they contributed significantly to
the constituent-quark mass through Chiral-Symmetry-Breaking.  

For the higher mass scalars predicted by the  \cite{Kokoski:1985is} model, we associate the $\epsilon_{s}(1360)$ with the $f_{0}(1500)$, where the $f_{0}(1510)$ 
is predicted to be an artifact resulting from the $\eta\eta^{'}$ threshold, while the $\epsilon(1780)$ and $\epsilon_{s}(1990)$ are associated with the $f_{0}(1710)$ and 
$f_{0}(2020)$ respectively. For the strange scalars, the excited scalar $\kappa(1890)$ is associated with the $K_{0}^{*}(1980)$. These additional scalar states
are expected to have some residual contribution to the low mass scalar threshold effects within the Flux-Tube Breaking Model.

In   \cite{Isgur:1988vm}, there were two issues limiting the implementation of the dispersion relation in the nominal fit. The first issue was related to the strong $s$ dependence of $a_{1}(1260)$.  It was found that by including the  $f_{0}(1370)$,
that the $a_{1}(1260)$ line-shape was stabilized, which validates the conjecture from 
 \cite{Isgur:1988vm} that the inclusion of additional resonances will stabilize the line-shape. The second limiting factor in investigating the impact of the 
dispersion relations to the line-shape of the $a_{1}(1260)$ resonance 
was the convergence of the integrations.  From numerical studies, an upper bound from the truncation error on $\hat{m}(s)$ and $\bar{m}(s)$ for the dispersion integral 
can be estimated to be $\ll 0.1\%$. However,
the largest source of uncertainty in the computation of the dispersion relation integrals is the numerical stability of $\Gamma(s)$ in terms of the
sampling density. For $0.25GeV^{2}/c^{4}<s<m_{\tau}^{2}$ the numerical error related to the sampling and numerical integration technique is $\le 0.6MeV/c^{2}$ for $\Gamma(s)$ 
and $\le 0.2\%$ for $\hat{m}(s)$ and $\bar{m}(s)$.
Figure \ref{fig:IMRConverge} illustrates the numerical stability related to the sampling and numerical integration technique of the integral for  
an upper limit of integration of $3\times m_{\tau}^{2}$, $10\times m_{\tau}^{2}$, $20\times m_{\tau}^{2}$, $40\times m_{\tau}^{2}$ and $80\times m_{\tau}^{2}$.

\subsubsection{Strange Decay Modes \label{sec:IMRstrange}}
The strange axial-vector contribution to the Flux-Tube Breaking Model  \cite{Isgur:1988vm,Godfrey:1985xj,Kokoski:1985is} is produced through the vector $K^{*}(892)$
$K_{1}(1270)$ and $K_{1}(1400)$ resonances, a mixture of the $K_{A}$(triplet [$1^{3}P_{1}$]) and $K_{B}$(singlet [$1^{1}P_{1}$]) states
, and the scalar, $K_{0}^{*}(1430)$, and pseudo-scalar contribution, $K^{*}(1460)$. Higher excitations are currently neglected.   
The vector resonance $K^{-,*}(892)$ proceeds through the $K^{-}\pi^{0}$ and $\bar{K}^{0}\pi^{-}$ channels. 
The most significant axial-vector decay
modes: $ K^{*}(892)\pi$, $K\rho(770)$,$K^{-}f_{0}(1370)$, $K_{0}^{*}(1430)\pi$ and
$K^{-}\omega(782)$ are included in the decay model which accounts for all of the measured fully hadronic modes  \cite{PDG2020}\footnote{The radiative decay 
mode $K_{1}^{-}\to K^{-}\gamma$, which has only been observed  \cite{PDG2020}, and $K_{A}\to K_{B}\pi$ mode, which has not been observed, are currently not included.}. 
For the $K_{1}^{-}\to\omega(782)K^{-}$ and $K^{-}(1400)\to\omega(782)K^{-}$, since the decay width 
of the $\omega(782)$ is comparable to the detector resolution, we compute the decay width for the four-body phase space integral utilizing a change of variables for the
narrow Breit-Wigner distribution  \cite{Nugent:2013hxa,Jadach:1993hs}.  
The amplitude of the decay $\omega(782)\to(\rho\pi)^{0}\to\pi^{+}\pi^{-}\pi^{0}$ is calculated in the context of the
normalized harmonic-oscillator $^{3}P_{0}$ amplitudes\footnote{The normalized harmonic-oscillator $^{3}P_{0}$ amplitudes, Amplitude/A, are defined in
Appendix B of   \cite{Kokoski:1985is}.}, within the context of Flux-Tube Breaking model  \cite{Kokoski:1985is}
to be $-\frac{9}{5}P_{1}$, where the parity flip from the Levi-Civita tensor in the vertex Form-Factor is taken into account \cite{PICH1987561}.
\footnote{The total width is determined by scaling the
measured branching fractions \cite{PDG2018}, and Clebsch-Gordan Coefficients are included.}.

 In contrast to hadronic production where the traditional mixing between the $K_{1}$
singlet and triplet states is applied  \cite{Godfrey:1985xj},
the $K_{1}$ singlet is a second class current and therefore predicted to be
suppressed in the decay of $\tau$ leptons  \cite{Suzuki:1993}.  The contribution of the $K_{1}$ singlet state is expected to be proportional
to the $SU(3)_{f}$ symmetry breaking effect, $|\delta_{K_{1}}|\sim\frac{m_{s}-m_{u}}{(m_{s}+m_{u})}=0.25$  \cite{Suzuki:1993}\footnote{The
$SU(3)_{f}$ symmetry breaking in   \cite{Suzuki:1993} includes the relative $\sqrt{2}$ factor
between the amplitudes of the ${}^{1}P_{1}$ and ${}^{3}P_{1}$ decay amplitudes \cite{Blundell:1995au}, which is already included in the definition of the $K_{A}$ and $K_{B}$ 
states  \cite{Kokoski:1985is}. We also note that there is a phase difference between  \cite{Blundell:1995au} %\cite[Eq. 18]{Blundell:1995au}
 and Equation \ref{eq:K1mixing} due to the definition of
the $K_{A}$ and $K_{B}$ amplitudes in  \cite{Kokoski:1985is}.} within the non-relativistic static quark limit. In general, the value of $|\delta_{K_{1}}|$ is 
model dependant and most likely $s$ dependant, $\delta_{K_{1}} \to \delta_{K_{1}}(s)$. 
Given sufficient accuracy, the ratio of $m_{s}-m_{u}$ to $m_{s}+m_{d}$ may provide discrimination between the various quark models and is complimentary to the more complete
determination of the quark and gluon kinetic and potential energy tensor  \cite{Ji:1995a,Ji:1995b,SoLID}.
The symmetry breaking term for the
singlet/triplet $K_{1}$ states, $\delta_{K_{1}}$, is implemented as a free parameter to allow the impact of the suppression to be studied, where
\begin{equation}
\resizebox{0.4\textwidth}{!}{$
\left[
\begin{array}{l}
|K_{1}(1270)>\\ 
|K_{1}(1400)>
\end{array}
\right]
=
\left[
\begin{array}{ll}
\delta_{K_{1}}cos(\theta_{K_{1}}) & sin(\theta_{K_{1}}) \\
-\delta_{K_{1}}sin(\theta_{K_{1}}) & cos(\theta_{K_{1}}) \\

\end{array}
\right]
\left[
\begin{array}{r}
|K_{B} \left(1^1P_{1}\right)> \\
|K_{A}\left( 1^3P_{1}\right)>
\end{array}
\right].\label{eq:K1mixing}
$}
\end{equation}

\noindent In this formalism the decay amplitudes of the physical $K_{1}$ states are directly related to the relative fraction of  
${}^{1}P_{1}$ and  ${}^{3}P_{1}$ states produced which may explain the variation in the $K_{1}(1270)$ mass and widths reported in  \cite{PDG2020}, with $\theta_{K_{1}}$ being
independent of the production mechanism, unlike in \cite{Blundell:1995au} where the Heavy-Quark limit is applied.
In   \cite{Asner:2000nx}, the CLEO experiment used the relative branching fractions of $\tau^{-}\to K^{*}(892)\pi^{-}\nu_{\tau}$
and $\tau^{-}\to \rho^{0}(770)K^{-}\nu_{\tau}$ to determine the mixing angle between the $K_{1}$ singlet and triplet states($\theta_{K_{1}}$).
However, in the Flux-Tube Breaking Model  \cite{Isgur:1988vm}, as illustrated by the  
decay mode dependent hadronic widths of the $K_{1}(1270)$ resonance presented in Figure \ref{fig:IMRK1}, 
there is also a strong correlation on the mass dependence decay widths of the $K_{1}(1270)$ and $K_{1}(1400)$ resonances which was neglected in   \cite{Asner:2000nx}.
Table \ref{table:IMRK1Gbar} presents the 
corresponding width at the pole-mass, 
$\Gamma(m_{res}^{2})$, for common values of the $K_{1}$ mixing angle and the $SU(3)_{f}$ symmetry breaking factor found in the 
literature  \cite{Suzuki:1993,Asner:2000nx,Godfrey:1985xj} and the corresponding measured values  \cite{PDG2020}. 
From this, it can be seen that there is also a strong correlation 
on the mass dependence decay widths of the $K_{1}(1270)$ and $K_{1}(1400)$ resonances which was neglected in  \cite{Asner:2000nx}. Inclusion of the mass dependant decay
widths for each of the individual decay modes 
in addition to the mode dependant amplitudes\footnote{It is important to note that the decay amplitudes in % \cite[Table II]{Kokoski:1985is} 
\cite{Kokoski:1985is} depend on both the intermediate
resonances and if they are produced through a S-wave or D-wave.}
would allow both $\theta_{K_{1}}$ and $\delta_{K_{1}}$ to be measured simultaneously, modulo an overall 
sign ambiguity.
 The sign ambiguity can be resolved by additionally measuring a production mechanism of the 
$K_{1}(1270)$ and $K_{1}(1400)$ states which is not produced through a mechanism that involves the $SU(3)_{f}$ symmetry breaking factor. 
In the context of $\tau$ decays, the excited $K^{*}(1680)$ state into $K_{1}(1270)\pi^{0}$ and 
$K_{1}(1400)\pi^{0}$, see % \cite[Table II]{Kokoski:1985is}
\cite{Kokoski:1985is} or  Sections \ref{sec:Phem}, is one of two 
primary process through which $(K\pi\pi\pi)^{-}$ proceeds. Thus with
the expected luminosity at BELLE-II  \cite{abe2010belle,Kou_2019}, $\tau$ decays provide an opportunity, ambient model dependent, to over constraint the mixing angle of the 
$K_{1}$ singlet and triplet states and the $SU(3)_{f}$ suppression factor in the strange system.
The invariant mass of the $K_{1}(1270)$ and $K_{1}(1400)$ are
presented for the ChRL models  \cite{Finkemeier:1995sr,Decker:1992kj}, the experimental CLEO model  \cite{Asner:2000nx} and the Flux-Tube Breaking model
based on  \cite{Isgur:1988vm,Suzuki:1993,Godfrey:1985xj,Kokoski:1985is} for  the expected mixing angles $\pm33/\pm57$  \cite{Suzuki:1993,Godfrey:1985xj}
 in Figure \ref{fig:IMRK1}.

The scalar/pseudo-scalar covariant amplitudes for the $K_{0}(1430)$ and $K(1460)$ resonances are constructed from the vertex amplitudes formalism from
Table \ref{table:IMRVertices} for amplitudes defined in  % \cite[Table III]{Kokoski:1985is}. 
 \cite{Kokoski:1985is}.
The $K_{0}(1430)$ decays include the $K\pi$, $K\eta$ and $K\eta{\prime}(958)$ final states  \cite{Kokoski:1985is}. The impact of mass dependent width and the associated
dispersion relations on the line-shape of the $K_{0}(1430)$ has already been described in Section \ref{sec:IMRnonstrange} and is summarized in Figure \ref{fig:scalar_IMR}.
In the three body decays which dominate the $K(1460)$ the impact of the threshold effect and dispersion relations is less significant.
The $K(1460)$ primarily decays through an intermediate vector resonance, $K^{*}(892)\pi$, $K^{*}(892)\eta$, $K\rho(770)$, $K\phi(1020)$ and $K\omega(782)$, but also
includes contributions from $K_{0}^{*}(1430)\pi$   \cite{PDG2020,Kokoski:1985is}.
From Isospin symmetry and  %\cite[Table IV]{Kokoski:1985is},
 \cite{Kokoski:1985is}, the $K(1460)\to K_{0}^{*}(1430)\pi$ amplitude is $\left(\frac{27}{8}\right)S_{2}$.

\subsubsection{Overview of Flux-Tube Breaking Models}
The predicted branching ratios for the decay modes included in the Flux-Tube Breaking Model can be found in Table \ref{table:IMRBR}. For $\tau^{-}\to\rho^{-}\nu_{\tau}$,
the $\alpha_{\rho}=1$ predictions are more consistent with the measured branching ratios  \cite{PDG2020}, in particular for
the $\tau^{-}\to K^{-}K^{0}\nu_{\tau}$ branching ratio. This is consistent with the expectations from the numerous precision measurements of the $\rho$ meson 
 \cite{PhysRevLett.103.231801,Davier:2020dy,Davier:2003pw,Aloisio:2004bu,Akhmetshin:2001ig,achasov:2006vp,Dumm_2013,Fujikawa_2008}. 
For the $\tau^{-} \to K^{*,-}(892) \nu_{\tau}$ there appears to be a slight preference for the $\alpha_{K^{*}(892)}=0$,  
the pure-resonant time ordered contribution, however the excited $K^{*}$ states and a complete line-shape comparison to the data 
need to be included before drawing conclusions. 
The line-shape of the $a_{1}(1260)$ tends to be more consistent with $\alpha_{a_{1}(1260)}=0$, or a purely resonant state after applying the dispersion relations. 
Unlike the previous resonances, the $K_{1}(1270)$ and $K_{1}(1400)$ decay modes in 
Table \ref{table:IMRBR}, are presented in terms of the $K_{1}$ mixing angle and the $SU(3)_{f}$ symmetry breaking factor, $\delta_{K_{1}}$.
The simulated branching ratios tend to be significantly higher than the world averages \cite{PDG2020}, a problem that is also seen in ChRL models. 
This may indicate that $f_{K_{1}}$ is lower than in Table \ref{table:IMRCouplingConstants}, which would not be unexpected given the model dependence of 
$f_{K_{1}}$ in  \cite{Blundell:1995au}. The relative magnitude of the $K^{*}(892)\pi$ and $K\rho(770)$ amplitudes is strongly dependant on the $SU(3)_{f}$ 
suppression factor, $\delta_{K_{1}}$, and these results may indicate a magnitude other than $|\delta_{K_{1}}|=0.25$. However, contributions from the 
excited $K^{*}$ resonances should be included since they are expected to impact the ratio of $K^{*}(892)\pi$ to $K\rho(770)$ production.
Based on the predicted branching ratios in Table \ref{table:IMRBR}, the resonant widths at the pole-mass for the $K_{1}(1270)$ and $K_{1}(1400)$, 
seen in Table \ref{table:IMRK1Gbar}, and Figure \ref{fig:IMRK1} the preferred $K_{1}$ mixing angle and the $SU(3)_{f}$ symmetry breaking factor are $\theta_{K_{1}}=\pm33$ 
and $\delta_{K_{1}}=\mp0.25$ respectively\footnote{The mixing angle between the singlet and triplet
states of the $K_{1}$ from  \cite{Suzuki:1993} are consistent with the angle from  \cite{Godfrey:1985xj}.} In Figure \ref{fig:IMRK1}, the  
$K_{1}(1270)$ and $K_{1}(1400)$  pole-masses come from the world averages \cite{PDG2020}. Based on the comparison of distributions in Figure \ref{fig:IMRK1}, the 
pole-masses for the $K_{1}(1270)$ state may be higher than the world averages \cite{PDG2020}, as was predicted in \cite{Godfrey:1985xj}. \footnote{This discrepancy could be related 
to the model dependant interpretation of the data, in particular, fixed-width Breit-Wigner and variable width Breit-Wigner constructed with dispersion relations are not expected to 
have the same pole-mass or width.}. Once confronted with experimental data, this Flux-Tube Breaking model will provide the 
opportunity to further test the 
underlying assumptions of the QCD models and to extract the associated physical parameters, including the pole-mass, string breaking constant and 
$K_{1}$ mixing angle in conjunction with the $SU(3)_{f}$ symmetry breaking factor $\frac{m_{s}-m_{u}}{m_{s}+mu_{u}}$.

\subsection{Summary and Comparison of Hadronic Models \label{sec:Hadcompare}} 

The Flux-Tube Breaking Model incorporates the more detailed decay and spin structure of the strange and non-strange decay modes, while the 
Chiral-Resonance-Lagrangian Models form a more complete set of final state decay modes.
In contrast, the phenomenological models focus on specific decay modes primarily from individual 
experimental measurements which were developed to investigate model assumptions.
The Gounaris-Sakuria  \cite{Lees:2012cj,Gounaris:1968mw} convention is typically used in measurements of $e^{+}e^{-}\to \rho^{0}(\gamma)$, while the K\"uhn-Santamaria
 \cite{Kuhn:1990ad} convention has been employed in $\tau$ decays  \cite{Jadach:1993hs}.
The simulated hadronic invariant mass distribution for $\tau^{-} \to \rho^{-}\nu_{\tau}\to \pi^{-}\pi^{0}\nu_{\tau}$ is presented in Figure \ref{fig:rho_compare}, for
the K\"uhn-Santamaria Model  \cite{Lees:2012cj,Kuhn:1990ad},  Gounaris-Sakuria Model   \cite{Lees:2012cj,Gounaris:1968mw}, ChRL Model 
 \cite{Kuhn:1990ad} with the parameterization corrected to the measured values  \cite{Lees:2012cj} and QCD Flux-Tube Breaking Model
  \cite{Isgur:1988vm,Kokoski:1985is}. Above the $\rho$ mass, the most apparent discrepancy is due to the Flux-Tube Breaking Model missing the $\rho^{\prime}(1450)$ and 
$\rho^{\prime\prime}(1700)$ resonances. However, a more physically significant discrepancy between the K\"uhn- Santamaria Model  \cite{Lees:2012cj,Kuhn:1990ad} 
and Gounaris-Sakuria Model   \cite{Lees:2012cj,Gounaris:1968mw} model is related to the normalization requirement for the vector and axial-vector propagators at 
$s=0$\footnote{ChRL Models, 
such as the K\"uhn-Santamaria Model  \cite{Kuhn:1990ad}, normalize the propagator so that $|P(s=0)|^{2}=1$. This generates a difference in the logarithmic dependence 
of the dispersion relations which can be observed in the mass spectra, particularly in the low mass region.}. In  \cite{Lees:2012cj} and  \cite{Fujikawa_2008} the 
Gounaris-Sakuria Model  \cite{Gounaris:1968mw} was observed to have good agreement with the data in the low-mass region. 
The Gounaris-Sakuria  \cite{Lees:2012cj,Gounaris:1968mw} convention is typically used in measurements of $e^{+}e^{-}\to \rho^{0}(\gamma)$, while the K\"uhn-Santamaria
 \cite{Kuhn:1990ad} convention has been employed in $\tau$ decays  \cite{Jadach:1993hs}.
Furthermore, in the same low mass regions, the Flux-Tube Breaking Model has an enhancement relative to the 
K\"uhn-Santamaria Model  \cite{Lees:2012cj,Kuhn:1990ad},  Gounaris-Sakuria Model   \cite{Lees:2012cj,Gounaris:1968mw} which is a consequence of including the 
vertex Form-Factors\footnote{In contrast to the other models  \cite{Gounaris:1968mw,Kuhn:1990ad,Dumm_2013,Gonzalex:2019sm}, the Flux-Tube Breaking Model has yet to be 
fitted/tuned to the data.}. More recently, ChRL models supplemented with dispersion relations by means of the Omnes equation  \cite{Gonzalex:2019sm,Gonz_lez_Sol_s_2019} 
when tuned to the BELLE data  \cite{Fujikawa_2008} also provide an adequate description of the low mass region.  

For the $\tau\to 3\pi\nu_{\tau}$, five categories of hadronic models are included in the simulation. As in the $\rho$ decay mode a simplified
Chiral-Resonance Model \cite{Decker:1992kj}\footnote{The simplified
Chiral-Resonance Model is modified using coefficients from   \cite{Decker:1992kj} to match the numerical values in TAUOLA  \cite{Jadach:1993hs}.},
the K\"uhn-Santamaria  \cite{Lees:2012cj,Kuhn:1990ad}, the K\"uhn-Santamaria modified with the
Gounaris-Sakuria convention  \cite{Lees:2012cj,Gounaris:1968mw}, a Flux-Tube Breaking Model  \cite{Isgur:1988vm} and the phenomenological models from the CLEO experiment 
 \cite{CLEO3pi} and Feindt  \cite{Feindt:1990ev} are included. The  K\"uhn-Santamaria  \cite{Kuhn:1990ad}, Flux-Tube Breaking Model  \cite{Isgur:1988vm} and 
CLEO Model  \cite{CLEO3pi} have been the 
primary models for investigating the $a_{1}(1260)$  \cite{PDG2020}, although more recently, the improved ChRL Model  \cite{Nugent:2013hxa} was 
fit to the BABAR data  \cite{Tau2012}. Dispersion relations have only been included in the nominal fit for the  K\"uhn-Santamaria  \cite{Kuhn:1990ad} 
and ChRL Model  \cite{Nugent:2013hxa,Shekhovtsova_2012}. Although the dispersion relations have a significant impact on the resonance line-shape, 
as was mentioned previously, one of the main conclusions from the CLEO measurement  \cite{CLEO3pi} and the fitting of the ChRL Model  \cite{Nugent:2013hxa,Tau2012}
was the need for additional resonances to be included to describe the data. The line-shape of the $f_{0}(1370)$ with the ${^{3}P_{0}}$ Flux-Tube Breaking Model 
bares a striking resemblance to the discrepancies from the pull 
plots for the fitted distributions that were associated with multiple missing resonances in  \cite{Nugent:2013hxa} and may explain the discrepancies found at high 
mass in  \cite{CLEO3pi,Schmidtler,Opal3pi,Argus3pi,Delphi3pi}.  Thus the Flux-Tube Breaking Model, modified to include the $f_{0}(1370)$, 
is qualitatively supported by the experimental
data and should not be considered excluded. Additional resonances, in particular the $f_{2}(1270)$ and $a_{1}(1640)$, have been reported in
 \cite{CLEO3pi} and are expected to play a non-negligible contribution and further support this conclusion. 
A comparison of the K\"uhn-Santamaria  \cite{Lees:2012cj,Kuhn:1990ad} and the K\"uhn-Santamaria modified with the
Gounaris-Sakuria convention  \cite{Lees:2012cj,Gounaris:1968mw} illustrate that the convention of the dispersion relation used in the model can
have a significant impact. The Flux-Tube Breaking Model  \cite{Isgur:1988vm}, include contributions from the vertex factors in the decay structure of the
mesons  \cite{Isgur:1988vm} as well as a complete description of the scalar sector. To illustrate the significance of the vertex projector,
a variant of the Feindt Model  \cite{Feindt:1990ev} with the Chiral-Resonance Model  \cite{Decker:1992kj} Form-Factors is presented in Figure \ref{fig:a1_compare}.
Additional phenomena neglected in these models, such as the ``off-shell'' $s$
dependence of the meson couplings  \cite{Isgur:1988vm}, can have a non-trivial impact on the observed mass distributions.

Thus far, we have not explicitly compared the angular momentum states of the hadronic resonances in terms of the covariant amplitudes. It is clear from Section 
\ref{sec:ChRL}-\ref{sec:FluxTube}, that the Form-Factor formalism \cite{Isgur:1988vm,Peskin:1995ev} for constructing the covariant amplitude in terms of 
vertex operators is not unique. Instead, the
vertex Form-Factor depends on the assumptions on which properties are assumed to be conserved at the vertex \cite{Isgur:1988vm}.
When there is a single scattering wave-amplitude for a given vertex, the
vertex factor should reproduce the well defined theoretical angular distributions for the angular momentum states corresponding to the 
quantum number of the incoming and outgoing states \cite{MarkusWagner}. For interactions where this is not the case, such as $A\to V+P$, the Born level diagrams do not
necessarily correspond to the angular momentum states, but instead can be an admixture of these states \cite{Isgur:1988vm,Feindt:1990ev}.
To illustrate this point, Figure \ref{fig:coschi_omegah} presents the cosine of the angle between the normal to the plane of the $\omega$ meson and the pion/kaon,  
$cos(\chi_{\omega h})$.
The $\tau^{-}\to \pi^{-} \omega \nu_{\tau}$ for the CLEO Model  \cite{Edwards:1999fj,Jadach:1993hs}, Novosibirsk  Model \cite{BONDAR2002139}  and Decker-Mirkes Model 
 \cite{Decker:1992jy} \footnote{For the Decker-Mirkes Model Vector ($\tau^{-}\to\pi^{-}\omega\nu_{\tau}$) the covariant amplitude is constructed with the spin 1 propagator 
following the generic vertex/propagator 
formalism from \ref{sec:Phem}. Except for the normalization related to the vertex factor, this covariant amplitude is equivalent to the Flux-Tube Breaking model 
 \cite{Isgur:1988vm,Godfrey:1985xj,Kokoski:1985is}.} are predicted to be in a
$J^{P}(l=1)=1^{-}$ state, with an angular distribution $1-cos^{2}(\chi_{\omega h})$. The CLEO Model \cite{Edwards:1999fj,Jadach:1993hs} and Decker-Mirkes 
Model  \cite{Decker:1992jy}, which differ by the inclusion of the propagator, are consistent with the theoretical prediction as would be expected for a transverse vertex, 
however, the hadronic current from the Novosibirsk Model \cite{BONDAR2002139} is not consistent with the theoretical prediction\footnote{This discrepancy appears to 
be related to an incorrect covariant/contravariant index in the Levi-Civita tensor  \cite[Eq. 17]{Bondar:1999} which is propagated to  \cite[Eq. 25]{BONDAR2002139}.}.
For the $\tau^{-}\to K_{1}^{-}(1270)\to K^{-}\omega\nu_{\tau} \to K^{-}\pi^{-}\pi^{+}\pi^{0}\nu_{\tau}$ system,
the $A\to VP$ vertex of the $K^{-}\omega$ is constructed using the Feindt Model \cite{Feindt:1990ev}, Decker-Mirkes Model  \cite{Decker:1992jy},
Novosibirsk Model   \cite{Bondar:1999,Castro:2011zd} and Flux-Tube Breaking Model  \cite{Isgur:1988vm,Godfrey:1985xj,Kokoski:1985is}
 parameterization. 
The Decker-Mirkes Model \cite{Decker:1992jy} contains both Born level amplitudes corresponding to two axial-vector currents, while the
Novosibirsk   \cite{Bondar:1999,Castro:2011zd} formalism contains only one axial-vector Born level current.
Neither the Decker-Mirkes  \cite{Decker:1992jy} or  Novosibirsk  \cite{Bondar:1999,Castro:2011zd}  amplitudes have been matched to the 
S-wave $J^{P}(l=0)=1^{+}$ and D-wave $J^{P}(l=2)=1^{+}$ distributions.
Both the S-wave and D-wave Feindt  \cite{Feindt:1990ev} and Flux-Tube Breaking Model  \cite{Isgur:1988vm,Godfrey:1985xj,Kokoski:1985is} contributions for the 
$\tau^{-}\to K_{1}^{-}(1270)\nu_{\tau}\to K^{-}\omega\nu_{\tau} \to K^{-}\pi^{-}\pi^{+}\pi^{0}\nu_{\tau}$ decay are consistent with the
theoretical prediction for $J^{P}(l=0)=1^{+}$ and $J^{P}(l=2)=1^{+}$ respectively. For the  Flux-Tube Breaking Model  \cite{Isgur:1988vm,Godfrey:1985xj,Kokoski:1985is},
the S-wave and D-wave components are defined in terms of the vertex amplitudes $A=[1S_{1},0D_{1}]$ and $A=[0S_{1},1D_{1}]$. The  pseudo-scalar state production in the
Flux-Tube Breaking Model \cite{Isgur:1988vm,Godfrey:1985xj,Kokoski:1985is}, $\tau^{-}\to K^{-}(1460)\nu_{\tau}\to K^{-}\omega \nu_{\tau}$, 
is found to be consistent with the expected theoretical $J^{P}(l=1)=0^{-}$ prediction.

\section{QED Spin Dynamic and Leptonic Decays \label{sec:QEDSpin}}

In $e^{+}e^{-}\to\tau^{+}\tau^{-}(\gamma)$ interactions, the polarizations of the $\tau$-pairs are produced in an entangled state.
 The normalized probability distribution of the $\tau$ lepton polarizations can be determined from the spin dependent 
matrix-elements corresponding to the $e^{+}e^{-}\to\tau^{+}\tau^{-}(\gamma)$ interactions being simulated. From the decomposition of the spin-average sum matrix element,
the normalized probability for the longitudinal polarization may be written

\begin{equation}
\resizebox{0.375\textwidth}{!}{$
 P_{\lambda_{k},\lambda_{l}} = \frac{\sum\limits_{i}^{}\sum\limits_{j}^{} \rho_{\lambda_{i},\lambda_{j}}\sum\limits_{m,...,n}^{}|\mathcal{M}_{\lambda_{i},\lambda_{j}\lambda_{k}\lambda_{l},\lambda_{m},...,\lambda_{n}}|^{2}}{|\bar{{\mathcal M}}|^{2}},
$}
\label{eq:qedpolar}
\end{equation}

\noindent where $\mathcal{M}_{\lambda_{i},\lambda_{j}\lambda_{k}\lambda_{l},...\lambda_{n}}$ is the matrix element with polarization states 
$\lambda$, $\rho_{\lambda_{i},\lambda_{j}}$ is the tensor of initial-state spin-average weights defined using the initial helicity state, and $|\bar{{\mathcal M}}|^{2}$ is the spin-average summed matrix 
element \cite{Halzen:1984mc}. To include transverse spin 
correlations, the normalized probability must be determined in terms of the modified Altarelli-Parisi  Density Function \cite{Collins:1987cp,Knowles:1988hu,Jadach:1984}. 
For a known initial and final state, 
the normalized spin probability corresponding to the modified Altarelli-Parisi  Density Function is 

\begin{equation}
\resizebox{0.375\textwidth}{!}{$
\begin{array}{ll}
  P = \frac{\rho_{\lambda_{j},\lambda_{j}}}{|\bar{\mathcal{M}}|^{2}}\times \mathcal{M}_{\lambda_{i},\lambda_{j}\lambda_{k}\lambda_{l},...,\lambda_{n}}^{}\mathcal{M}_{\lambda_{i},\lambda_{j}\lambda_{k}^{\
'}\lambda_{l}^{'},...,\lambda_{n}}^{*}
\times \prod\limits_{\alpha=k}^{n} D_{\lambda_{\alpha},\lambda_{\alpha}^{'}}
\label{eq:qedpolarAP}
\end{array}
$}
\end{equation}

\noindent where $D_{\lambda_{\alpha}\lambda_{\alpha}^{'}}=\frac{1}{|\bar{\mathcal{M}}^{(D)}|^{2}} \mathcal{M}_{\lambda_{\alpha},\lambda_{\beta}\lambda_{\mu},...,\lambda_{\nu}}^{(D)}\mathcal{M}_{\lambda_{\alpha}^{'},\lambda_{\beta}^{'}\lambda_{\mu}^{'},...,\lambda_{\nu}^{'}}^{(D)*}$ for the $\tau$ leptons and $D_{\lambda_{\alpha}\lambda_{\alpha}^{'}}=\delta_{\lambda_{\alpha}\lambda_{\alpha}^{'}}$ for the photons which are stable \cite{Collins:1987cp,Knowles:1988hu}. A $SU(2)\to SO(3)$ transformation \cite{Jadach:1984}, is then applied to
convert the modified Altarelli-Parisi Density Function \cite{Collins:1987cp,Knowles:1988hu,Jadach:1984}, a set of complex matrices, into real spin vectors accompanied 
by the normalization and a real spin probability density function, which are more suitable for computing the probability. This is accomplished by using the completeness-relation and spin 
projection operators

\begin{equation}
\resizebox{0.375\textwidth}{!}{$
\begin{array}{lclcl}
  \sum\limits_{s=1,2}^{}u^{(s)}(p)\bar{u}^{(s)}(p)&=&p\!\!\!/+m&=&2m\Lambda_{+} \\
 \sum\limits_{s=1,2}^{}v^{(s)}(p)\bar{v}^{(s)}(p)&=&p\!\!\!/-m&=&-2m\Lambda_{-} 
\label{qe:qedpolarCR}
\end{array}
$}
\end{equation}

\noindent \cite{Halzen:1984mc}. The resulting spin vectors are defined relative to the $\tau^{-}$/$\tau^{+}$ along the z-axis with $\phi=0$. Taking into account the 
azimuthal angle of the respective $\tau^{+}\tau^{-}$ pair relative to the $e^{+}e^{-}$ frame which the spin probability density matrix is computed in, the spin 
vectors reference frames are converted such that the  $\tau^{-}$($\tau^{+}$) is defined along the +z(-z)-axis with $\phi_{\tau^{-}}=0$.
The spin vectors and associated normalization are 
then contracted with the spin probability density matrix to compute the probability of the given spin configuration.  Equations \ref{eq:qedpolar}, 
\ref{eq:qedpolarAP} and \ref{qe:qedpolarCR} are applied to
construct a factorized as well as a non-factorized algorithm for embedding spin correlations in the simulation. In the factorized approach,

\begin{enumerate}
\item  The $e^{+}e^{-}\to\ l^{+}l^{-}(\gamma)$ decay is simulated by means of the spin-average summed matrix element and exponentiation procedure presented in Section \ref{sec:QED}.
\item  Using probability distribution in Equation \ref{eq:qedpolar}, a random die is thrown to determine the polarization state 
that the $\tau^{+}\tau^{-}$ pair collapse into. Each $\tau$ is then decayed with the helicity state corresponding to the chosen $P_{\lambda_{k},\lambda_{l}}$, 
where the Lorentz invariants of the matrix element, (see Section \ref{sec:Tau}), is applied to allow all the calculations to be performed in a common reference 
frame, the $e^{+}e^{-}$ centre-of-mass reference frame.  This provides an efficient and consistent mechanism to include the longitudinal polarization in the decays of 
$\tau$ leptons with and without radiative effect.  \footnote{This formulation of the longitudinal polarization is mathematically equivalent to the decay acting 
as a filter with the helicity state being an observable and collapsing the correlated wavefunctions into the corresponding eigenstates, 
meaning there is no orthogonal spin information.}. For users who only wish to include 
longitudinal spin correlations the algorithm can be stopped here.  
\item After both $\tau$ leptons have been decayed, the transverse spin effects are included through an accept/reject algorithm where the $\tau$ decay products are 
rotated by a random angle about the longitudinal axis of the respective $\tau$ lepton in the $e^{+}e^{-}$ centre-of-mass reference frame with the probability of 
the acceptance/rejection being based on Equation \ref{eq:qedpolarAP} after being converting into the spin vectors with the associated normalization and 
the spin probability density matrix. By applying a random transverse rotation, the computationally intensive matrix element calculations is
substantially reduced. 
\end{enumerate}

\noindent In the non-factorized approach, 

\begin{enumerate}
\item  The $e^{+}e^{-}\to\ l^{+}l^{-}(\gamma)$ decay is simulated by means of the spin-average summed matrix element and exponentiation procedure presented in Section 
\ref{sec:QED}.
\item Both $\tau$ leptons are decayed in an unpolarized state using the spin-average summed matrix element.
\item The longitudinal and transverse spin correlations are included through an accept/reject algorithm similar to the former method, however, 
the $\tau$ decay products are rotated by a transverse and longitudinal random angle in the centre-of-mass frame for the respective $\tau$ lepton before being boosted 
into the $e^{+}e^{-}$ centre-of-mass reference frame. Again, the probability of the acceptance/rejection being based on Equation \ref{eq:qedpolarAP} after being 
converting into the spin vectors with the associated normalization and the spin probability density matrix. This method is 
slightly less efficient than the factorized approach but is more similar to the algorithm in \cite{Jadach:1984}.  
\end{enumerate}
 
\noindent Figure \ref{fig:tauxPolar} illustrates the above methodology for the simulated polarization in $e^{+}e^{-}\to\tau^{+}\tau^{-}(\gamma)$ for unpolarized 
$e^{+}e^{-}\to \tau^{-}(\to\pi^{-}\nu_{\tau})\tau^{+}(\to\pi^{+}\nu_{\tau})(\gamma)$ events and
 $e^{+}e^{-}\to \tau^{-}(\to\pi^{-}\nu_{\tau})\tau^{+}(\to\pi^{+}\nu_{\tau})(\gamma)$ with polarized electrons similar to the polarization expected at the 
BELLE-II Experiment  \cite{abe2010belle,Kou_2019}, where
Type IV Exponentiation has been applied to subtract the divergences associated with radiative corrections  \cite{Yennie:1961}. Figure \ref{fig:tauxRadPolarcutoff} 
presents a more detailed picture of polarization in terms of the soft-photon cut-off and the Born and LO radiative corrections. As illustrated in Figure \ref{fig:tauxRadPolarcutoff}, the $M_{\pi^{+}\pi^{-}}$ observable 
for the polarization, there is a clear dependency on the number of emitted photons and the corresponding energy of the emissions, however, within the statistical
precision of the simulation the Born+LO distributions are consistent. This can be attributed to the polarization distribution asymptotically converging to an 
infra-red safe limit. The convergence and accuracy of the polarization sensitive observables, in addition to the differential cross-section, needs to be 
checked at higher orders, given that the number of photons is not an infra-red safe observable
\footnote{Classically, this problem was first considered by J. Lamour in  \cite{Lamour_1987_II} particularly in terms of the corresponding
  problem of the electron falling into the proton as $t\to\infty$ due to Poisson (exponential) emission of radiation. The lack of photon emission was resolved through  
  quantization and the wavelength of the electron. In contrast, for the linear case of photon emission in scattering, which is analogous to the circular case in many ways, 
  the photon radiation is formulated very differently in terms infra-red observables and the minimum quantities based in the experimental configuration for 
  measuring the photon emission  \cite{Peskin:1995ev,Yennie:1961}. The minimum photon energy, which is somewhat arbitrary in this picture, corresponds 
  to the maximum allowable time (wavelength) for the electromagnetic interaction  \cite{Peskin:1995ev}. The lack of symmetry between the solutions 
  in terms of the wavelengths of the emitted photon(s) and source particles is peculiar, however the correct behaviour is quite essential to the underpinning 
  of the theoretical formulation of Quantum Mechanics. 
}. 

\section{Conclusion\label{sec:Conclusion}}
We have presented a new MC-Generator for simulating $e^{+}e^{-}\to\mu^{+}\mu^{-}(\gamma)$ and $e^{+}e^{-}\to\tau^{+}\tau^{-}(\gamma)$ for center-of mass-energies 
from threshold up to $\sim 10.58GeV$. The running of $\alpha_{QED}$ is implemented at LO incorporating both the leptonic and hadronic vacuum polarizations with the
infra-red divergences being subtracted through the multiplicative exponentiation procedure  \cite{Peskin:1995ev,Yennie:1961}. Several analytic formulations 
of the exponentiation procedure, from only subtracting the divergences associated with the soft-photon and virtual photon  \cite{Peskin:1995ev}, to a complete 
subtraction  \cite{Schwinger:1998}  
which takes into account the non-divergent component of the vacuum polarization and the spin $1/2$ component of the magnetic Form-Factor  \cite{Yennie:1961},
%by equating the ${\mathcal O}(\alpha)$ term in the exponential series $e^{-{\mathcal O}(\alpha)}=1-{\mathcal O}(\alpha)+\frac{{\mathcal O}(\alpha)^{2}}{2!}...$ 
% \cite{Peskin:1995ev} to the full LO subtraction  \cite{Schwinger:1998}, 
are included in the MC generator.  
The $e^{+}e^{-}\to\mu^{+}\mu^{-}(\gamma)$ and $e^{+}e^{-}\to\tau^{+}\tau^{-}(\gamma)$  
collisions are simulated up to LO radiative corrections including the initial and final state interference. Extensions to include higher order corrections are 
anticipated to be forthcoming and will allow for a theoretically consistent simulation with $n_{\gamma}\ge 2$\footnote{The current state of the art MC simulations, 
{\tt KK2F}  \cite{kk2f} which also uses the Yennie-Frautschi-Suura Exponentiation formalism, {\tt BabaYaga@NLO} 
\cite{Balossini_2006,Balossini_2008,CarloniCalame_2004,CarloniCalame_2001,CarloniCalame_2000} which is NLO with higher orders contributions from Parton Matching using 
the Sudakov Form-Factor \cite{Balossini_2006,Balossini_2008} and {\tt PHOKHARA}  \cite{Binner:1999bt,Czyz:2000wh,Campanario:2013uea} which
is limited to events with 1 or 2 photons due to the approach applied for the infra-red subtractions  \cite{Campanario:2013uea,Balossini_2008}.}.
Such improvements are crucial for reaching the theoretical precision required by BELLE-II.

Two fundamentally different theoretical 
constructs are incorporated for the simulation of hadronic $\tau$ decays,
 namely ChRL Models and Chromoelectric Flux-Tube Breaking Model, constructed  in the context of the
${^{3}P_{0}}$ quark model and ``strong lattice formation'' \cite{Kokoski:1985is,Isgur:1983wj,Isgur:1984bm}. Where the latter theoretical construct provides a framework which is
compatible with lattice QCD. 
The Flux-Tube Breaking Model contains a more complete simulation of the decay structure for the $a_{1}(1260)$ and $K_{1}(1270/1400)$ resonances, 
in particular, in terms of the S-wave/D-wave structure, as well as the scalar and pseudo-scalar intermediate resonances. This will allow for a more detailed investigation 
of the hadronic angular structure in $\tau$ decays, which is essential for an experimentally driven longitudinal subtraction of the spectral density functions for 
the extraction of $\alpha_{s}$, the strange quark mass, $m_{s}$, and the CKM matrix element, $|V_{us}|$,  \cite{Maltman_2001}. Moreover, the S-wave/D-wave structure 
from the Feindt Model and Flux-Tube Breaking Model for the axial-vector current will allow for a physically motivated assessment of the efficiency dependence on angular structure in the 
hadronic decays.
With the increased statistics expected from the BELLE-II Experiment  \cite{abe2010belle,Kou_2019},  there is an opportunity for a  more in-depth investigation into the
nature of the lowest scalar ``states'', namely the $\sigma/f_{0}(500)$, $f_{0}(980)$ and $\kappa/K_{0}^{*}(700)$, and thereby the constituent quark mass. The 
Flux-Tube Breaking Model, a ${}^{3}P_{0}$ model, provides an alternative hypothesis to the existence of physical low mass scalar resonances which are 
predicted in many models to be related to the generation of the constituent mass  \cite{PDG2020}.
In future, theoretical predictions in $\tau$ decays for the models which predict the existence of the  $\sigma/f_{0}(500)$, $f_{0}(980)$ and $\kappa/K_{0}^{*}(700)$  in
Linear $\sigma$ Models  \cite{Tornqvist_1999,Napsuciale_2004,Napsuciale_2004b,Ishida_1999,Sadron_1999,Fariborz_2014,Black_1999},
Chiral-Lagrangian Models  \cite{PhysRevD.59.074001,PEL_EZ_2004,Dabado_1997,Oller_2003}, 
Non-linear Sigma Models \cite{GellMann:1960,Skyrme:1962vq,Skyrme:1962vh,Speight:2018,Nambu:1961},
Tetra-quark Model  \cite{Alford_2000,Maiani_2004,_t_Hooft_2008}, Molecular Models  \cite{Phys.Rev.130.776,Kalashnikova_2005}, 
%Skyrme Model \cite{Skyrme:1962vq,Skyrme:1962vh,Speight:2018} and Nambu-Jona-Lasinio Model  \cite{Nambu:1961}, 
Flux-Tube Breaking Model \cite{Isgur:1988vm,Isgur:1983wj,Isgur:1984bm,Kokoski:1985is,Godfrey:1985xj},
Bag Model  \cite{Chodos:1974a,Chodos:1974b,DeGrand:1975}, String Models \cite{Maldacena:1997re}, 
 {\it etc}, will be required to fully elucidate the nature of the low mass scalar state.
Moreover, the Flux-Tube Breaking Model provides a model dependent frame-work for the simultaneous extraction of the mixing angle between the $K_{1}$ 
singlet and triplet states, $\theta_{K_{1}}$, and the $SU(3)_{f}$ breaking factors, $\delta_{K_{1}}$. With adequate precision, the $SU(3)_{f}$ breaking factors 
$\delta_{K_{1}}=\frac{m_{s}-m_{u}}{m_{s}+m_{u}}$ can provide complimentary data to the quark and gluon kinetic and potential 
energy tensor  \cite{Ji:1995a,Ji:1995b,SoLID} and
has the potential to distinguish between the various QCD models.
In future, it is expected that the missing excited vector and axial-vector resonances will be included. 
This is particularly important in terms of having an independent prediction of the Wess-Zumnio anomaly given the issues found with the simulation in {\tt TAUOLA} 
(see Section \ref{sec:ChRL}), and for measured modes such as $\tau^{-}\to \phi K^{-}\nu_{\tau}$ and $\tau^{-}\to \phi\pi^{-}\nu_{\tau}$\footnote{The decay mode 
$\tau^{-}\to \phi\pi^{-}\nu_{\tau}$ is a OZI  \cite{Zweig:1964jf,Okubo:1963fa,Iizuka:1966} suppressed modes which may provide a particularly 
interesting opportunity to measure $\alpha_{s}$ below the QCD 
perturbative threshold using Lattice QCD. Neither the $a_{1}(1260)$ or excited $\rho$ mesons are known to decay into $\phi(1020)\pi$. However, given the non-zero 
$\omega-\phi$ mixing and that the excited $\rho$ mesons decay into $\omega(782)\pi$ it would be interesting to determine if the $\phi(1020)\pi$ is produced from a vector
or axial-vector state.}.   
In addition to the implementation of the ChRL and Flux-Tube Breaking Models, phenomenological models, mainly developed by experimental 
collaborations, have been implemented for individual decay modes. This includes a new phenomenological model which predicts 
the $h_{1}^{0}(1170)$ and $h_{1}^{0}(1415)$ can be observed in $\tau$ decays through the decays of excited $\rho$ mesons. 
In addition to the implementation of additional hadronic models to investigate the fundamental assumption in the QCD models, 
further improvement will be made by including the emission of hard photons and the associated Electro-weak corrections 
\cite{PhysRevLett.61.1815,PhysRevD.42.3888,Erler:2002mv}.
This would facilitate measurements of the Michel-Parameters \cite{Stahl:2000aq,Michel:1949qe,PhysRev.106.170,Bacino:1979fz,Behrends:1985pm,PhysRevD.36.1971,Janssen:1989wg,Albrecht:1990zj,Albrecht:1993fr,Albrecht:1994nm,Albrecht:1997gn,Abe:1997dy,physrevd.56.5320,Ackerstaff:1999,Seager:1999uf,Wunsch:1999uh,Oberhof:2015hea,Abdesselam,Epifanov} and allow the investigation of the model dependant assumptions of photon emissions from the hadronic interactions \cite{LopezCastro:2001apj,FloresTlalpa:2006gz,FloresTlalpa:2006gf,FloresTlalpa:2006gs,Lees:2015qna}
produced at a charged-weak vertex. 
The MC generator presented here provides a consistent framework for investing  $e^{+}e^{-}\to\mu^{+}\mu^{-}(\gamma)$ and $e^{+}e^{-}\to\tau^{+}\tau^{-}(\gamma)$
for energies up to $\sim 10.58GeV$ centre-of-mass energies with multiple theoretical frameworks for modelling the hadronic structure in $\tau$ to allow for a more
precise investigation of the hadronic structure and to answer open questions in low energy QCD.

\section*{Acknowledgement}
 I would like to thank Alberto Luisiani for suggesting that I use the 
exponentiation procedure from  \cite{Yennie:1961} for removing the infra-red divergences. 
I would also like to thank Zbigniew W\c as for the discussion on 
numerical integration in particular practical change in variables for the integration of the Breit-Wigner distribution in regards to our paper 
 \cite{Nugent:2013hxa}.
GCC Version 4.8.5 was used for compilation and the plots are generated using the external program GNUPlot  \cite{gnuplot4.2}.
Additional references employed in the development of this MC program include  \cite{Abramowitz:1964,WehShen:2016,Gill:1982,Nyborg:1970,gituliar_2018,joquiere,watson,Lewis,Morris,Krillov_1995}.

\footnotesize
\bibliography{paper}
\normalsize

\input{paper_fig.tex}

\clearpage
\input{paper_tab.tex}

\end{document}

%% file: paper_fig.tex
\begin{figure*}[tb]
\begin{center}
  \resizebox{260pt}{185pt}{
    \includegraphics{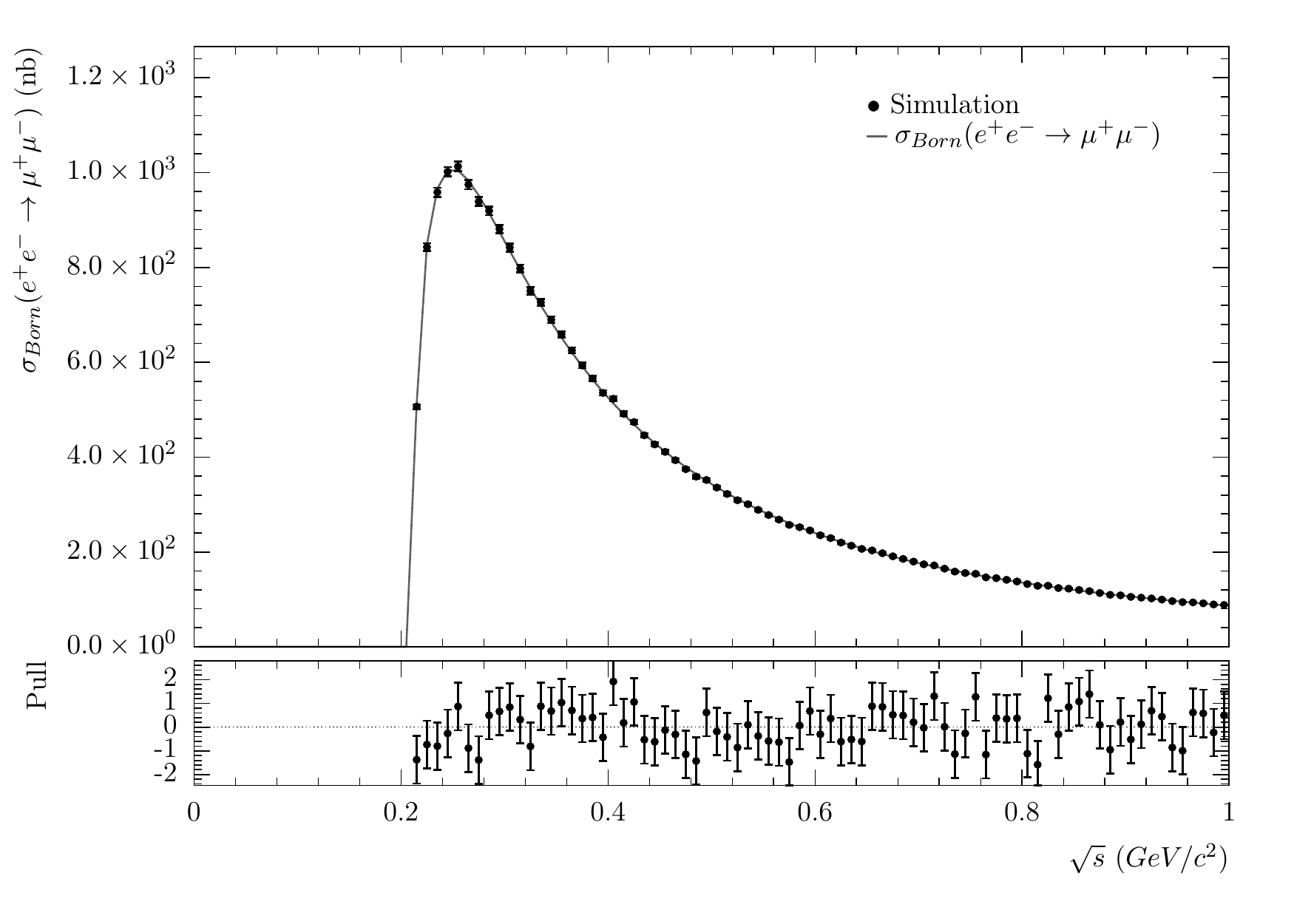}
  }
  \caption{A comparison of the simulation of the $\sigma(e^{+}e^{-}\to\mu^{+}\mu^{-})$ at Born level to the Born level theoretical
    prediction  \cite{Smith1994117} as a function of $s$. The simulation includes 1 million events per data-point. The
    pull plot shows that the simulation and the theoretical predication  \cite{Smith1994117} are statistically consistent
    at the per-mill level.\label{fig:CS_mumu}}
\end{center}
\end{figure*}

\begin{figure*}[tbp]
\begin{center}
  \resizebox{260pt}{185pt}{
    \includegraphics{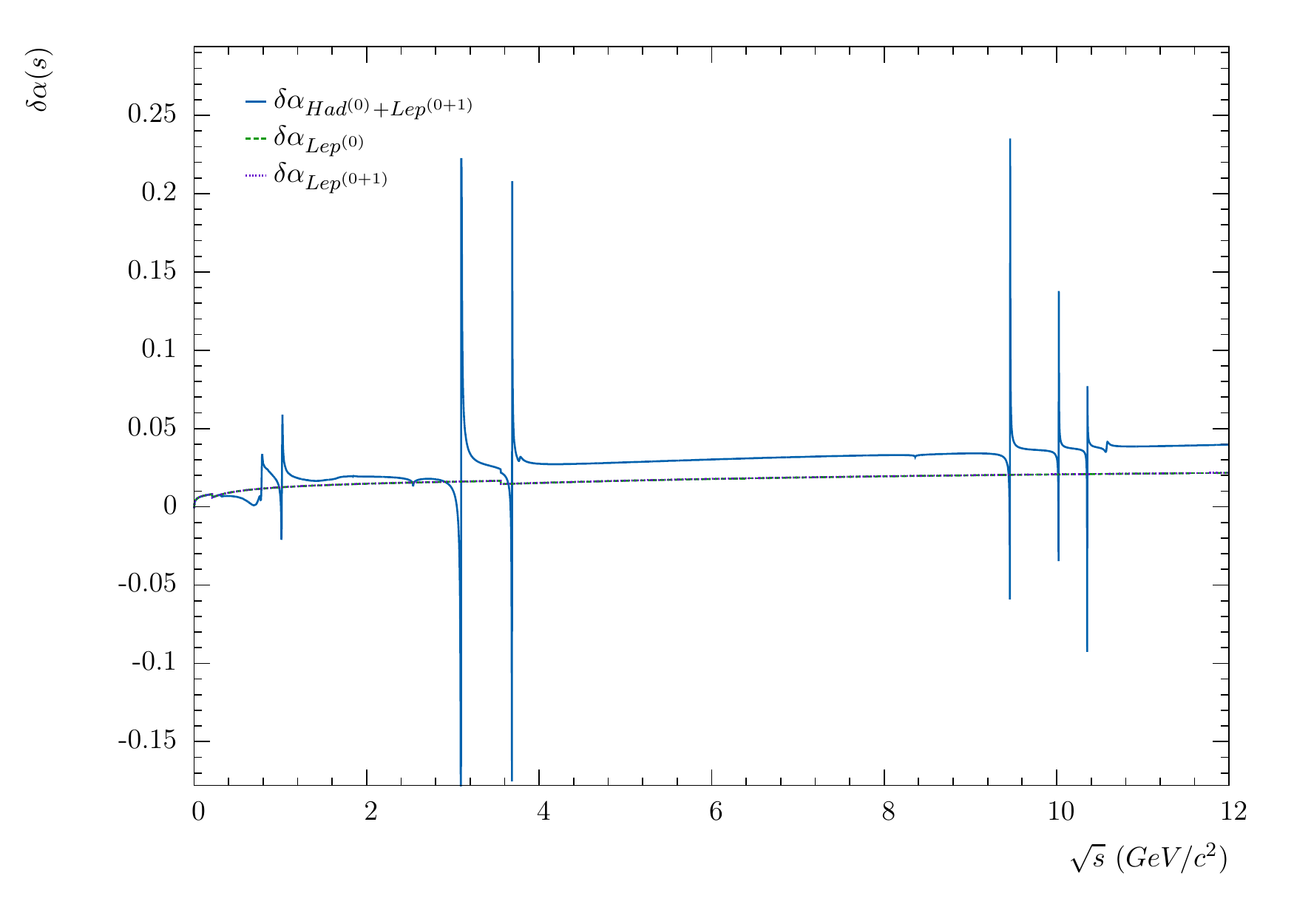}
  }
  \caption{$\delta\alpha$ for the running of $\alpha(s)$ as a function of $\sqrt{s}$ for $-q^{2}>0$. 
$\delta\alpha_{Had^{(0)}+Lep^{(0+1)}}$ includes the LO Hadronic and LO+NLO Leptonic vacuum polarization (Solid) while $\delta\alpha_{Lep^{(0)}}$ (Dashed) and 
$\delta\alpha_{Lep^{(0+1)}}$ (Fine-Dotted) include the 
LO and LO+NLO Leptonic vacuum polarization respectively. This distribution can be compared to  \cite[Fig. 5.18]{Jegerlehner} and \cite[Sec. 6.4]{MC:2010}.
\label{fig:RunningOfAlpha}}
\end{center}
\end{figure*}

\begin{figure*}[tbp]
\begin{center}
  \resizebox{260pt}{185pt}{
    \includegraphics{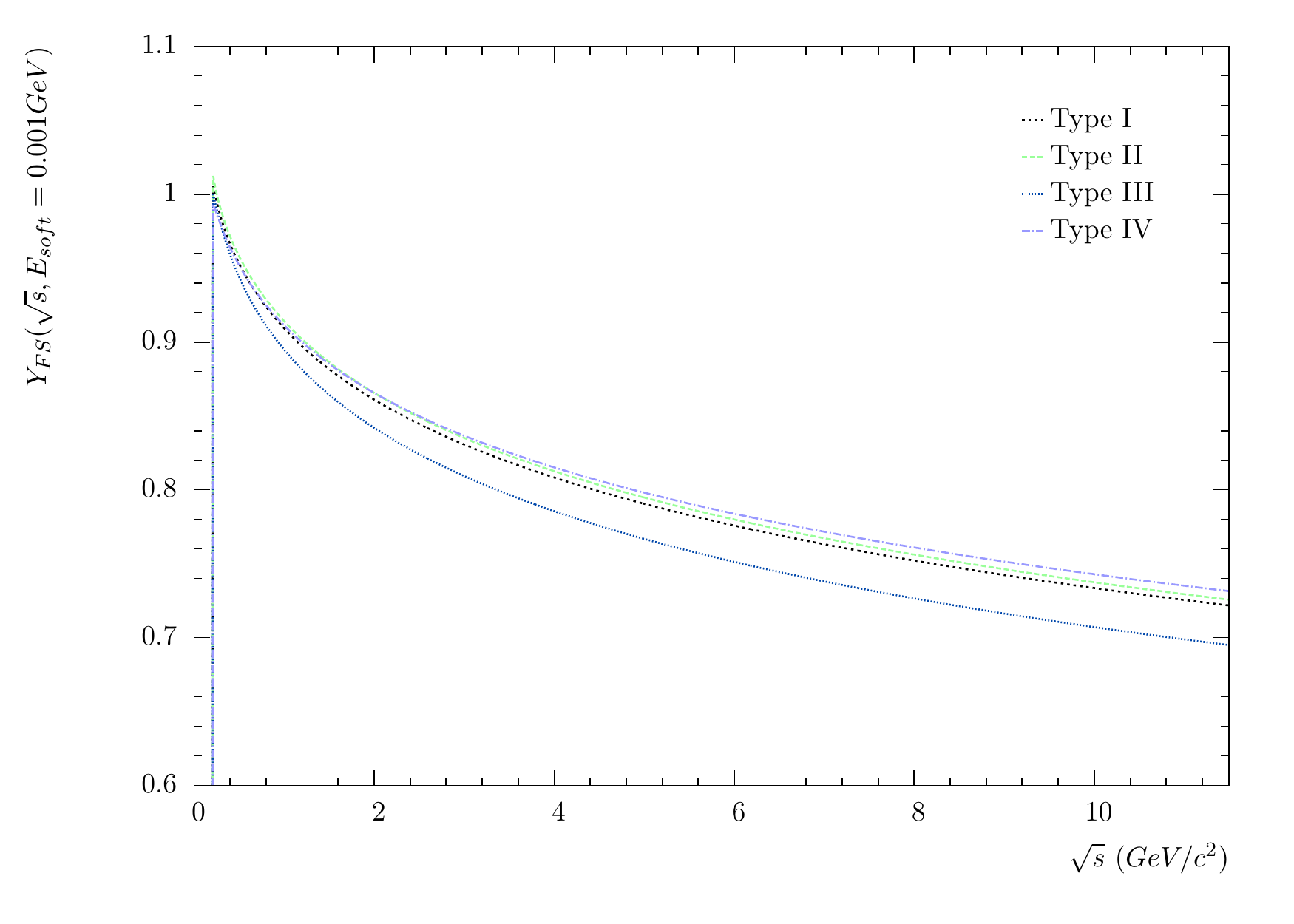}
  }
  \caption{A comparison of the YFS Exponentiation Form-Factors for the Final-State as a function of $\sqrt{s}$.\label{fig:YFS}}
\end{center}
\end{figure*}
\noindent

\begin{figure*}[tbp]
\begin{center}
  \resizebox{260pt}{185pt}{
    \includegraphics{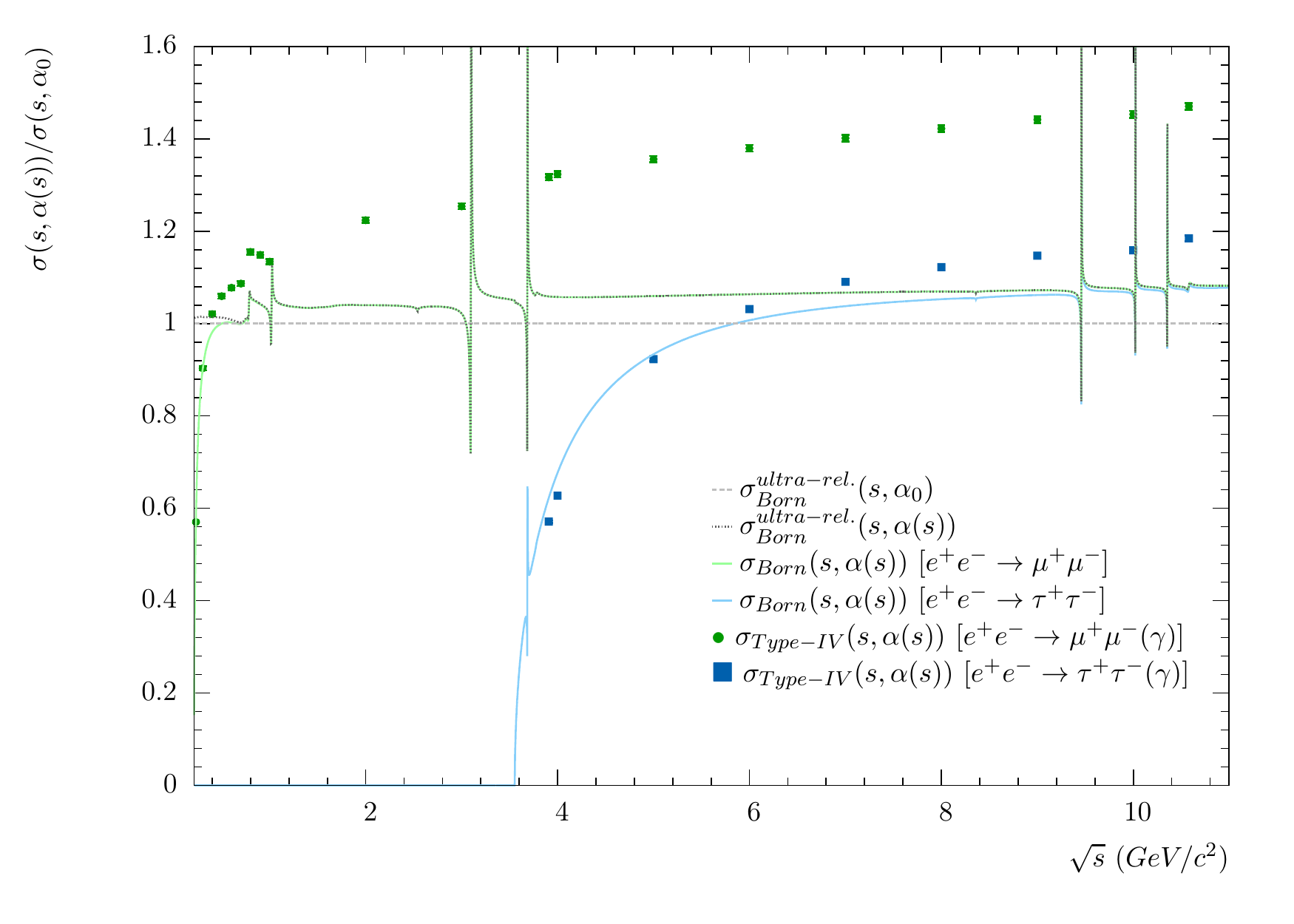}
  }
  \caption{The cross-section as a function of invariant mass, $\sqrt{s}$, for the $e^{+}e^{-}\to \mu^{+}\mu^{-}(\gamma)$ and $e^{+}e^{-}\to \tau^{+}\tau^{-}(\gamma)$
interactions normalized to the ultra-relativistic approximation without the vacuum polarization. The data points represent the LO $e^{+}e^{-}\to \mu^{+}\mu^{-}(\gamma)$ 
and $e^{+}e^{-}\to \tau^{+}\tau^{-}(\gamma)$ normalized cross-sections with Type-IV Exponentiation and are simulated to a statistical precision of $0.5\%$. The 
soft-photon cut-off is taken as 20\% of $\sqrt{s}-m_{thres}$. The theoretical predictions for the Born cross-section with and without vacuum polarization for the
ultra-relativistic approximation \cite{Griffiths:1987tj} as well as the Born cross-section including final-state masses \cite{Smith1994117} are presented. This 
illustrates the impact of the vacuum polarization on
the cross-section and that the $e^{+}e^{-}\to \tau^{+}\tau^{-}(\gamma)$ production is not yet fully ultra-relativistic at $\sqrt{s}=10.58GeV$.
 \label{fig:CSvsMass}}
\end{center}
\end{figure*}
\noindent

\begin{figure*}[tbp]
\begin{center}
  \resizebox{260pt}{185pt}{
    \includegraphics{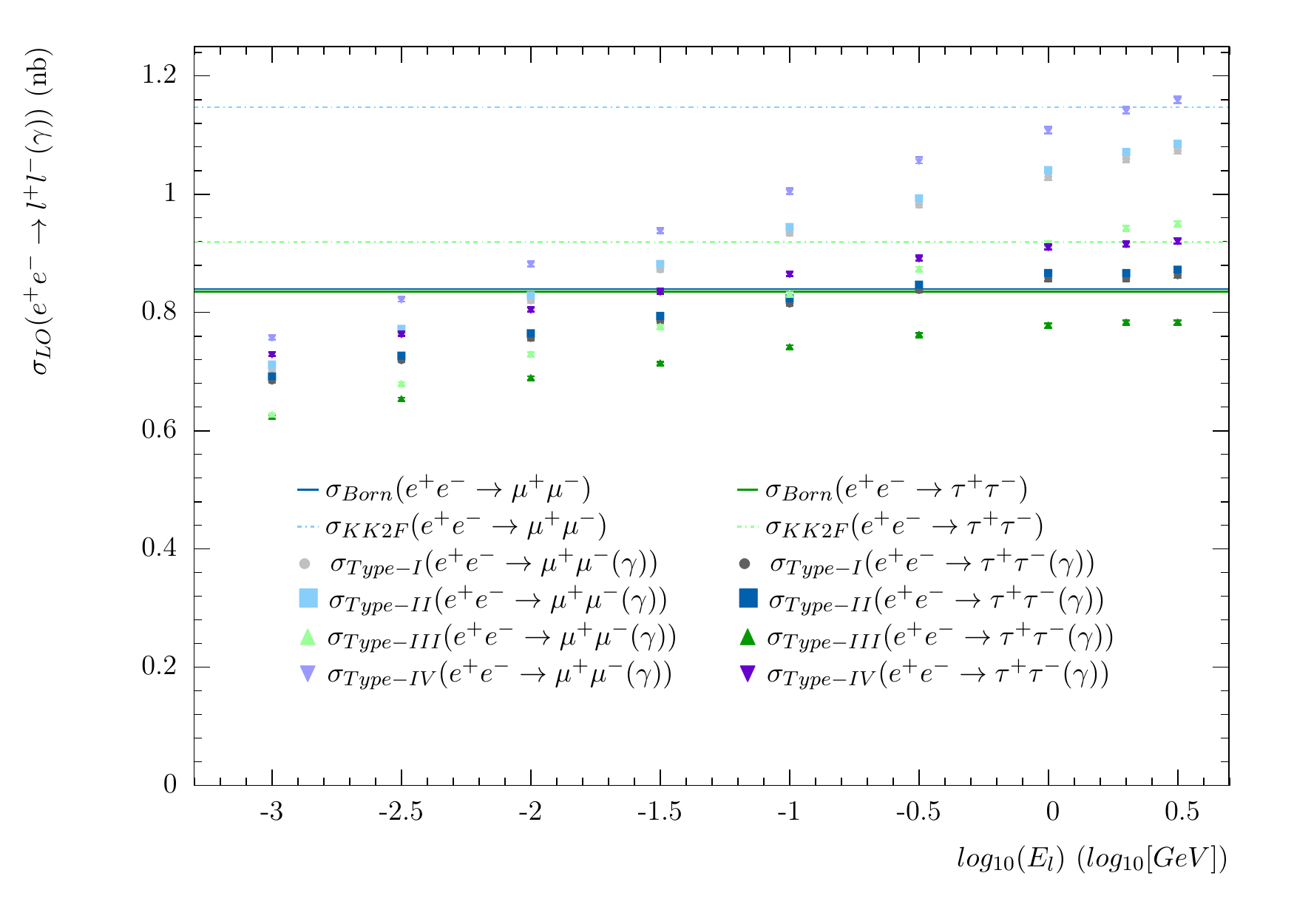}
  }
  \resizebox{260pt}{185pt}{
    \includegraphics{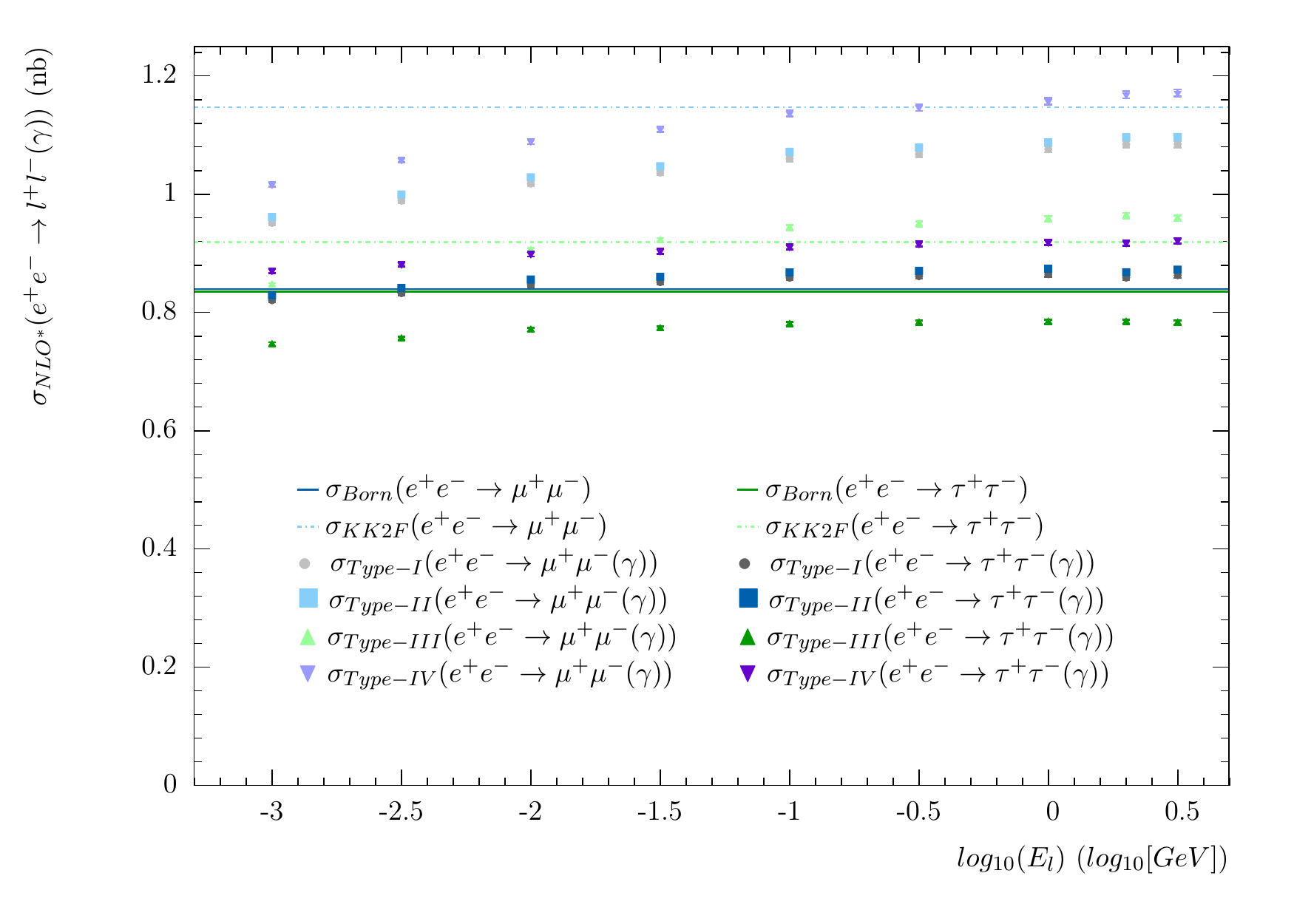}
  }
  \caption{The cross-section as a function of the soft-photon cut-off, $E_{l}$ for the $e^{+}e^{-}\to \mu^{+}\mu^{-}(\gamma)$ and $e^{+}e^{-}\to \tau^{+}\tau^{-}(\gamma)$
interactions at LO (top) and NLO$^{*}$ (bottom) at $\sqrt{s}=10.58GeV$. The cross-section for the YSF Exponentiation Form-Factors from Section \ref{sec:YFS} 
are compared to the expected Born and the KK2F  \cite{Banerjee_2008} predictions.
The LO calculation include the $\mathcal{M}_{0}^{1}$ and $\mathcal{M}_{1}^{1}$ diagrams as defined in Section \ref{sec:YFS}, while the NLO$^{*}$ also includes the 
real photon emissions terms at NLO, $\mathcal{M}_{2}^{1}$. The NLO$^{*}$ is calculated using weighted events.  The statistical precision of the simulated points 
is $0.5\%$.  \label{fig:CSvsEl}}
\end{center}
\end{figure*}
\noindent

\begin{figure*}[tb]
\begin{center}
  \resizebox{260pt}{185pt}{
    \includegraphics{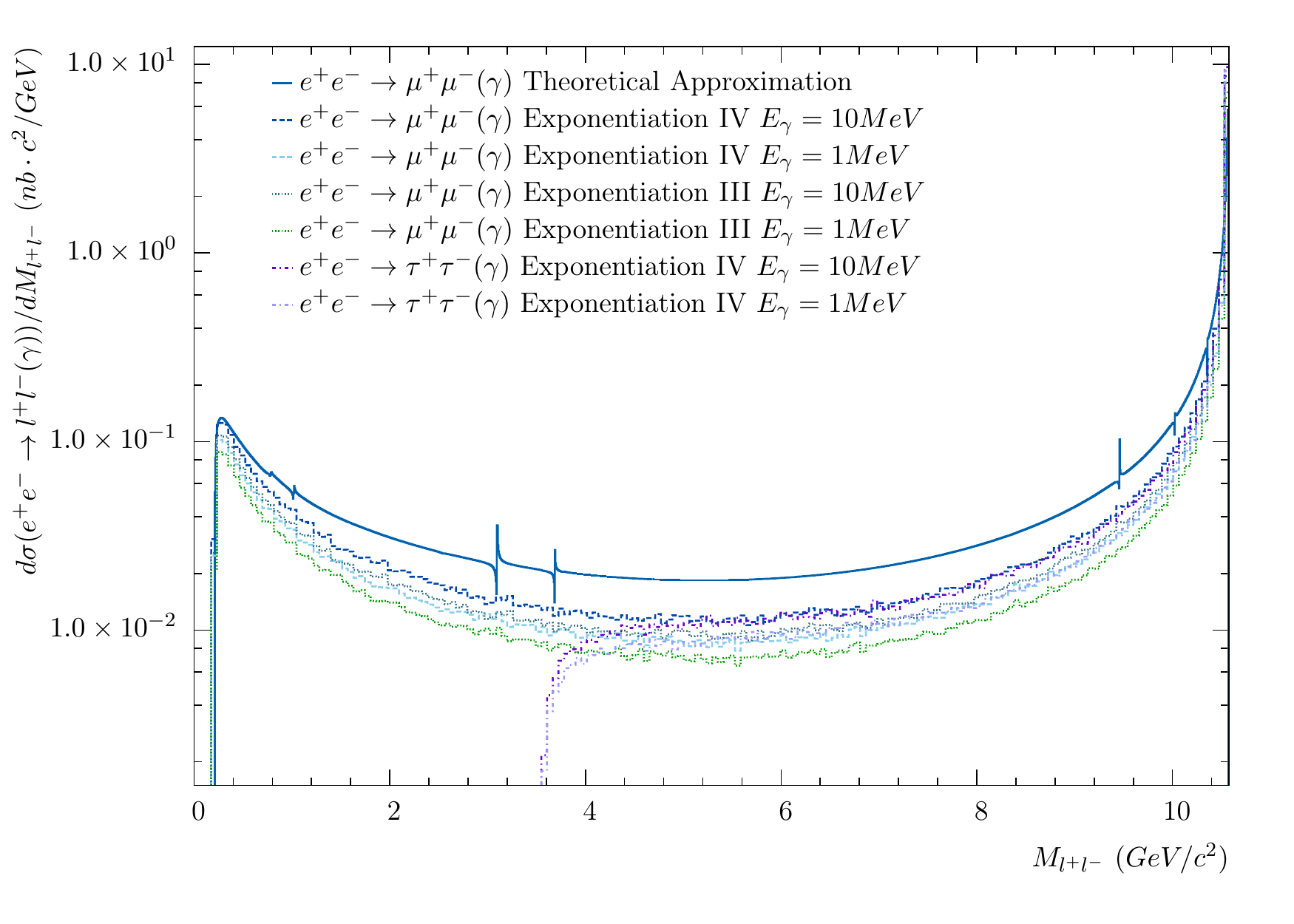}
  }
  \caption{The differential cross-section of $\frac{d\sigma(e^{+}e^{-}\to l^{+}l^{-}(\gamma))}{dM_{l^{+}l^{-}}\hfill}$  for $l=\mu$ and $l=\tau$ compared to the
theoretical predictions
from  \cite{Aubert_2004} modified with corrections for additional initial and final state radiation and the Coulomb potential
 \cite{PhysRevD_72_114019,PhysRevLett.103.231801}. The simulation includes the LO ``on-shell'' leptonic and hadronic contributions to the running of the coupling
constant $\alpha$ with Type III for $l=\mu$ and Type IV Exponentiation for $l=\mu$ and $l=\tau$. Both soft-photon cuts of $10MeV$ and $1MeV$ are presented.\label{fig:CS_mumu_Rad}}
\end{center}
\end{figure*}

\begin{figure*}[tbp]
\begin{center}
  \resizebox{421pt}{150pt}{ % 520 185
    \includegraphics{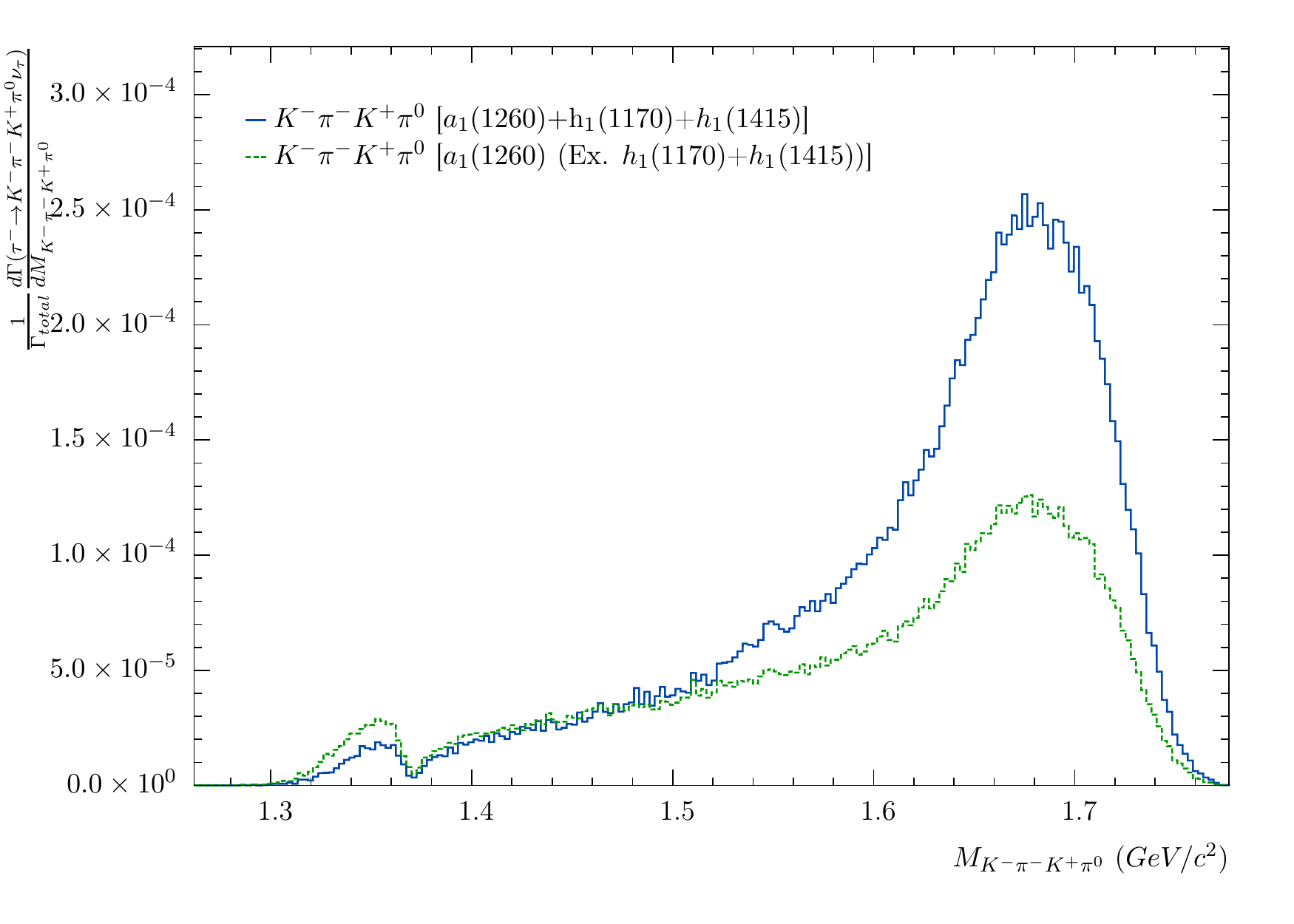}
    \includegraphics{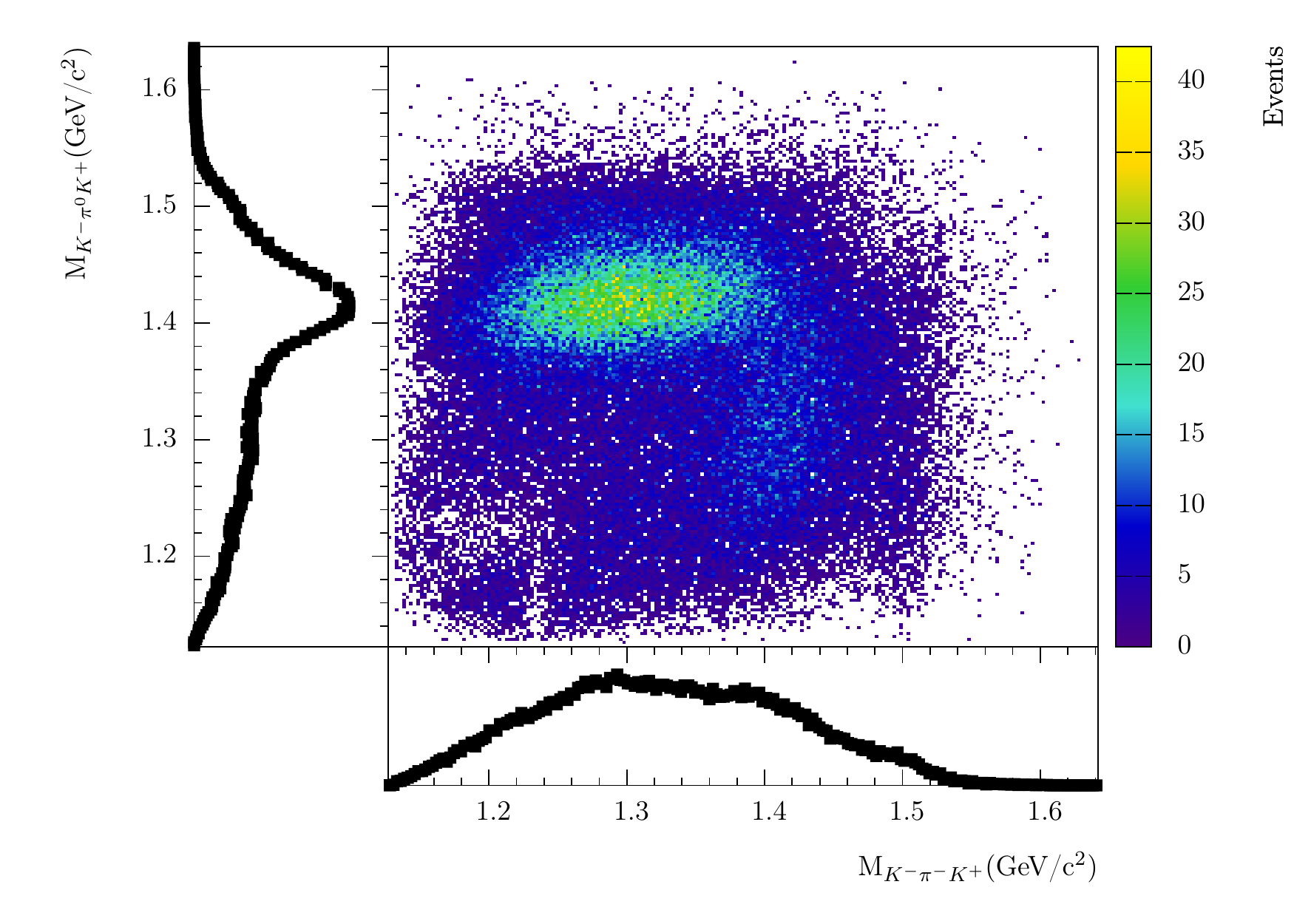}
  }
  \resizebox{421pt}{150pt}{
    \includegraphics{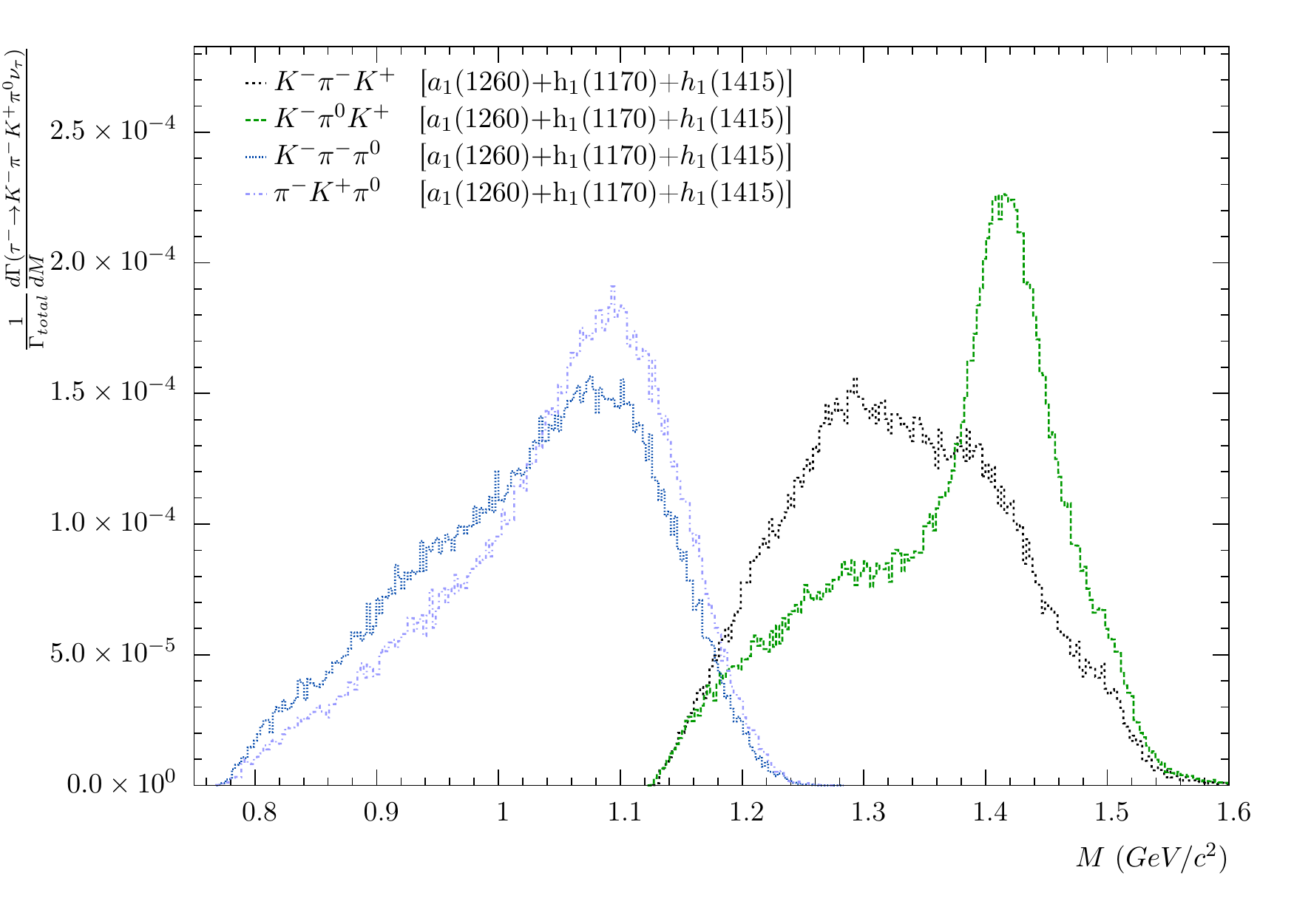}
    \includegraphics{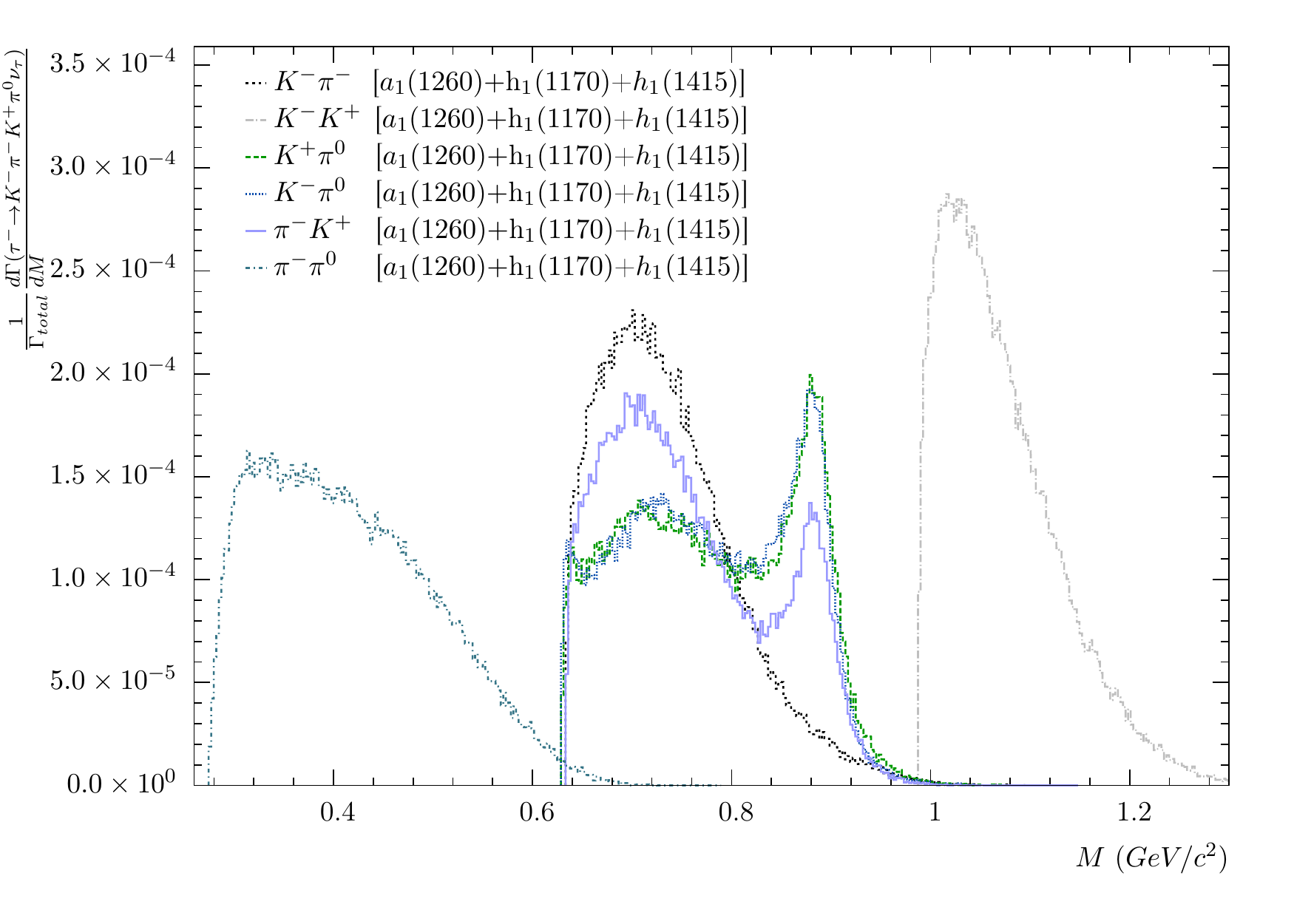}
  }
  \resizebox{421pt}{150pt}{
    \includegraphics{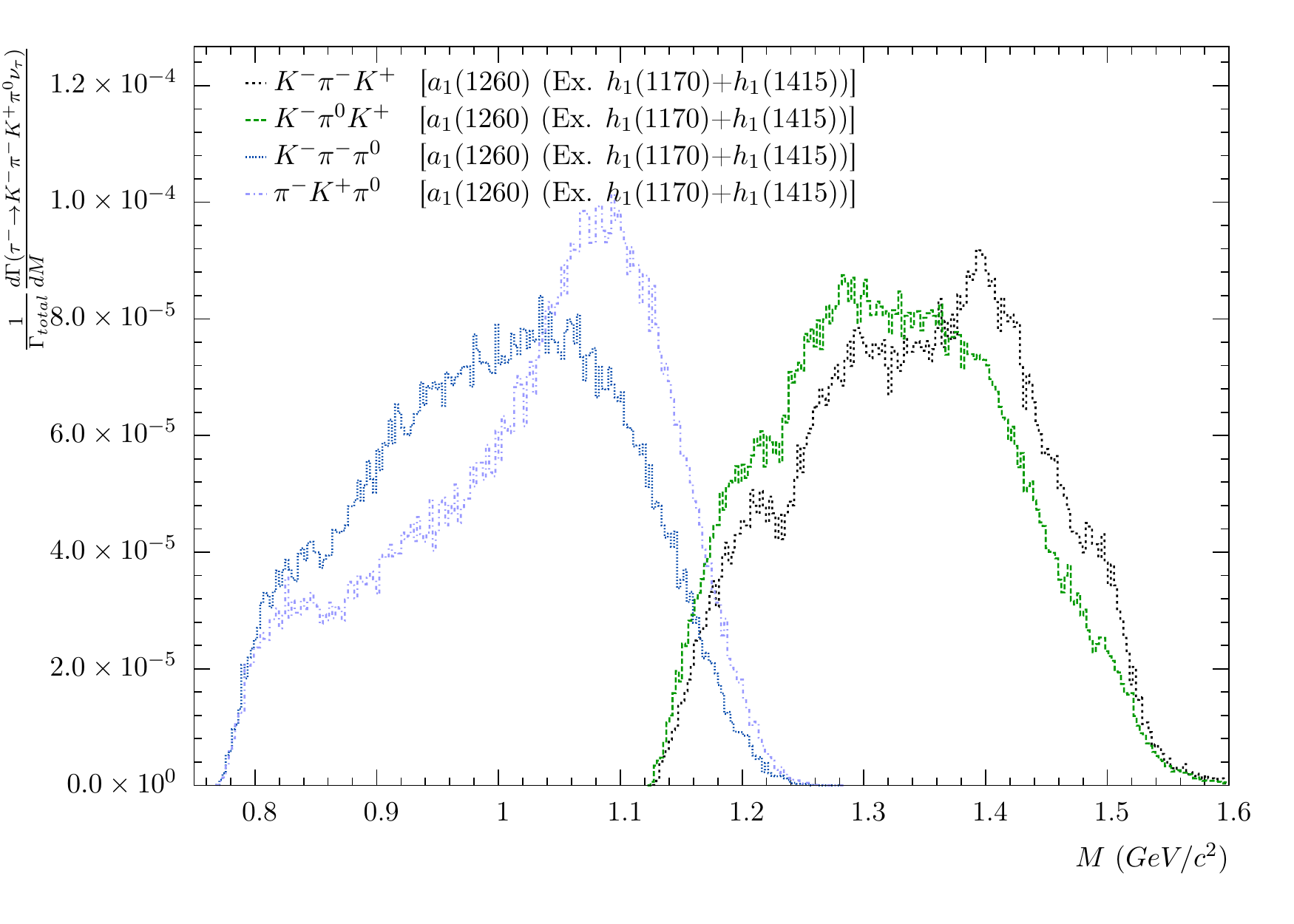}
    \includegraphics{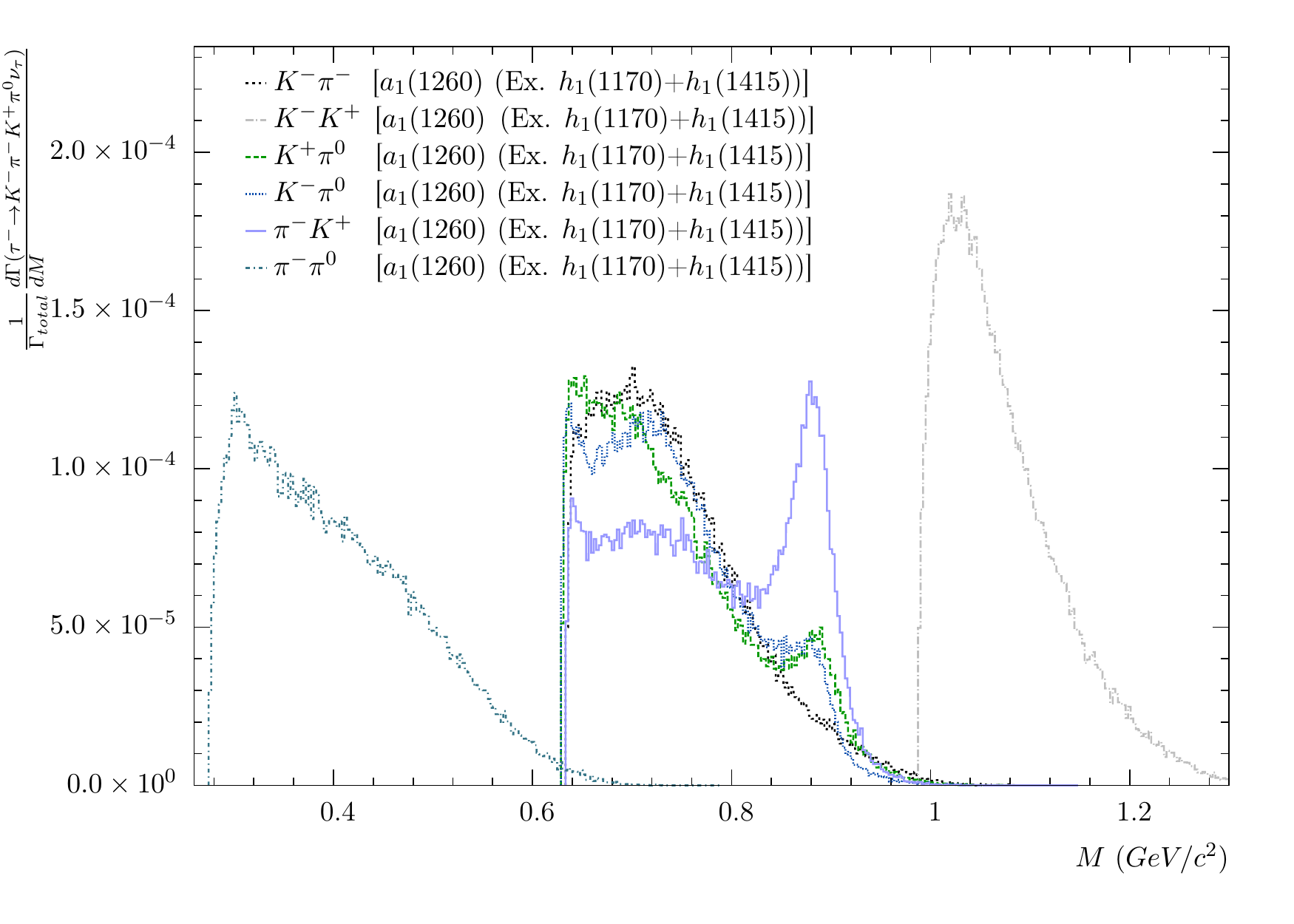}
  }
\end{center}
  \caption{The $K\pi^{-}K\pi^{0}$ invariant mass (Top-Left), the $K^{+}\pi^{0}K^{-}$ verses $K^{+}\pi^{-}K^{-}$ invariant mass distributions with mass projections 
 with the $h_{1}(1170)$ and $h_{1}(1415)$ included (Top-Right),
 the three-body  (Centre-Left) and the two-body invariant mass spectra with the $h_{1}(1170)$ and $h_{1}(1415)$ included (Centre-Right)  and 
 the three-body (Bottom-Left) and the two-body invariant mass spectra without the $h_{1}(1170)$ and $h_{1}(1415)$ included (Bottom-Right) for 
 the $\tau^{-}\to K\pi^{-}K\pi^{0}\nu_{\tau}$ 
 Phenomenological Model. \label{fig:phemKpiKpi0}   }
\end{figure*}

\clearpage

\begin{figure*}[tbp]
\begin{center}
\resizebox{260pt}{185pt}{
    \includegraphics{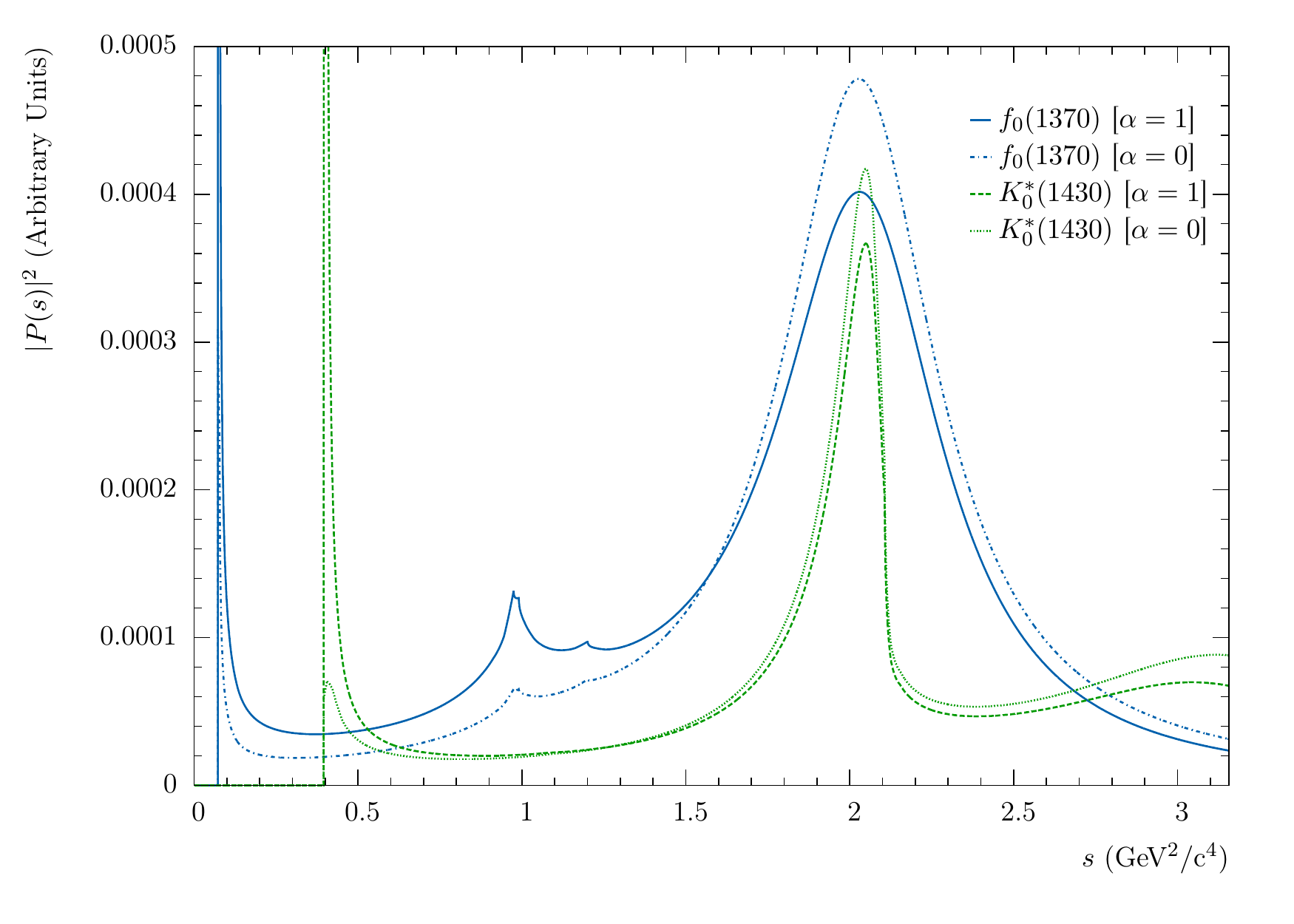}
  }
  \caption{The Breit-Wigner distributions after including the mass dependence of the decay width and the running of the mass for the $f_{0}(1370)$ and $K_{0}^{*}(1430)$ 
within the frame-work of the Flux-Tube Breaking
Model  \cite{Isgur:1988vm,Kokoski:1985is} with an integration cut-off of $20\times m_{\tau}^{2}$ to ensure convergence. The normalization of the distributions is for
presentation purposes only. \label{fig:scalar_IMR}}
\end{center}
\end{figure*}

\begin{figure*}[tbp]
\begin{center}
\resizebox{260pt}{185pt}{
    \includegraphics{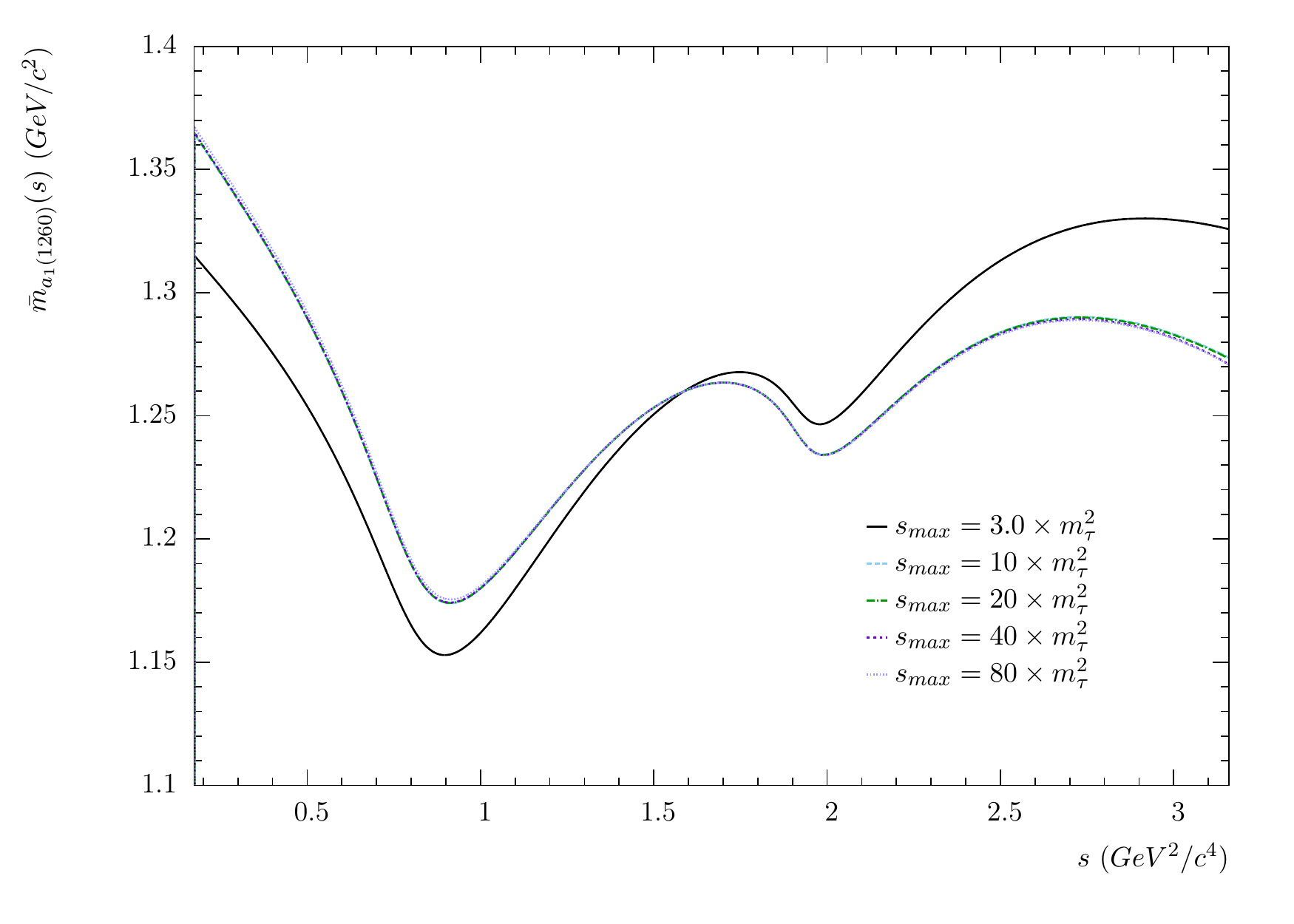}
  }
  \caption{The $\bar{m}(s)$ distribution as a function of increasing  $s_{max}$. The $3\times m_{\tau}^{2}$ distribution is similar to the distribution presented in
  \cite[Fig. 9]{Isgur:1988vm}.  By $s_{max}\sim 20\times m_{\tau}^{2}$ the integral has converged to $<0.1\%$, while higher values of $s_{max}\sim 20\times m_{\tau}^{2}$ become more sensitive to the numerical stability of $\Gamma(s)$ in terms of the sampling density.% and extrapolation.
 \label{fig:IMRConverge}}
\end{center}
\end{figure*}

\begin{figure*}[tbp]
\resizebox{520pt}{185pt}{
    \includegraphics{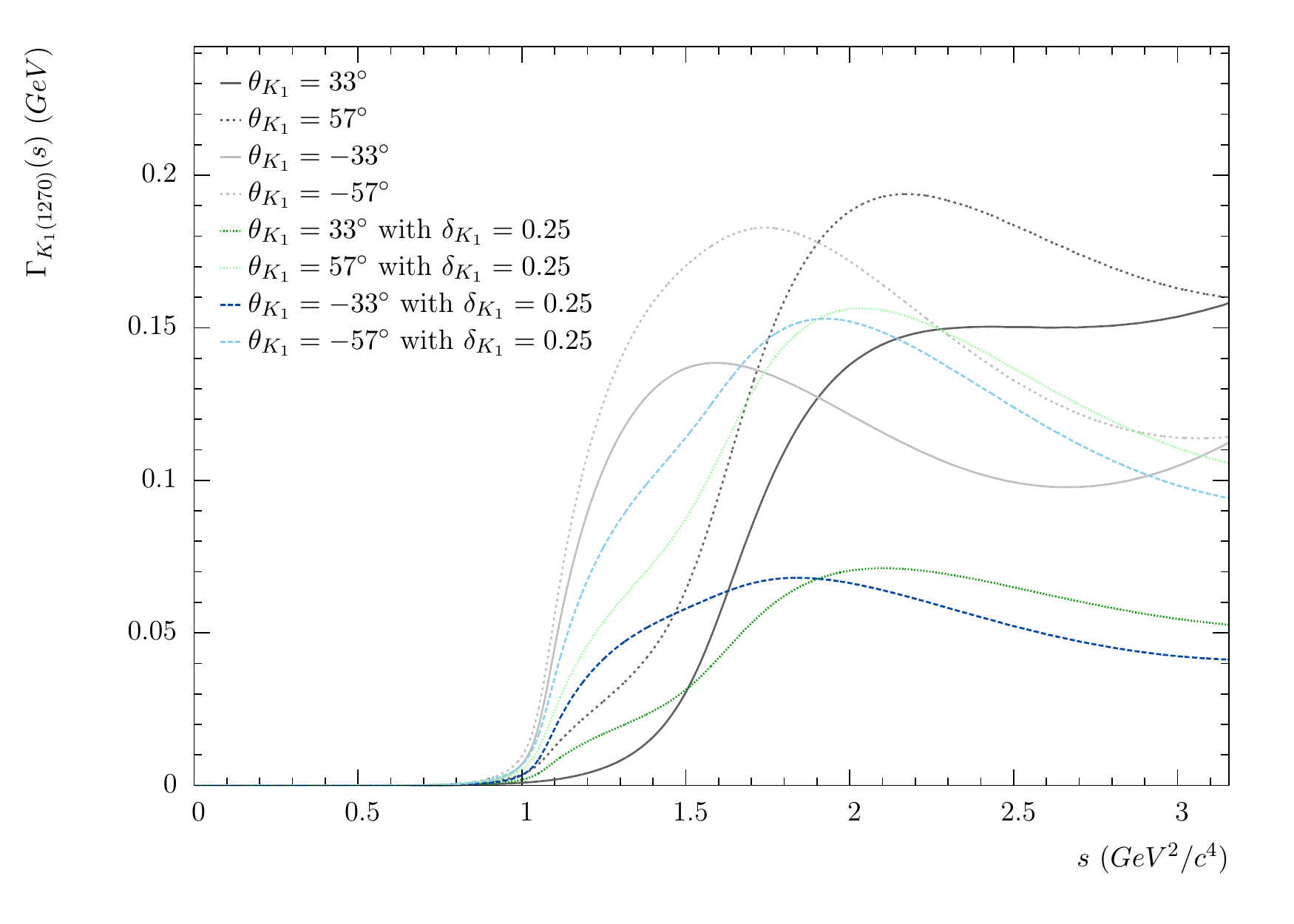}
    \includegraphics{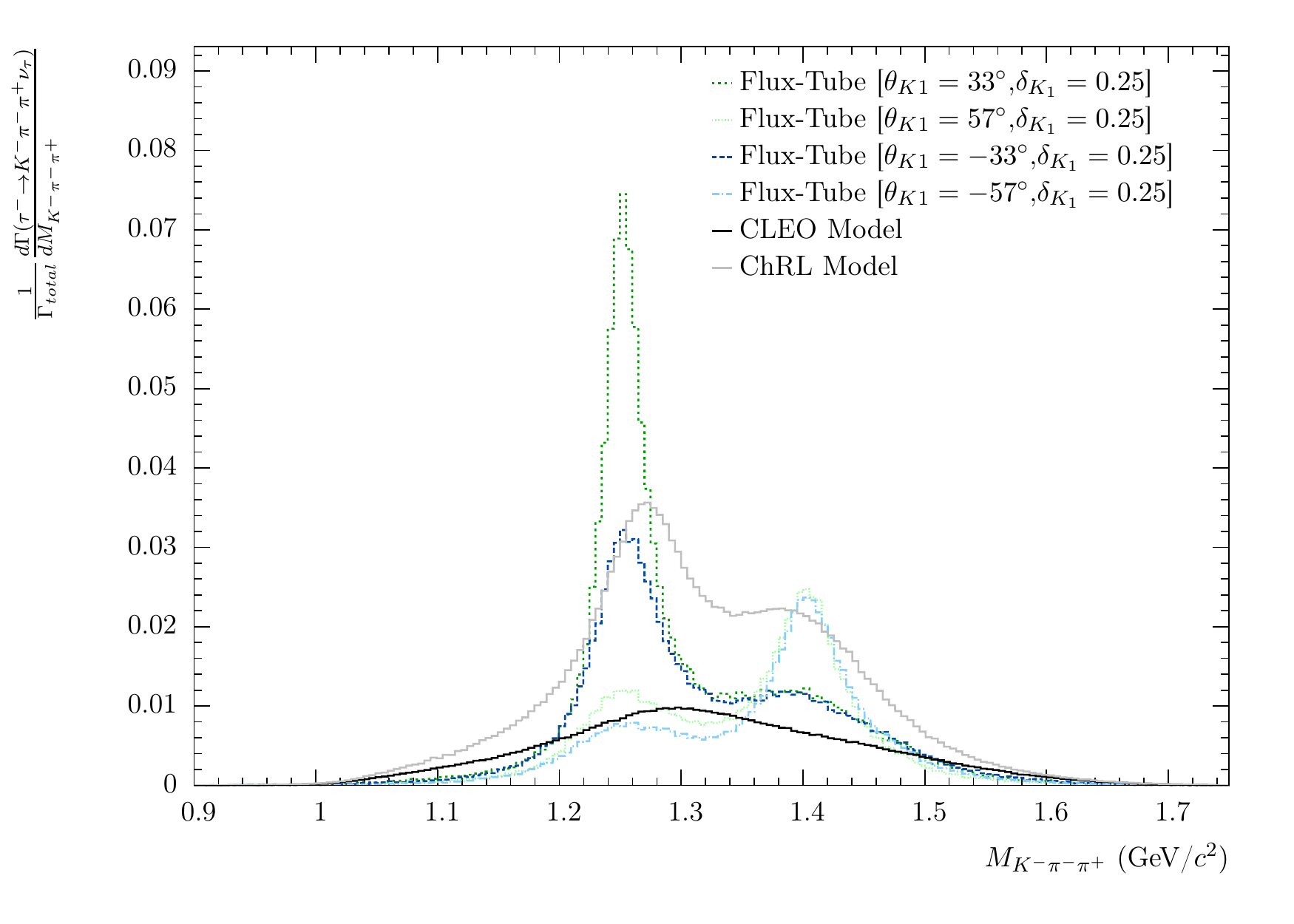}
}
\resizebox{520pt}{185pt}{
  \includegraphics{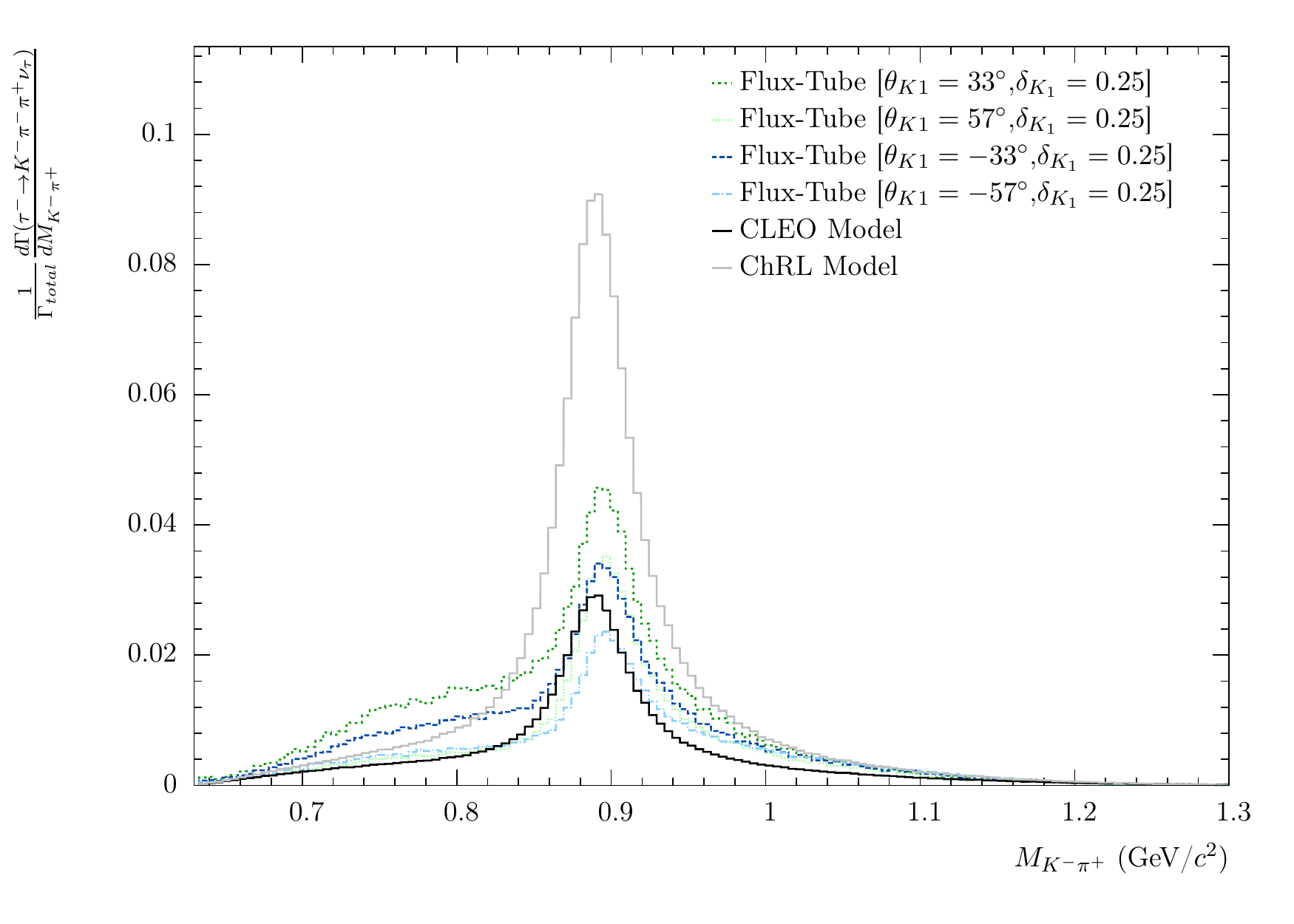}
  \includegraphics{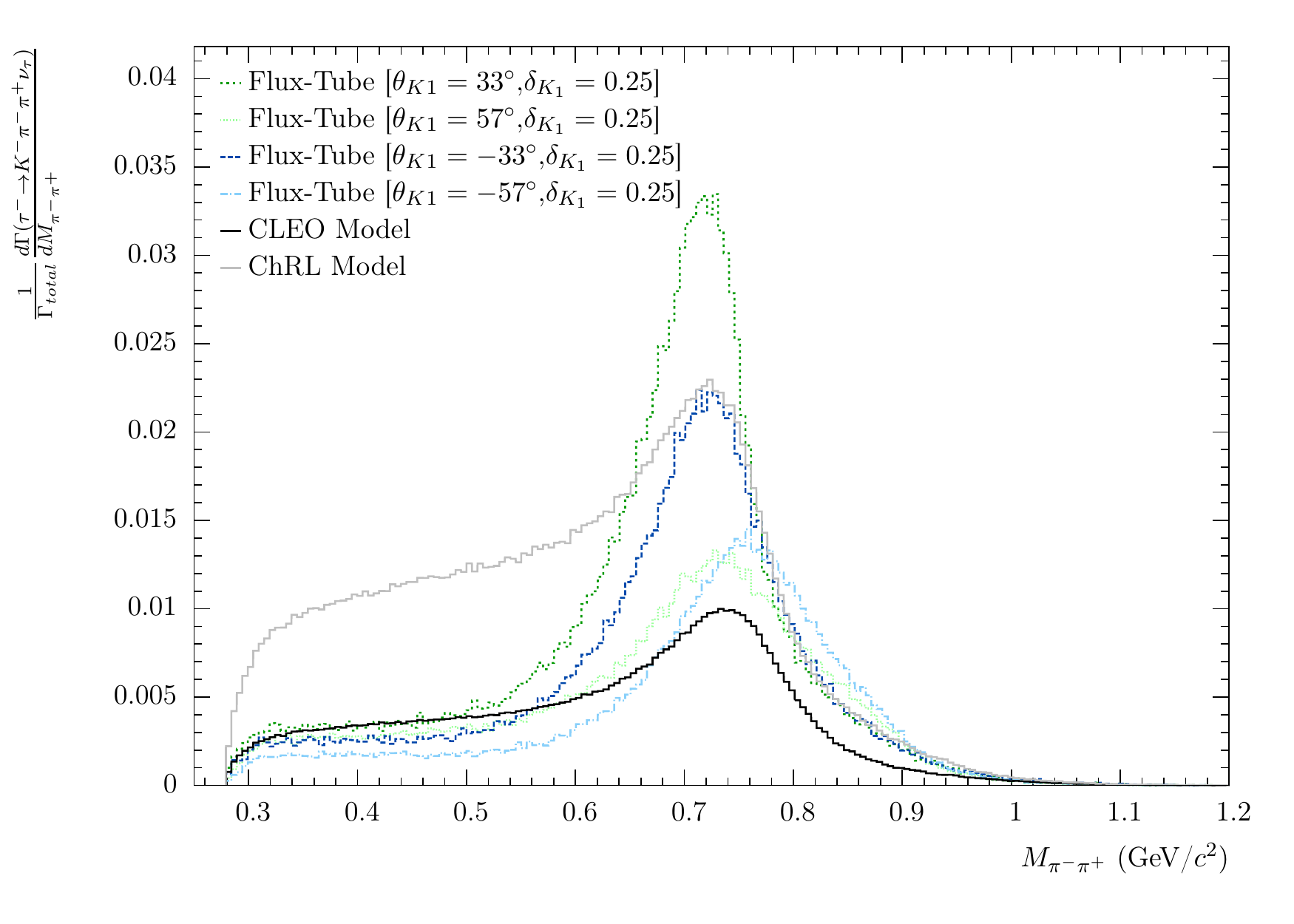}
}
\caption{The mass dependence of the $K_{1}(1270)$ decay width, $\Gamma(s)$, in the Flux-Tube Breaking Model in terms of 
$\theta_{K_{1}}$ and the $SU(3)_{f}$ suppression term $\delta_{K_{1}}$ (Top-Left). 
The decay width, $\Gamma_{K_{1}(1270)}(s)$ is presented for $\theta_{K_{1}}=\pm33^{\circ}/\pm57^{\circ}$ without the $SU(3)_{f}$ suppression and 
$\theta_{K_{1}}=\pm33^{\circ}/\pm57^{\circ}$ with the $SU(3)_{f}$ suppression factor $\delta_{K_{1}}=0.25$. In this figure, it is 
important to note that the $K^{*}(892)\pi$ partial decay width begins contributing to the total decay width at a much lower mass than 
the  $K\rho(770)$ and $K\omega(782)$ partial decay widths. Thus the line-shape of the $K_{1}(1260)$ (and also $K_{1}(1400)$) resonance(s) are strongly correlated to the
relative partial widths of the $K_{1}\to K^{*}(892)\pi$ to $K_{1}\to K\rho(770)$ and $K_{1}\to K\omega(782)$ decay modes. A comparison of the invariant mass distribution 
of the $K^{-}\pi^{-}\pi^{+}$ (Top-Right), $K^{-}\pi^{+}$ (Bottom-Left)  and $\pi^{-}\pi^{+}$ (Bottom-Right) for the ChRL \cite{Finkemeier:1995sr} Model,
the CLEO Model \cite{Asner:2000nx} and the Flux-Tube Breaking Model \cite{Isgur:1988vm,Godfrey:1985xj,Kokoski:1985is} 
for $\theta_{K_{1}}=\pm33^{\circ}/\pm57^{\circ}$ with $\delta_{K_{1}}=0.25$. For presentation purposes, the CLEO Model \cite{Asner:2000nx} is normalized to the
world average \cite{PDG2020}.  
 \label{fig:IMRK1}}
\end{figure*}

\begin{figure*}[tbp]
\begin{center}
  \resizebox{260pt}{185pt}{
    \includegraphics{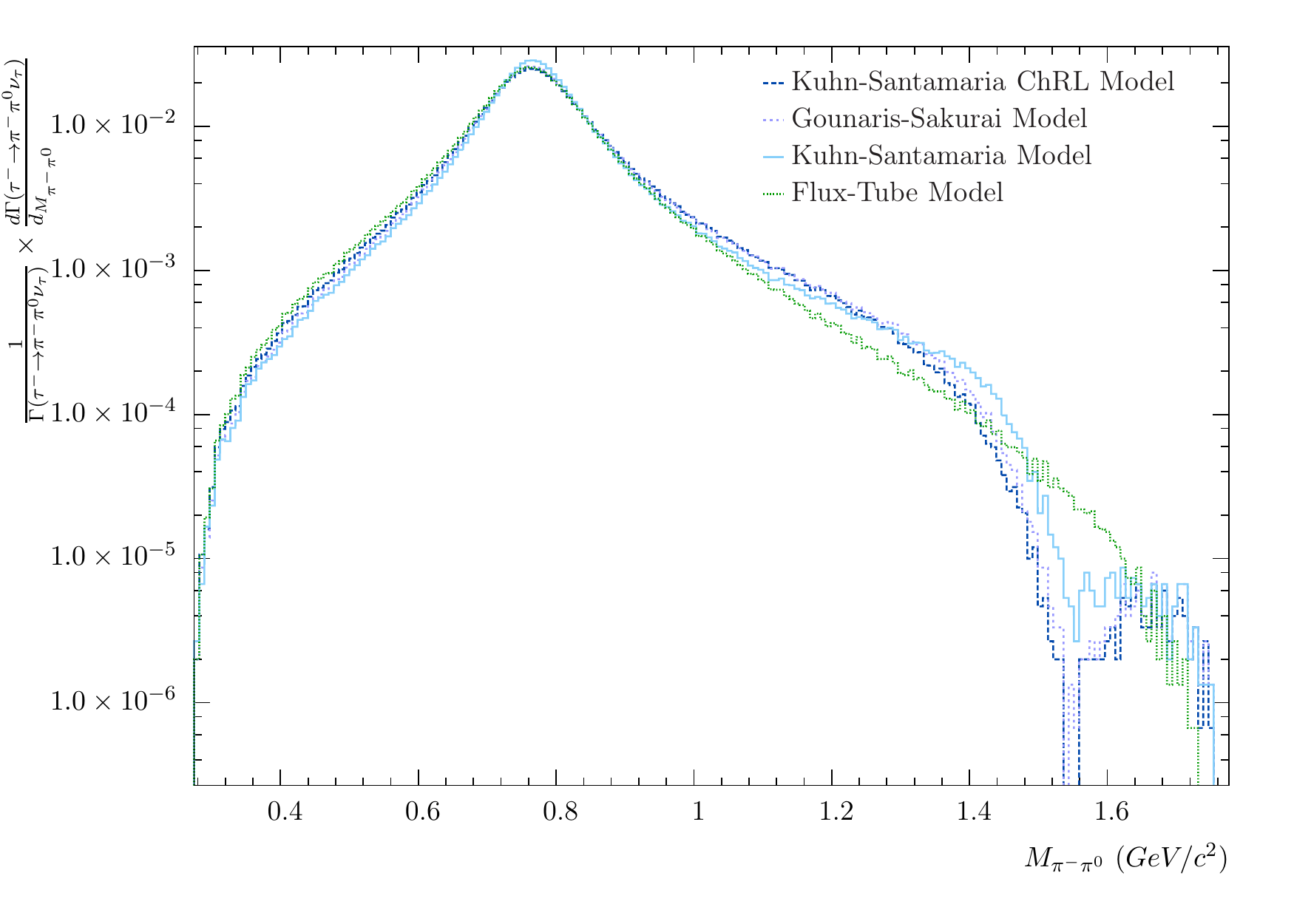}
  }
  \caption{A comparison of the $\tau^{-}\to\rho^{-}\nu_{\tau}$ hadronic $\pi^{-}\pi^{+}$ invariant mass for the simplified K\"uhn-Santamaria ChRL Model 
 \cite{Lees:2012cj,Kuhn:1990ad,Decker:1992kj}, the Gounaris-Sakuria Model   \cite{Lees:2012cj,Gounaris:1968mw},  the K\"uhn-Santamaria Model  \cite{Lees:2012cj,Kuhn:1990ad} and the Flux-Tube Breaking Model   \cite{Isgur:1988vm,Kokoski:1985is}. 
 \label{fig:rho_compare}}
\end{center}
\end{figure*}

\begin{figure*}[tbp]
  \resizebox{520pt}{185pt}{
    \includegraphics{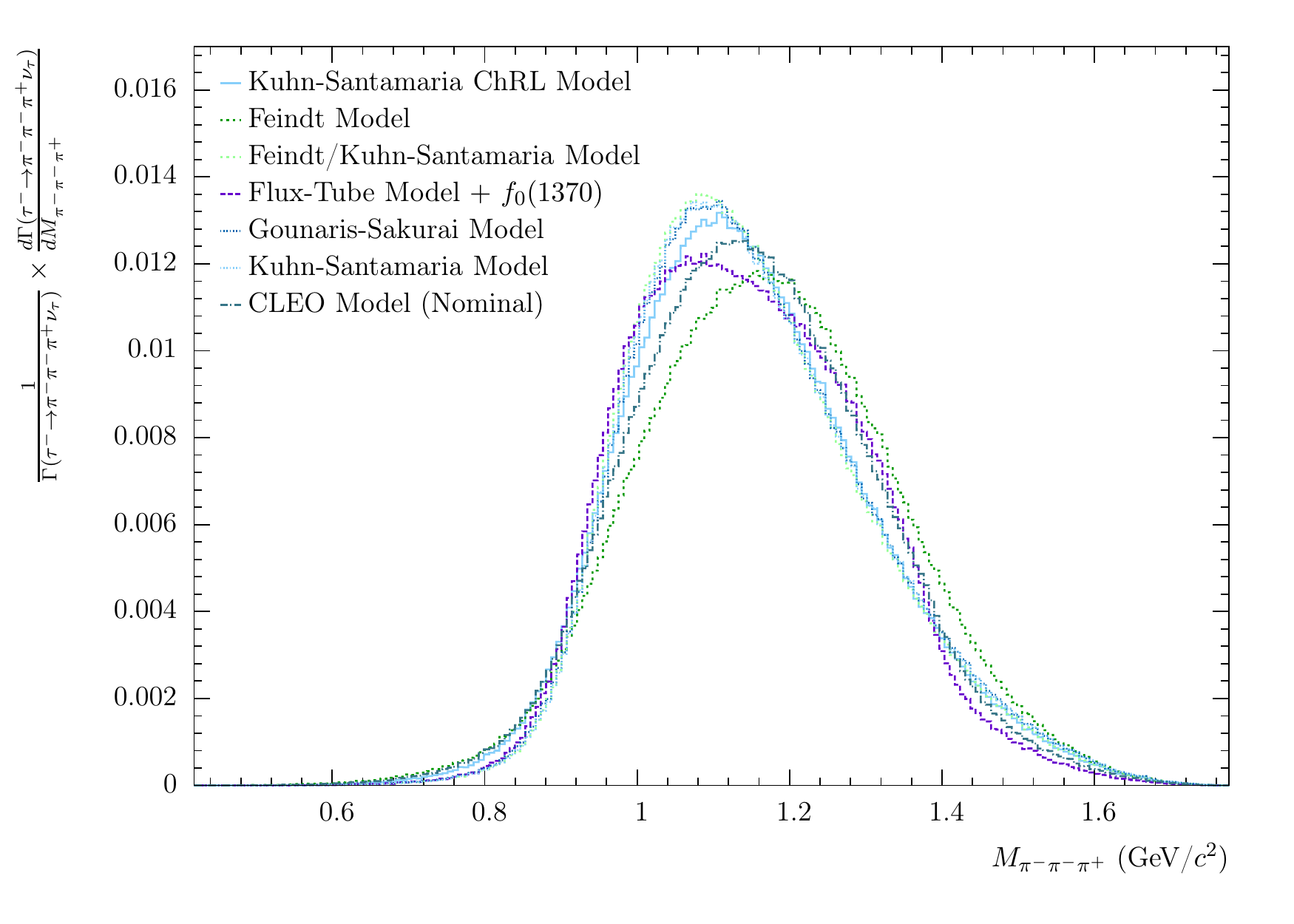}
    \includegraphics{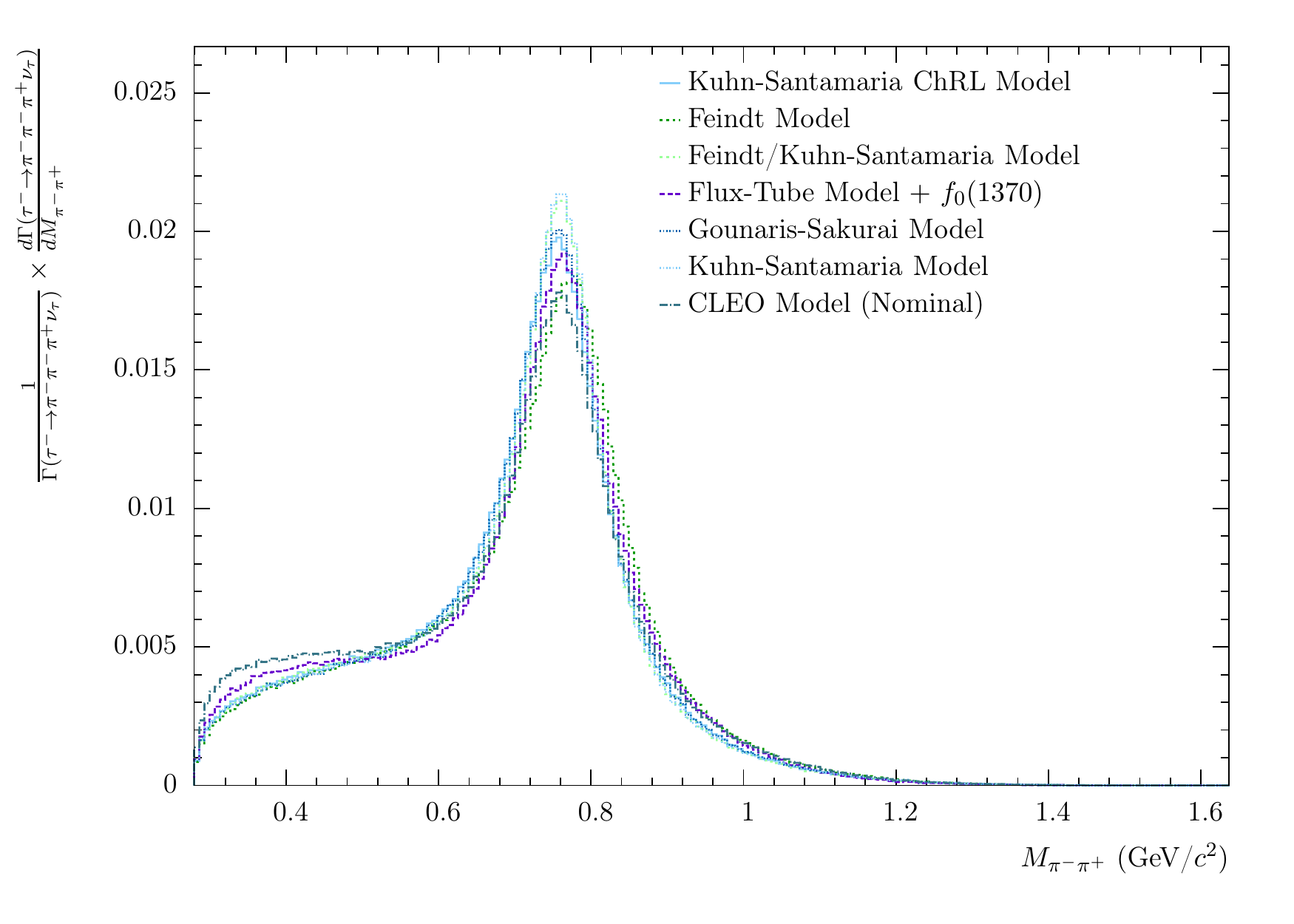}
  }
  \caption{A comparison of the $\tau^{-}\to a_{1}(1260)\nu_{\tau}$ hadronic $\pi^{-}\pi^{-}\pi^{+}$ (Left) and $\pi^{-}\pi^{+}$ (Right) invariant masses for the simplified K\"uhn-Santamaria ChRL Model  \cite{Lees:2012cj,Kuhn:1990ad}, the Feindt Model \cite{Feindt:1990ev}, the Feindt current \cite{Feindt:1990ev} with the K\"uhn-Santamaria Model Form-Factor \cite{Lees:2012cj,Kuhn:1990ad}, the Flux-Tube Breaking Model   \cite{Isgur:1988vm,Kokoski:1985is},  the Gounaris-Sakuria Model  \cite{Lees:2012cj,Gounaris:1968mw}, the  K\"uhn-Santamaria Model  \cite{Lees:2012cj,Kuhn:1990ad} and the CLEO Model \cite{CLEO3pi}. \label{fig:a1_compare}}
\end{figure*}
\clearpage

\begin{figure*}[tbp]
\begin{center}
  \resizebox{260pt}{185pt}{
    \includegraphics{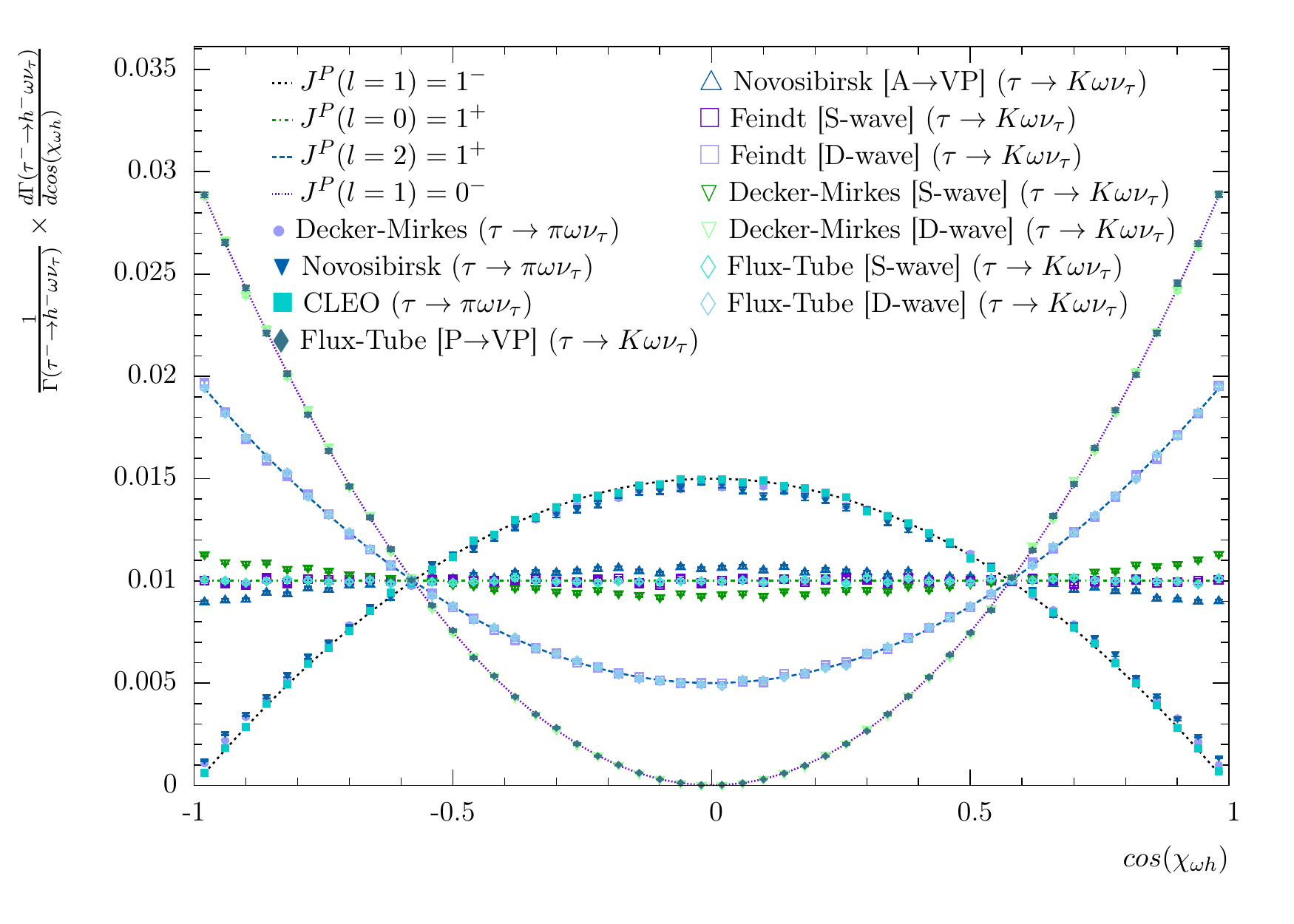}
  }
  \caption{A comparison of the theoretical curves for the $J^{P}(l=1)=1^{-}$, $J^{P}(l=0)=1^{+}$, $J^{P}(l=2)=1^{+}$, and $J^{P}(l=1)=0^{-}$  \cite{Edwards:1999fj} and the 
simulated distributions of $cos(\chi_{\omega h})$ where $h$ is either a pion or kaon. The simulated models for $\tau^{-}\to \omega\pi^{-}\nu_{\tau}$, which proceed 
through a
vector current ($J^{P}(l=1)=1^{-}$), include the CLEO Model  \cite{Edwards:1999fj}, the Novosibirsk Model  \cite{BONDAR2002139} and the Decker and Mirkes Model 
 \cite{Decker:1992jy}. 
Selection cuts are applied to reject the region of phase-space where both $\omega(782)$ resonances peak, since the interference between the two resonances 
may bias the angular distribution. The simulated models for $\tau^{-}\to \omega K^{-}\nu_{\tau}$  proceed through the axial vector current, $J^{P}(l=0)=1^{+}$, 
$J^{P}(l=2)=1^{+}$, for the Feindt Model  \cite{Feindt:1990ev}, the Decker-Mirkes Model  \cite{Decker:1992jy}, the Novosibirsk Model  \cite{BONDAR2002139},
 and Flux-Tube Breaking Model 
 \cite{Isgur:1988vm,Godfrey:1985xj,Kokoski:1985is}. The S-wave and D-wave for the Flux-Tube Breaking Model are defined in terms of the vertex amplitudes 
as $A=[1S_{1},0D_{1}]$ and $A=[0S_{1},1D_{1}]$ respectively, in conjunction with the associated vertex factors $A\to VP(f)$ and $A\to VP(g)$ from 
Table \ref{table:IMRVertices}. The Flux-Tube Breaking Model also contains the pseudo-scalar state $[P\to VP]$,  ($J^{P}(l=1)=0^{-}$) defined as $A=P_{4}$ 
in association with the vertex factor $P\to VP$ from Table \ref{table:IMRVertices}, $K(1460)$ which decays to  
$\tau^{-}\to \omega K^{-}\nu_{\tau}$  \cite{Isgur:1988vm,Godfrey:1985xj,Kokoski:1985is}.   
 \label{fig:coschi_omegah}}
\end{center}
\end{figure*}

\begin{figure*}[tbp]
\begin{center}
  \resizebox{260pt}{185pt}{
    \includegraphics{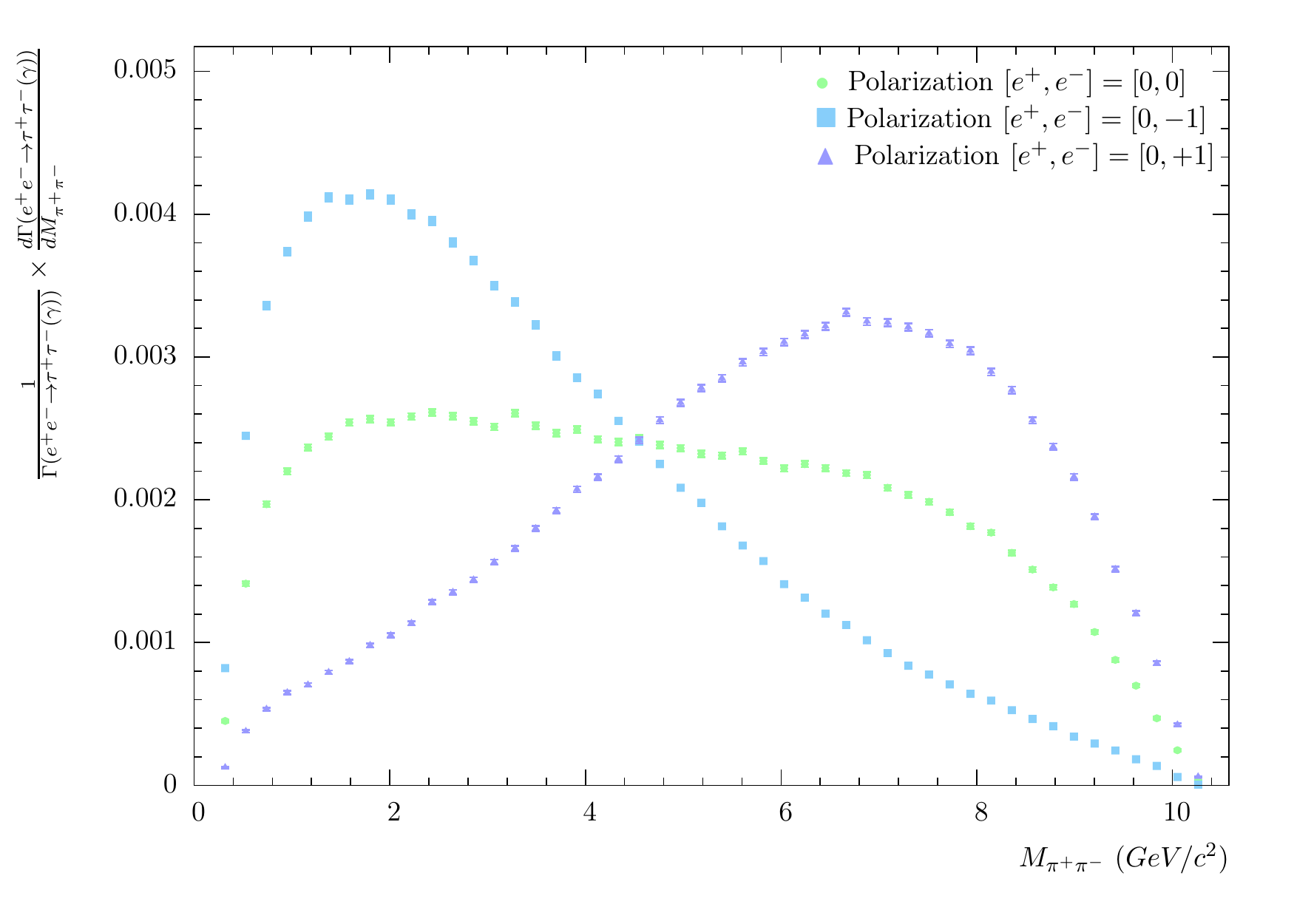}
  }
  \caption{The $M_{\pi^{+}\pi^{-}}$ distribution \cite{Pierzchala:2001gc} for unpolarized (Circle), negatively polarized electron beam (Square) and positively polarized electron beam (Triangular) 
simulated $e^{+}e^{-}\to \tau^{+}(\to\pi^{+}\nu_{\tau})\tau^{-}(\to\pi^{-}\nu_{\tau})(\gamma)$ events with Type IV Exponentiation
and a Soft-Photon Cut-Off of $E_{\gamma}=0.01$GeV. The requirement $cos(\theta_{\pi^{\pm}-e^{\pm}})>0$ has been applied to both pions.\label{fig:tauxPolar}}
\end{center}
\end{figure*}

\begin{figure*}[tbp]
\begin{center}
  \resizebox{260pt}{185pt}{
    \includegraphics{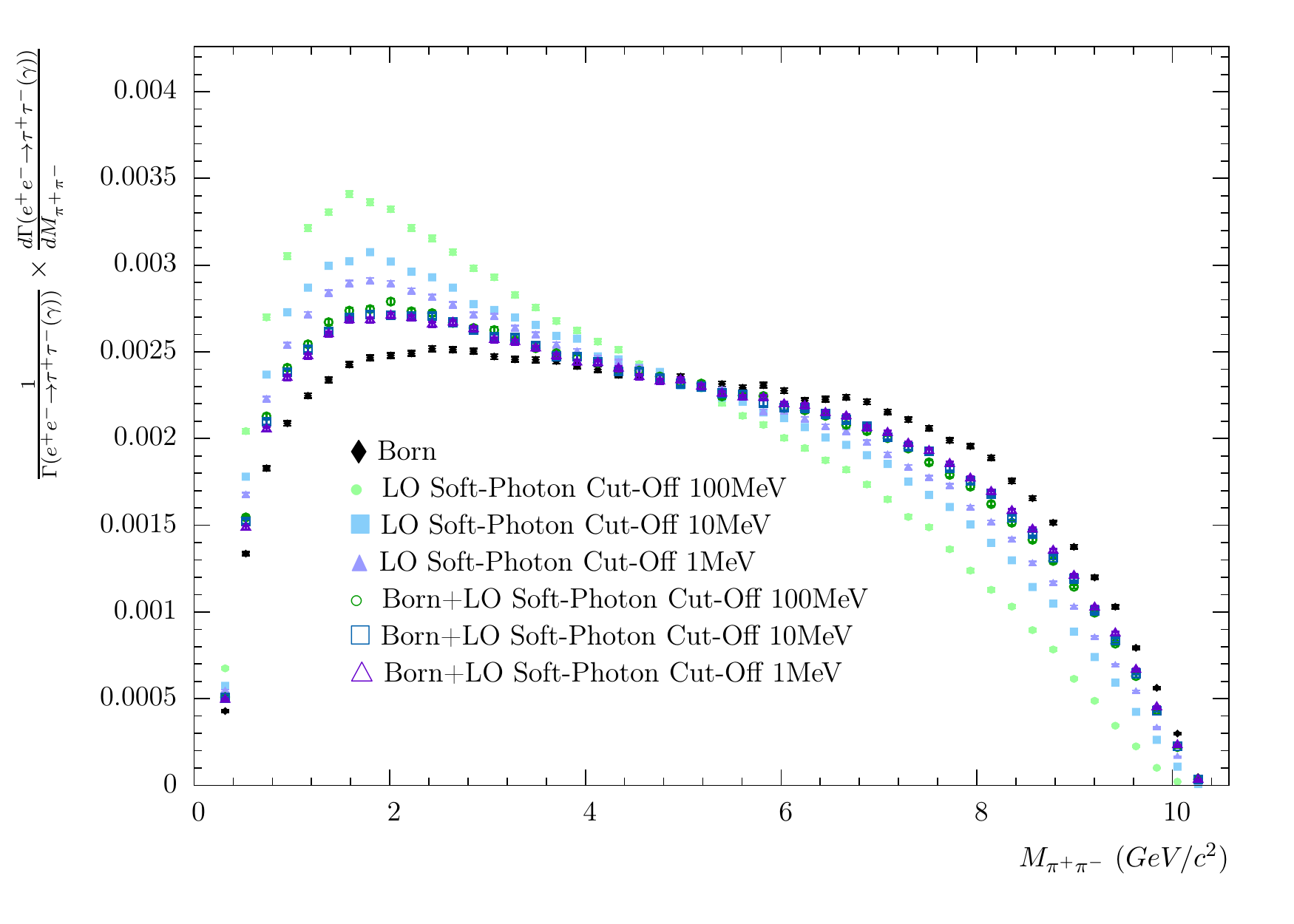}
  }
  \caption{The  $M_{\pi^{+}\pi^{-}}$ distribution  \cite{Pierzchala:2001gc} for the unpolarized simulated $e^{+}e^{-}\to \tau^{+}(\to\pi^{+}\nu_{\tau})\tau^{-}(\to\pi^{-}\nu_{\tau})$ 
events, Born level, (Diamond), 
for simulated  $e^{+}e^{-}\to \tau^{+}(\to\pi^{+}\nu_{\tau})\tau^{-}(\to\pi^{-}\nu_{\tau})\gamma$ events, LO radiative corrections, (Solid) 
and for simulated inclusive $e^{+}e^{-}\to \tau^{+}(\to\pi^{+}\nu_{\tau})\tau^{-}(\to\pi^{-}\nu_{\tau})(\gamma)$ events, Born+LO, (Open). The 
$e^{+}e^{-}\to \tau^{+}(\to\pi^{+}\nu_{\tau})\tau^{-}(\to\pi^{-}\nu_{\tau})\gamma$  and 
$e^{+}e^{-}\to \tau^{+}(\to\pi^{+}\nu_{\tau})\tau^{-}(\to\pi^{-}\nu_{\tau})(\gamma)$ events are evaluated using Type IV Exponentiation with a
soft-photon cut-off of $E_{\gamma}=0.1$GeV (Circle), $E_{\gamma}=0.01$GeV (Square) and $E_{\gamma}=0.001$GeV (Triangle).
\label{fig:tauxRadPolarcutoff}}
\end{center}
\end{figure*}

%% file: paper_tab.tex
\begin{table*}[b]
\begin{center}
\caption{The branching ratio predicted by the ChRL Models  \cite{Finkemeier:1995sr,Decker:1992kj,Kuhn:2006nw}.
The branching ratios are determined from the simulated decay rate normalized by the world-average of the
$\tau$ lepton life-time \cite{PDG2020} and $\hbar$. The contribution from the Wess-Zumino anomaly has been separated out for the  \cite{Finkemeier:1995sr,Decker:1992kj}
 models for comparison with the published results. The simulated samples have a statistical uncertainty of 0.3\%. ${}^{\dagger}$ indicates that the modes have been 
measured without explicit particle identification.  
\label{table:ChRLBR}}
\scriptsize
\begin{tabular}{p{2.5cm}p{2.1cm}p{2.1cm}p{2.1cm}p{2.1cm}p{2.1cm}p{2.5cm}}
\hline
Decay Mode & $\Gamma_{x}/\Gamma_{total}$  \cite{Kuhn:1990ad,Lees:2012cj} & $\Gamma_{x}/\Gamma_{total}$  \cite{Decker:1992kj} &  $\Gamma_{x,a}/\Gamma_{total}$  \cite{Decker:1992kj} & $\Gamma_{x}/\Gamma_{total}$  \cite{Finkemeier:1995sr} &  $\Gamma_{x,a}/\Gamma_{total}$  \cite{Finkemeier:1995sr} & $\Gamma_{x}/\Gamma_{total}$ PDG  \cite{PDG2020} \\
\hline
$\tau^{-}\to\pi^{-}\pi^{0}\nu_{\tau}$ & $(2.203)\times 10^{-1}$ &  & \multicolumn{1}{c}{$-$} &$(2.080)\times 10^{-1}$ & \multicolumn{1}{c}{$-$} 
&$(25.46\pm0.09)\times10^{-2}$\\
$\tau^{-}\to K^{-}\pi^{0}\nu_{\tau}$ &\multicolumn{1}{c}{$-$} & \multicolumn{1}{c}{$-$} & \multicolumn{1}{c}{$-$} &$(5.152)\times 10^{-3}$ & \multicolumn{1}{c}{$-$} 
&$(4.33\pm0.15)\times10^{-3}$\\
$\tau^{-}\to \bar{K}^{0}\pi^{-}\nu_{\tau}$ &\multicolumn{1}{c}{$-$} & \multicolumn{1}{c}{$-$} & \multicolumn{1}{c}{$-$} 
&$(9.502)\times 10^{-3}$ & \multicolumn{1}{c}{$-$} &$(8.38\pm0.14)\times10^{-3}$ \\
$\tau^{-}\to K^{-}K^{0}\nu_{\tau}$ &  $(9.014)\times 10^{-4}$ & \multicolumn{1}{c}{$-$} &  \multicolumn{1}{c}{$-$}& $(1.623)\times 10^{-3}$& \multicolumn{1}{c}{$-$} 
&$(1.486\pm0.034)\times10^{-3}$\\
$\tau^{-}\to K^{-}\eta\nu_{\tau}$ &  \multicolumn{1}{c}{$-$}&   \multicolumn{1}{c}{$-$}&   \multicolumn{1}{c}{$-$}& $(1.739)\times10^{-4}$ &   \multicolumn{1}{c}{$-$}& $(1.55\pm0.08)\times10^{-4}$\\ 
$\tau^{-}\to K^{-}\eta'(958)\nu_{\tau}$  &  \multicolumn{1}{c}{$-$}&   \multicolumn{1}{c}{$-$}&   \multicolumn{1}{c}{$-$}& $(9.155)\times10^{-5}$ &   \multicolumn{1}{c}{$-$} & $<2.4\times10^{-6}$ 90\% CL \\
$\tau^{-}\to\pi^{-}\pi^{-}\pi^{+}\nu_{\tau}$ & $(12.87)\times 10^{-2}$  & $(12.96)\times 10^{-2}$ &0.0 & \multicolumn{1}{c}{$-$} & \multicolumn{1}{c}{$-$} &$(9.02\pm0.05)\times10^{-2}$\\
$\tau^{-}\to\pi^{0}\pi^{0}\pi^{-}\nu_{\tau}$ & $(12.93)\times 10^{-2}$ & $(13.23)\times 10^{-2}$ &0.0 &\multicolumn{1}{c}{$-$}& \multicolumn{1}{c}{$-$} &$(9.26\pm 0.10)\times10^{-2}$ \\
$\tau^{-}\to K^{-}\pi^{0}\pi^{0}\nu_{\tau}$ &\multicolumn{1}{c}{$-$} &  $(1.128)\times 10^{-3}$ &0.0& $(1.540)\times10^{-3}$ & $(1.256)\times10^{-4}$ 
&$(6.5\pm2.2)\times10^{-4}$\\
$\tau^{-}\to K^{-}\pi^{-}\pi^{+}\nu_{\tau}$ & \multicolumn{1}{c}{$-$} &$(5.788)\times 10^{-3}$& $(3.697)\times10^{-5}$ 
&$(7.637)\times10^{-3}$ & $(4.079)\times10^{-4}$ &$(2.93\pm0.07)\times10^{-3}$\\
$\tau^{-}\to \pi^{-}\bar{K}^{0}\pi^{0}\nu_{\tau}$ &\multicolumn{1}{c}{$-$} & $(5.810)\times10^{-3}$ & $(7.141)\times 10^{-5}$ 
&$(9.298)\times 10^{-3}$& $(5.567)\times10^{-4}$ &$(3.82\pm0.13)\times 10^{-3}$ \\
$\tau^{-}\to K^{-}\pi^{-}K^{+}\nu_{\tau}$ &\multicolumn{1}{c}{$-$} & $(1.021)\times10^{-3}$ & $(3.692)\times10^{-4}$ 
&$(2.082)\times 10^{-3}$ & $(8.505)\times10^{-4}$ &$(1.435\pm0.027)\times10^{-3}$\\
$\tau^{-}\to K^{0}\pi^{-}\bar{K}^{0}\nu_{\tau}$ &\multicolumn{1}{c}{$-$} & $(9.428)\times10^{-4}$ & $(3.392)\times10^{-4}$ 
&$(1.919)\times 10^{-3}$ & $(7.833)\times10^{-4}$ &$(1.55\pm0.24)\times 10^{-3}$\\
$\tau^{-}\to K_{S}^{0}\pi^{-}K_{S}^{0}\nu_{\tau}$ &\multicolumn{1}{c}{$-$} &\multicolumn{1}{c}{$-$} &\multicolumn{1}{c}{$-$} 
&$(4.651)\times 10^{-4}$ & $(1.410)\times10^{-4}$ &$(2.34\pm0.06)\times10^{-4}$\\
$\tau^{-}\to K_{S}^{0}\pi^{-}K_{L}^{0}\nu_{\tau}$ &\multicolumn{1}{c}{$-$} &\multicolumn{1}{c}{$-$} &\multicolumn{1}{c}{$-$} 
&$(9.870)\times 10^{-4}$ & $(5.028)\times10^{-4}$ &$(1.08\pm0.24)\times10^{-3}$\\
$\tau^{-}\to K_{L}^{0}\pi^{-}K_{L}^{0}\nu_{\tau}$ &\multicolumn{1}{c}{$-$} &\multicolumn{1}{c}{$-$} &\multicolumn{1}{c}{$-$} 
&$(4.651)\times 10^{-4}$ & $(1.410)\times10^{-4}$ &$(2.35\pm0.06)\times10^{-4}$\\
$\tau^{-}\to K^{-}\pi^{0}K^{0}\nu_{\tau}$ &\multicolumn{1}{c}{$-$} & $(1.424)\times 10^{-4}$&0.0 &$(1.630)\times 10^{-3}$ & $(5.987)\times10^{-4}$   %%%%% check
&$(1.50\pm0.07)\times 10^{-3}$\\
$\tau^{-}\to K^{-}K^{-}K^{+}\nu_{\tau}$ &\multicolumn{1}{c}{$-$} & $(7.542)\times 10^{-7}$&0.0& $(1.508)\times10^{-5}$ & $(3.190)\times10^{-8}$ & 
$(2.2\pm0.8)\times 10^{-5}$\\
$\tau^{-}\to \eta\pi^{-}\pi^{0}\nu_{\tau}$ &\multicolumn{1}{c}{$-$} & $(1.335)\times 10^{-3}$& $(1.335)\times 10^{-3}$
&\multicolumn{1}{c}{$-$} & \multicolumn{1}{c}{$-$}& $(1.39\pm0.07)\times10^{-3}$ \\
\hline
Decay Mode & & $\Gamma_{x}/\Gamma_{total}$  \cite{Kuhn:2006nw} &  $\Gamma_{x}/(R\times\Gamma_{total})$  \cite{Kuhn:2006nw}& & & $\Gamma_{x}/\Gamma_{total}$ PDG  \cite{PDG2020} \\
\hline
\multicolumn{2}{l}{$\tau^{-}\to \pi^{-}\pi^{-}\pi^{-}\pi^{+}\pi^{+}\nu_{\tau}$ } & $ (1.911)\times10^{-3}$ & $(9.554)\times10^{-4}$ & & & $(7.75\pm0.30)\times 10^{-4}$\\
\multicolumn{2}{l}{$\tau^{-}\to \pi^{-}\pi^{-}\pi^{+}\pi^{0}\pi^{0}\nu_{\tau}$ } & $(7.7587)\times 10^{-3}$ &  $(5.7468)\times 10^{-3}$ & &  & ${(4.95\pm0.31)\times10^{-3}}^{\dagger}$\\
\multicolumn{2}{l}{$\tau^{-}\to \pi^{-}\pi^{-}\pi^{+}\pi^{0}\pi^{0}\nu_{\tau}$ [Excl. $\omega(782)$] } & $(2.982)\times10^{-3}$ & $(9.994)\times10^{-4}$ & & & ${(1.0\pm0.4)\times10^{-3}}^{\dagger}$\\
\multicolumn{2}{l}{$\tau^{-}\to \pi^{-}\pi^{-}\pi^{+}\pi^{0}\pi^{0}\nu_{\tau}$ [$\omega(782)$ only]} &  $(4.753)\times 10^{-3}$ &\multicolumn{1}{c}{$-$} & & & ${(4.1\pm0.4)\times10^{-3}}^{\dagger}$\\
\multicolumn{2}{l}{$\tau^{-}\to \pi^{-}\pi^{0}\pi^{0}\pi^{0}\pi^{0}\nu_{\tau}$ } & $(8.018)\times10^{-4}$ & $(8.018)\times10^{-4}$& & & ${(1.1\pm0.4)\times10^{-3}}^{\dagger}$ \\
\hline
\end{tabular}
\end{center}
\end{table*}

\clearpage

\begin{table*}[t]
\begin{center}
\caption{The numerical constants for the phenomenological models for $\tau^{-}\to K^{-}\pi^{-}\pi^{-}\pi^{0}\nu_{\tau}$ and
 $\tau^{-}\to K^{-}\pi^{-}K^{-}\pi^{0}\nu_{\tau}$ decays.
${}^{\dagger}$ implies that the value was extrapolated from the reference.
\label{table:SDwave}}
\scriptsize
\begin{tabular}{p{2.75cm}p{2cm}p{2cm}p{0.5cm}}
\hline 
Decay & $D/S$ Ratio & &Ref \\
\hline
$K^{*}(1680)\to K_{1}\pi$ & $1/2$ & &\\
$K_{1}(1270)\to K^{*}\pi$ & $1.0\pm0.7$  &&  \cite{PDG2020} \\
$K_{1}(1400)\to K^{*}\pi$ & $0.04\pm0.01$ &&  \cite{PDG2020} \\
$K_{1}(1270)\to K\rho$ & $1.0\pm0.7$  &&  \cite{PDG2020}${}^{\dagger}$ \\
$K_{1}(1270)\to K\omega(782)$ & $1.0\pm0.7$  &&  \cite{PDG2020}${}^{\dagger}$ \\
$\rho(1700)\to a_{1}(1260)\pi$  & $1/2$ & &\\
$\rho(1700)\to h_{1}^{0}(1170)\pi$  & $1/2$ & &\\
$\rho(1700)\to h_{1}^{0}(1415)\pi$  & $0$ & &\\
$a_{1}(1260)\to\rho(770)\pi$  & $-0.062\pm0.20$ &&  \cite{PDG2020}  \\
$a_{1}(1260)\to K^{*}(892)K$  & $-0.062\pm0.20$ &&  \cite{PDG2020}${}^{\dagger}$  \\
$h_{1}^{0}(1170)\to \rho(770)\pi$  & $1/8$ &&  \cite{Kokoski:1985is}  \\
$h_{1}^{0}(1415)\to K^{*}(892)K$  & $1/8$ &&  \cite{Kokoski:1985is}   \\
\hline
Meson  & Mass ($GeV/c^{2}$)& Width ($GeV/c^{2}$)&  Ref \\
\hline
$K^{*}(1680)$ &1.714 & 0.323 &  \cite{Finkemeier:1995sr}\\
$K^{*}(1410)$ &1.412 &0.227  &  \cite{Finkemeier:1995sr}\\
$K_{1}(1270)$ & 1.270 & 0.090 &  \cite{Finkemeier:1995sr}\\
$K_{1}(1400)$ & 1.402 & 0.174&  \cite{Finkemeier:1995sr}\\
$\omega(782)$ & 0.782& 0.00843&  \cite{Finkemeier:1995sr}\\
$\rho(1770)$  & 1.750 & 0.120 & \cite{Finkemeier:1995sr}\\
$a_{1}(1260)$  & 1.251 & 0.475 & \cite{Finkemeier:1995sr}\\
$h_{1}^{0}(1170)$  & $1.166\pm0.006$& $0.375\pm0.035$& \cite{PDG2020}\\
$h_{1}^{0}(1415)$  & $1.416\pm0.008$& $0.090\pm0.015$ & \cite{PDG2020}\\
$\rho(1450)$ & 1.370& 0.510 & \cite{Finkemeier:1995sr}\\
$\rho(770)$ & 0.773& 0.145 & \cite{Finkemeier:1995sr}\\
$K^{*}(892)$ & 0.892& 0.050 & \cite{Finkemeier:1995sr}\\
\hline
\end{tabular}
\end{center}
\end{table*}

\begin{table*}[b]
\begin{center}
\caption{The branching ratios predicted by the phenomenological models for $\tau^{-}\to K^{-}\pi^{-}\pi^{-}\pi^{0}\nu_{\tau}$ and
 $\tau^{-}\to K^{-}\pi^{-}K^{-}\pi^{0}\nu_{\tau}$ decays. The branching ratios are determined from the simulated decay rate normalized by the world-average of the
$\tau$ lepton life-time \cite{PDG2020} and $\hbar$. The simulated samples have a statistical uncertainty of 0.3\%.
\label{table:PhemBR}} 
\scriptsize
\begin{tabular}{p{3.6cm}p{1.75cm}p{2.25cm}}
\hline
Decay & $\Gamma_{x}/\Gamma_{total}$ &  $\Gamma_{x}/\Gamma_{total}$ PDG  \cite{PDG2020}  \\
\hline
$\tau^{-}\to K^{-}\pi^{-}\pi^{+}\pi^{0}\nu_{\tau}$ & $(7.890)\times 10^{-4}$  & $(7.9\pm1.2 )\times 10^{-4}$\\
$\tau^{-}\to K^{-}\omega(782)\nu_{\tau}$ & $(3.335)\times 10^{-4}$ &  $(4.1\pm0.9 )\times 10^{-4}$ \\
$\tau^{-}\to K^{-}\pi^{-}K^{+}\pi^{0}\nu_{\tau}$ & $(3.621)\times 10^{-5}$  & $(6.1\pm 1.8)\times 10^{-5}$\\
$\tau^{-}\to K^{-}\pi^{-}K^{+}\pi^{0}\nu_{\tau}$ [Ex. ${}^{1}P_{1}$] & $(2.237)\times 10^{-5}$ &  \multicolumn{1}{c}{$-$} \\
\hline
\end{tabular}
\end{center}
\end{table*}

\begin{table*}[t]
\begin{center}
\caption{The meson coupling constants for the Flux-Tube Breaking Model. The coupling constants for the vector mesons are based on an approximation from Chiral resonance 
models  \cite{Nugent:2013hxa}. The approximation for the $K_{1}(1270)$ and $K_{1}(1400)$ is related to the radius for the reduced mass of the quark anti-quark 
system forming the meson, since the coupling constant depends directly on the radius  \cite{Isgur:1988vm}. It is assumed that this is the primary effect for the 
difference between $f_{K}$ and $f_{\pi}$ and that a simple scaling can be applied to excited states compared to the more sophisticated pQCD estimate 
 \cite{Finkemeier:1997vn}. $f_{K_{1}}$  with the $SU(3)_{f}$ suppression is discussed in the context of mixing between the 
$K_{A}({}^{3}P_{1})$ and $K_{B}({}^{1}P_{1})$ states in Section \ref{sec:IMRstrange}. Within this formalism $f_{K_{1}}=f_{K_{A}({}^{3}P_{1})}=\frac{1}{\delta_{K_{1}}}f_{K_{B}({}^{1}P_{1})}=\frac{m_{s}+m_{u}}{m_{s}-m_{u}}f_{K_{B}({}^{1}P_{1})}$. Note: Within the $SU(3)_{f}$ and Heavy Quark limits where 
the $K_{1}(1270)$ can be approximated as the $K_{A}({}^{3}P_{1})$ state the pQCD and the simple ChRL scaling are approximately equivalent. 
\label{table:IMRCouplingConstants}}
\scriptsize
\begin{tabular}{p{5.3cm}p{2.5cm}}
\hline
Meson State  & Coupling Constant  \\
\hline
$\rho(770)$ & $\sqrt{2}f_{\pi}$  \cite{PDG2020,Nugent:2013hxa} \\
$a_{1}(1260)$ & 0.25  \cite{Isgur:1988vm}\\
$K^{*}(892)$ &  $\sqrt{2}f_{K}$  \cite{PDG2020,Nugent:2013hxa}\\
$K_{1}(1270)$ & 0.25$\frac{f_{K}}{f_{\pi}}$   \cite{PDG2020,Isgur:1988vm,Finkemeier:1997vn}\\
$K_{1}(1400)$ & 0.25$\frac{f_{K}}{f_{\pi}}$   \cite{PDG2020,Isgur:1988vm,Finkemeier:1997vn}\\
$K_{0}^{*}(1430)$ &  0.0225$\times\frac{m_{s}}{100MeV}$  \cite{Maltman_2001}\\
$K(1460)$ & 0.0214  \cite{Maltman_2001}\\
\hline
\end{tabular}
\end{center}
\end{table*}

\begin{table*}[b]
\begin{center}
\caption{A summary of the vertices implemented in the Flux-Tube Breaking Model for an initial hadronic state, A, with momentum $q^{\mu}$ and an intermediate 
hadronic state, B, with momentum $k^{\mu}$. The final state particles 
are given a momentum $p^{\mu}$. The corresponding vertex amplitudes $f_{A\to BC}$ are defined in  % \cite[Table III]{Kokoski:1985is}. 
\cite {Kokoski:1985is}. Following the procedure 
in \cite{Isgur:1988vm}, the Form-Factors are obtained from the amplitudes by the substitution of the appropriate Dalitz variables. 
$f_{A\to VP}$ and $g_{A\to VP}$ are determined from the S and D wave amplitudes and are defined in  \cite{Isgur:1988vm}.  % \cite[Eq. B11 \& B12]{Isgur:1988vm}. 
The $V\to VP$ is a vertex associated with the Wess-Zumino Anomaly  \cite{PICH1987561,Wess:1971yu,Donoghue:1996}. $|k_{i}|^{2}$ is the momentum in the CM frame for the 
intermediate particle. 
\label{table:IMRVertices}}
\scriptsize
\begin{tabular}{p{2cm}p{5.8cm}}
\hline
Interaction & Vertex \\
\hline
$A\to VP(f)$ & $f_{A\to VP}(q^{2},k^{2})\left[-g^{\mu\nu}+\frac{k^{\mu}k^{\nu}}{k^{2}}+\frac{q^{\mu}q^{\nu}}{q^{2}}-\frac{k\cdot q q^{\mu}k^{\nu}}{k^{2}q^{2}} \right] $\\
$A\to VP(g)$ &$g_{A\to VP}(q^{2},k^{2})\left[\frac{k\cdot q q^{\mu}}{q^{2}}-k^{\mu}\right]\left[q^{\nu}-\frac{k\cdot q k^{\nu}}{k^{2}}\right]$ \\
$A\to SP$    & $f_{A\to SP}(k^{2})\left[p_{l}^{\nu}-k^{\nu}\right]$ \\
$V\to VP$    & $f_{V\to VP}(q^{2},k^{2})\left[\frac{{\epsilon^{\mu\nu}}_{\beta\gamma}k^{\beta}p_{3}^{\gamma}}{|k_{i}|^{2}}\right]$ \\
$V\to PP$    & $-\imath f_{V\to PP}(k^{2})\left[p_{m}^{\nu}-p_{n}^{\nu}\right]$ \\
$P\to VP$    & $\imath f_{P\to VP} \left[q^{\nu}-\frac{k\cdot q  k^{\nu}}{k^{2}}\right]$ \\
$P\to SP$    & $-\imath f_{P\to SP}(q^{2},k^{2})$\\
$S\to PP$    & $-\imath f_{S\to PP}(k^{2})$ \\
\hline
\end{tabular}
\end{center}
\end{table*}

\begin{table*}[t]
\begin{center}
\caption{The $K_{1}(1270)$ and $K_{1}(1400)$ decay widths at the pole-mass, $\Gamma(m_{res})$, predicted in the Flux-Tube Breaking Model for the common values 
of the $K_{1}$ mixing angle and the $SU(3)_{f}$ symmetry breaking factor compared to the measured world values  \cite{PDG2020}. 
\label{table:IMRK1Gbar}}
\scriptsize
\begin{tabular}{p{1.5cm}p{1.5cm}p{2.0cm}p{2.0cm}}
\hline
$\theta_{K_{1}}$ & $\delta_{K_{1}}$ & \multicolumn{2}{c}{Width ($GeV/c^{2}$)} \\
& & $K_{1}(1270)$   & $K_{1}(1400)$ \\
\input{K1WidthTable.tex}
\hline
\multicolumn{2}{l}{$\tau$ Decays  \cite{PDG2020}} & $0.260_{-110}^{+120}$&  \\
\multicolumn{2}{l}{$K$ Beams  \cite{PDG2020}} & $0.090\pm0.008$&  \\
\multicolumn{2}{l}{$K$ Back-Scatter  \cite{PDG2020}} & $0.075\pm0.015$&  \\
\multicolumn{2}{l}{non-$K$ Beams   \cite{PDG2020}} & $0.120\pm0.008$&  \\
\multicolumn{2}{l}{Avg.   \cite{PDG2020}} & $0.090\pm0.020$& $0.174\pm0.013$ \\
\hline
\end{tabular}
\end{center}
\end{table*}

\begin{table*}[b]
\begin{center}
\caption{The Branching Ratio predicted by the Flux-Tube Model for the strange and non-strange decay modes. For the result presented here, we keep
$\alpha_{\rho(770)}=1$, $\alpha_{\omega}=1$, $\alpha_{f_{0}(1370)}=1$,  $\alpha_{\phi(1020)}=1$, $\alpha_{K^{*}(892)}=1$, 
$\alpha_{K(1460)}=1$, $\alpha_{K_{0}^{*}(1430)}=1$ (except for the $\tau^{-}\to K^{-}\eta\nu_{\tau}$  and $\tau^{-}\to K^{-}\eta^{'}(958)\nu_{\tau}$ channels, 
which currently only proceed through the $K_{0}^{*}(1430)$. This is signified by $S^{\dagger}$.).
The Branching Ratios are determined from the simulated decay rate normalized by the world-average of the
$\tau$ lepton life-time  \cite{PDG2020} and $\hbar$. The simulated samples have a statistical uncertainty of 0.3\%.
The primary production mechanism for the $\tau^{-}\to K^{-}\eta\nu_{\tau}$, $\tau^{-}\to K^{-}\eta^{'}(958)\nu_{\tau}$  , $\tau^{-}\to K^{-}K^{-}K^{+}\nu_{\tau}$  
and $\tau^{-}\to K^{-}K^{0}\bar{K}^{0}\nu_{\tau}$ include excited resonances which are currently not included in the results presented here.
\label{table:IMRBR}}
\scriptsize
\begin{tabular}{p{2.2cm}p{2.7cm}p{2.7cm}p{2.7cm}p{2.7cm}p{2.9cm}}
\hline
Decay Mode & $\Gamma_{x}/\Gamma_{total}$  & $\Gamma_{x}/\Gamma_{total}$ & $\Gamma_{x}/\Gamma_{total}$ & $\Gamma_{x}/\Gamma_{total}$  &  $\Gamma_{x}/\Gamma_{total}$ PDG  \cite{PDG2020}
 \\
 & $\alpha_{V/S^{\dagger}}=0$ & $\alpha_{V/S^{\dagger}}=1$ &  &   &  \\
\hline
$\tau^{-}\to\pi^{-}\pi^{0}\nu_{\tau}$ & $(2.816)\times10^{-1}$ & $(2.891)\times10^{-1}$ & & &$(25.46\pm0.09)\times 10^{-2}$\\
$\tau^{-}\to K^{-}\pi^{0}\nu_{\tau}$ &  $(4.104)\times10^{-3}$ & $(4.088)\times10^{-3}$ & & &$(4.33\pm0.15)\times 10^{-3}$\\
$\tau^{-}\to \bar{K}^{0}\pi^{-}\nu_{\tau}$ &$(7.933)\times10^{-3}$ & $(7.924)\times10^{-3}$ & &  &$(8.38\pm0.14)\times 10^{-3}$ \\
$\tau^{-}\to K^{-}K^{0}\nu_{\tau}$ & $(2.431)\times10^{-3}$ & $(1.505)\times10^{-3}$ & & &$(1.486\pm0.034)\times 10^{-3}$\\
$\tau^{-}\to K^{-}\eta\nu_{\tau}$ & $(2.115)\times10^{-7}$ & $(2.211)\times10^{-7}$ && &$(1.55\pm 0.08)\times 10^{-4}$ \\
$\tau^{-}\to K^{-}\eta^{'}(958)\nu_{\tau}$ & $(6.231)\times10^{-7}$ & $(6.206)\times10^{-7}$ && &$<(2.4)\times 10^{-6}$ $@90\%$ C.L.\\
\hline
 &\multicolumn{2}{c}{ [$\rho(770)$ + $K^{*}(892)$]} & \multicolumn{2}{c}{[$\rho(770)$ + $K^{*}(892)$ + $f_{0}(1370)$]} & PDG \cite{PDG2020} \\
 & $\alpha_{a_{1}(1260)}=0$  &  $\alpha_{a_{1}(1260)}=1$ &  $\alpha_{a_{1}(1260)}=0$ &  $\alpha_{a_{1}(1260)}=1$ & 
\input{A1_BR_Table.tex}
\hline
 &  $\theta_{K_{1}}=33^{\circ}$,$\delta_{K_{1}}=-0.25$ & $\theta_{K_{1}}=57^{\circ}$,$\delta_{K_{1}}=-0.25$ & 
$\theta_{K_{1}}=33^{\circ}$,$\delta_{K_{1}}=0.25$ & $\theta_{K_{1}}=57^{\circ}$,$\delta_{K_{1}}=0.25$   & PDG \cite{PDG2020}
\input{K1_BR_Table.tex}
\hline
\end{tabular}
\end{center}
\end{table*}

%% file: K1WidthTable.tex
\\ \hline 
$ -33^{\circ} $ & $ 1 $ & 0.1383 & 0.1657 \\ 
$ -45^{\circ} $ & $ 1 $ & 0.1616 & 0.1436 \\ 
$ -57^{\circ} $ & $ 1 $ & 0.1764 & 0.1205 \\ 
$ 33^{\circ} $ & $ 1 $ & 0.0469 & 0.1554 \\ 
$ 45^{\circ} $ & $ 1 $ & 0.0615 & 0.1323 \\ 
$ 57^{\circ} $ & $ 1 $ & 0.0848 & 0.1102 \\ 
$ -33^{\circ} $ & $ -0.25 $ & 0.0383 & 0.1363 \\ 
$ -45^{\circ} $ & $ -0.25 $ & 0.0686 & 0.0979 \\ 
$ -57^{\circ} $ & $ -0.25 $ & 0.1010 & 0.0597 \\ 
$ 33^{\circ} $ & $ -0.25 $ & 0.0612 & 0.1389 \\ 
$ 45^{\circ} $ & $ -0.25 $ & 0.0936 & 0.1007 \\ 
$ 57^{\circ} $ & $ -0.25 $ & 0.1239 & 0.0623 \\ 
$ -33^{\circ} $ & $ 0.25 $ & 0.0612 & 0.1389 \\ 
$ -45^{\circ} $ & $ 0.25 $ & 0.0936 & 0.1007 \\ 
$ -57^{\circ} $ & $ 0.25 $ & 0.1239 & 0.0623 \\ 
$ 33^{\circ} $ & $ 0.25 $ & 0.0383 & 0.1363 \\ 
$ 45^{\circ} $ & $ 0.25 $ & 0.0686 & 0.0979 \\ 
$ 57^{\circ} $ & $ 0.25 $ & 0.1010 & 0.0597 \\ 
\hline 

%% file: A1_BR_Table.tex
\\ \hline 
 $\pi^{-}\pi^{-}\pi^{+}$ & $(9.919)\times 10^{-2}$& $(1.144)\times 10^{-1}$& $(9.675)\times 10^{-2}$& $(1.177)\times 10^{-1}$ &$(9.02\pm0.05)\times 10^{-2}$\\  
 $\pi^{0}\pi^{0} \pi^{-}$ & $(1.009)\times 10^{-1}$& $(1.172)\times 10^{-1}$& $(9.819)\times 10^{-2}$& $(1.204)\times 10^{-1}$ &$(9.26\pm 0.10)\times 10^{-2}$\\  
 $K^{-}\pi^{-} K^{+}$ & $(1.621)\times 10^{-3}$& $(1.547)\times 10^{-3}$& $(1.489)\times 10^{-3}$& $(1.296)\times 10^{-3}$ &$(1.435\pm 0.027)\times 10^{-3}$\\  
 $K^{0}\pi^{-} \bar{K}^{0}$ & $(1.508)\times 10^{-3}$& $(1.440)\times 10^{-3}$& $(1.379)\times 10^{-3}$& $(1.207)\times 10^{-3}$ &$(1.55\pm0.24)\times 10^{-3}$\\  
 $K^{-}\pi^{0} K^{0}$ & $(1.440)\times 10^{-3}$& $(1.364)\times 10^{-3}$& $(1.459)\times 10^{-3}$& $(1.266)\times 10^{-3}$ &$(1.50\pm0.07)\times 10^{-3}$\\  

%% file: K1_BR_Table.tex
\\ \hline 
 $K^{-}\pi^{-}\pi^{+}$ & $(4.488)\times 10^{-3}$& $(3.162)\times 10^{-3}$& $(5.876)\times 10^{-3}$& $(3.534)\times 10^{-3}$ &$(2.93\pm0.07)\times10^{-3}$\\  
 $\pi^{-}K^{0}\pi^{0}$ & $(8.187)\times 10^{-3}$& $(5.905)\times 10^{-3}$& $(1.051)\times 10^{-2}$& $(5.650)\times 10^{-3}$ &$(3.82\pm0.13)\times 10^{-3}$\\  
 $\pi^{0}\pi^{0} K^{-}$ & $(5.473)\times 10^{-4}$& $(3.752)\times 10^{-4}$& $(7.862)\times 10^{-4}$& $(7.761)\times 10^{-4}$ &$(6.5\pm 2.2)\times 10^{-4}$\\  
 $K^{-}K^{-}K^{+}$ & $(3.500)\times 10^{-8}$& $(2.256)\times 10^{-8}$& $(3.222)\times 10^{-8}$& $(1.951)\times 10^{-8}$ &$(2.2\pm0.8)\times 10^{-5}$\\  
 $K^{-}K^{0}\bar{K}^{0}$ & $(2.832)\times 10^{-8}$& $(1.827)\times 10^{-8}$& $(2.612)\times 10^{-8}$& $(1.576)\times 10^{-8}$ &$<6.3 \times 10^{-7}$ $@90\%$ C.L.\\  
 $K^{-}\omega(782)$ & $(2.261)\times 10^{-3}$& $(2.547)\times 10^{-3}$& $(1.972)\times 10^{-3}$& $(1.546)\times 10^{-3}$ &$(4.1\pm0.9) \times 10^{-4}$\\  